\documentclass[aps, floatfix,nofootinbib,amsmath,amssymb,amsfonts,notitlepage, twocolumn,superscriptaddress]{revtex4-2} 

\usepackage{balance}
\usepackage[T1]{fontenc}
\usepackage[latin9]{inputenc}
\usepackage{lmodern} 
\usepackage{color}
\usepackage{bbold}
\usepackage{dsfont}
\usepackage{graphicx}
\usepackage[caption=false]{subfig}
\usepackage[colorlinks]{hyperref}
\usepackage[bottom]{footmisc}
\usepackage{enumerate}
\usepackage{float}
\usepackage{mathtools}
\usepackage{empheq}
\usepackage{tikz}
\usepackage{soul}
\usetikzlibrary{patterns,tikzmark, matrix,decorations.pathreplacing, calc, positioning,fit}

\usepackage{hf-tikz}

\usepackage[normalem]{ulem}
\DeclareMathAlphabet\mbc{OMS}{cmsy}{b}{n}
\usepackage{balance}

\usepackage{diagbox}

\begin{document}

\global\long\def\eqn#1{\begin{align}#1\end{align}}
\global\long\def\vec#1{\overrightarrow{#1}}
\global\long\def\ket#1{\left|#1\right\rangle }
\global\long\def\bra#1{\left\langle #1\right|}
\global\long\def\bkt#1{\left(#1\right)}
\global\long\def\sbkt#1{\left[#1\right]}
\global\long\def\cbkt#1{\left\{#1\right\}}
\global\long\def\abs#1{\left\vert#1\right\vert}
\global\long\def\cev#1{\overleftarrow{#1}}
\global\long\def\der#1#2{\frac{{d}#1}{{d}#2}}
\global\long\def\pard#1#2{\frac{{\partial}#1}{{\partial}#2}}
\global\long\def\re{\mathrm{Re}}
\global\long\def\im{\mathrm{Im}}
\global\long\def\dd{\mathrm{d}}
\global\long\def\ddd{\mathcal{D}}

\global\long\def\avg#1{\left\langle #1 \right\rangle}
\global\long\def\mr#1{\mathrm{#1}}
\global\long\def\mb#1{{\mathbf #1}}
\global\long\def\mc#1{\mathcal{#1}}
\global\long\def\tr{\mathrm{Tr}}

\global\long\def\nth{$n^{\mathrm{th}}$\,}
\global\long\def\mth{$m^{\mathrm{th}}$\,}
\global\long\def\non{\nonumber}

\newcommand{\orange}[1]{{\color{orange} {#1}}}
\newcommand{\cyan}[1]{{\color{cyan} {#1}}}
\newcommand{\teal}[1]{{\color{teal} {#1}}}
\newcommand{\blue}[1]{{\color{blue} {#1}}}
\newcommand{\yellow}[1]{{\color{yellow} {#1}}}
\newcommand{\green}[1]{{\color{green} {#1}}}
\newcommand{\red}[1]{{\color{red} {#1}}}
\newcommand{\purple}[1]{{\color{purple} {#1}}}

\global\long\def\todo#1{\cyan{{$\bigstar$ \orange{\bf\sc #1 }}$\bigstar$} }

\global\long\def\addref#1{\orange{{$\bigstar$ \cyan{\bf\sc Add reference }}$\bigstar$} }

\global\long\def\redflag#1{\Rflag{first} \red{\bf \sc #1}}

\title{Quantum Electrodynamics with Time-varying Dielectrics}
\author{Ashwith Prabhu}
\affiliation{Department of Physics, University of Illinois Urbana-Champaign, Urbana, IL, USA }
\author{Jennifer Parra-Contreras}
\affiliation{Department of Physics, University of Arizona, Tucson, AZ, USA}
\author{Elizabeth A. Goldschmidt}
\affiliation{Department of Physics, University of Illinois Urbana-Champaign, Urbana, IL, USA }
\author{Kanu Sinha}
\affiliation{Wyant College of Optical Sciences and Department of Physics, University of Arizona, Tucson, AZ, USA}

\begin{abstract}
We present a framework for quantization of electromagnetic field in the presence of dielectric media with time-varying optical properties. Considering a microscopic model for the dielectric as a collection of matter fields interacting with the electromagnetic environment, we allow for the possibility of dynamically varying light-matter coupling. We obtain the normal modes of the coupled light-matter degrees of freedom, showing that the corresponding creation and annihilation operators obey equal-time canonical commutation relations. We show that these normal modes can consequently couple to quantum emitters in the vicinity of dynamic dielectric media, and the resulting radiative properties of atoms are thus obtained. Our results are pertinent to time-varying boundary conditions realizable across a wide range of state-of-the-art physical platforms and timescales.
\end{abstract}

\maketitle

\section{Introduction}

The quantization of the electromagnetic field in dielectric environments is central to both fundamental quantum electrodynamics (QED) phenomena and modern quantum photonic devices. The quantum theory of light in media underlies our understanding in areas of study as diverse as nanophotonics\,\cite{NovotnyBook, Damico2019}, plasmonics\,\cite{Tame2013, Bozhevolnyi2017}, quantum fluctuation phenomena\,\cite{Milonni, Buhmann1, Buhmann2} and surface chemistry\,\cite{Flick18, Galego16, Hertzog19}. As a particular example, the need for a complete theory of quantum emitters in and near dielectric interfaces has exploded in recent decades as improvements in photonic engineering have pushed devices into the sub-wavelength regime\,\cite{Damico2019}. Such devices have broad applications that include generating and manipulating quantum light\,\cite{Senellart2017, LodahlRMP2015,Oshea2013,Johns2017, Lodahl2017, Vahala2003}, quantum sensing at the molecular scale\,\cite{Yu2021,Wang2022}, precision tests of fundamental physics\,\cite{Szigetti2020, Blanco2023, Carney2021}, and probing excitations in quantum materials\,\cite{Casola2018}.

Until recently, quantum theories of light in and near media have assumed  their optical response  to be stationary during all dynamics. However, several recent theoretical and experimental developments motivate the need for a detailed understanding of quantum emitters in time-varying media. Advances in photonics and material science have enabled the development of materials that can guide and manipulate electromagnetic waves, in ``unnatural'' ways not achievable with typical dielectrics. Such meta-materials, while diverse in their composition and applications, fundamentally consist of sub-wavelength structures that   
allow for spatial engineering of light waves\,\cite{liu2011metamaterials, chen2016review, zheludev2012metamaterials}. A relatively novel frontier in the avenue of meta-material research is the development of temporally modulated or time-varying dielectrics\,\cite{xiao2020active, engheta2023four, galiffi2022photonics}. A prime example of such exotic dielectrics is epsilon-near-zero (ENZ) meta-materials whose transmission properties can be altered on sub-picosecond timescales\,\cite{maas2013experimental}. Such dielectrics have applications in all-optical switching, non-reciprocal transmission, frequency conversion among others.\,\cite{neira2018all, diroll2016large, xie2020tunable, bohn2021all, sounas2017non, Zhou20}.   

Another exciting candidate for realizing time-varying dielectrics is atom arrays. Ordered atom arrays have been studied extensively lately, both experimentally and theoretically \,\cite{masson2020atomic, srakaew2023subwavelength, nandi2021coherent, rui2020subradiant}. Ordered arrays of a few thousand atoms can behave as mirrors with significant reflectance\,\cite{corzo2016large}. Such atom mirrors have transmission properties akin to bulk dielectrics whilst also retaining their quantum nature, in that the atoms can be driven coherently, allowing the state of atoms and consequently, their transmission properties to be altered on transient time scales\,\cite{su2019observation}. 
. 

Time-independent descriptions of quantum light are insufficient to capture the important features of these and other relevant physical systems. And, to the best of our knowledge, existing treatments of temporally-modulated meta-materials restrict themselves to the classical regime. In this work, we develop a general formalism that allows investigation of QED phenomena in the presence of time-varying dielectrics. As an application of the formalism, we study the spontaneous emission of a two level atom embedded in a time-varying dielectric. This formalism is suitable as a full description of quantum emitters in a dielectric environment with arbitrary temporal variation.

 \section{Quantization in dielectric media}
 Before we delve into time-varying media, it is instructive to briefly survey the QED description of a typical, time-invariant dielectric. In classical electromagnetism, the microscopic details of a dielectric are distilled into its susceptibility. The susceptibility effectively encapsulates the response of the dielectric to the external field  and can be integrated into Maxwell's equations to analyze the classical behavior of the EM field. Studying the quantum mechanical behavior of the EM field requires that we first canonically quantize the field.
Various schemes for quantization of EM field in the presence of dispersive  media  based on a macroscopic description of  dielectric properties have been developed\,\cite{ glauber1991quantum, watson1949phenomenological, drummond1990electromagnetic, deutsch1991paraxial, Raymer2020}.  Dispersion leads to temporal non-locality whose inclusion in the Lagrangian is non-trivial and dissipation results in the collapse of the commutation relations of the quantized field operators. This issue was remedied by representing the dielectric in terms of a harmonic polarization field coupled to a continuum of reservoir degrees of freedom. While such a microscopic model for matter was first described by Fano\,\cite{ PhysRev.103.1202} and Hopfield\,\cite{PhysRev.112.1555}, its application to the quantization of the electromagnetic field was detailed by Huttner and Barnett\,\cite{huttner1992quantization}. In the Huttner-Barnett model, the quantized macroscopic field variables, namely the electric, magnetic, and displacement fields are expressed in terms of a polariton field. This quantized polariton field arises canonically as a product of Fano-diagonalizing the Hamiltonian for the coupled matter and electromagnetic field. The Huttner-Barnett model has demonstrated enduring success due to its ability to hold for all linear, homogeneous, dispersive and dissipative dielectric media. It has been expanded upon to include inhomogeneity, anisotropy and even magnetism\,\cite{suttorp2004fano, kheirandish2008extension}. Alternate approaches for treating quantized fields in absorptive and dispersive media focus on devising Hamiltonians comprising solely of the photonic degrees of freedom with the matter degrees of freedom appearing only in the form of material susceptibility\,\cite{bhat2006hamiltonian, sipe2009photons} and including material properties via phenomenological noise currents using a Langevin-like approach\,\cite{Gruner96, Knoll}. However, until this work, temporal variations of dielectric properties have not been incorporated into such QED models.

\begin{figure}[t]
    \centering
    \includegraphics[width = 0.45\textwidth]{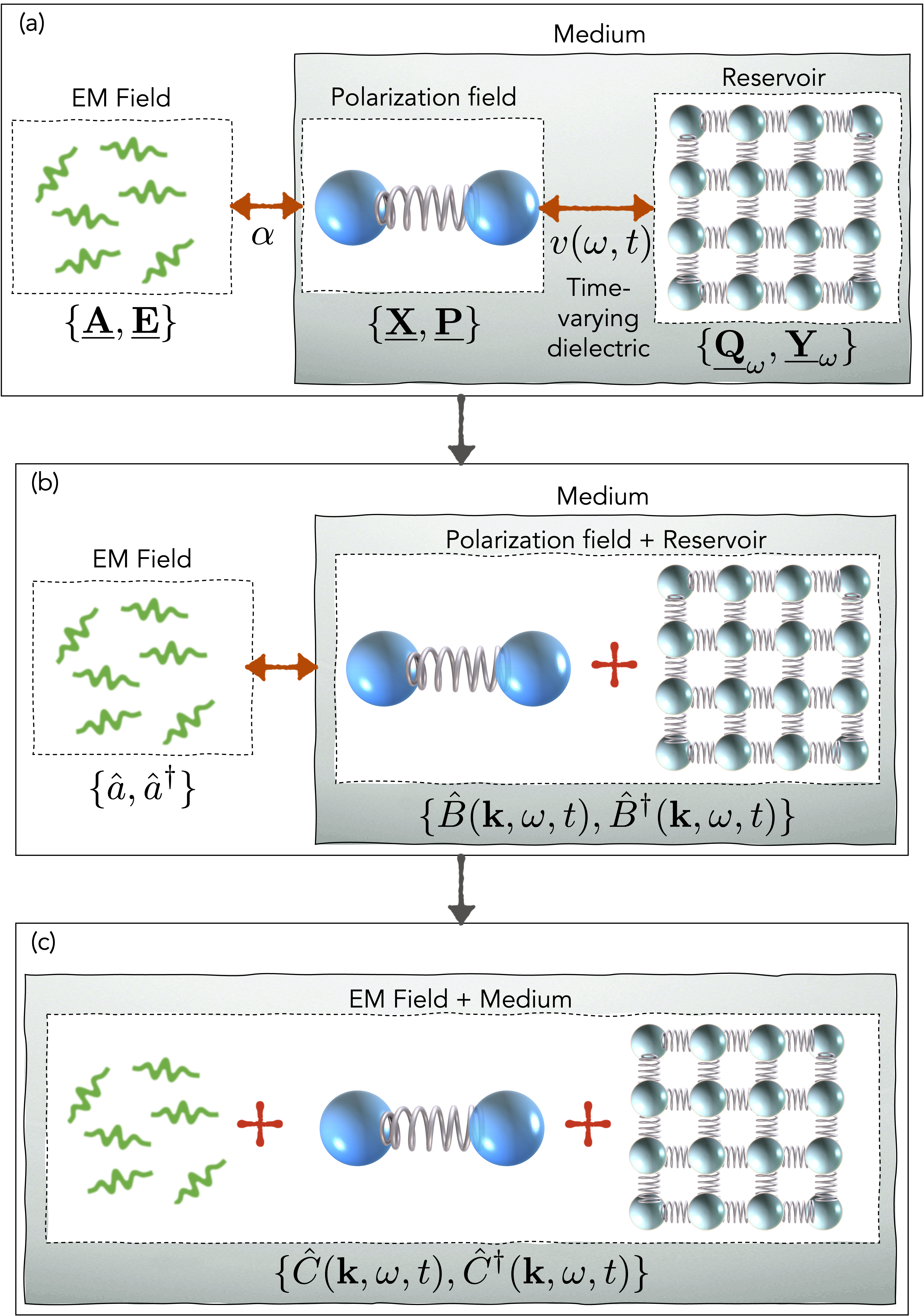}
    \caption{Schematic representation of the model for EM field quantization in a time-varying dielectric. }
    \label{Fig:sch}
\end{figure}

\section{Time-Varying 
Model}
In this section, our analysis closely follows the Huttner-Barnett model\,\cite{huttner1992quantization} and we adopt an identical system of notations. In the Huttner-Barnett model, the dielectric is modeled as a harmonic polarization field coupled to a continuum or reservoir of harmonic matter fields. The EM field interacts solely with the primary polarization field, as shown in Fig.\,\ref{Fig:sch}\,(a). The coupling between the primary polarization field and the reservoir fields ultimately accounts for the dissipation and dispersion induced by the dielectric. We encode the temporal variation of the dielectric into the model by introducing a time dependence on the coupling between the primary polarization field and the reservoir.

\subsection{Lagrangian}
We begin by introducing a Lagrangian for the EM field and the dielectric as modeled above. We opt for the Coulomb gauge for the vector potential since this allows us to decouple the longitudinal and transverse components the EM field\,\cite{huttner1992quantization}. We focus solely on the transverse components since the longitudinal components of the EM fields do not couple to the matter fields. In the interest of brevity, all field vectors are to be interpreted as transverse. The total Lagrangian is decomposed into the electromagnetic (em), matter (mat), reservoir (res), and interaction (int) Lagrangians, respectively, as:
\begin{align}
L= L_\mr{em}+ L_\mr{mat } + L_\mr{res} + L_\mr{int}.
\label{lagrangian}
\end{align}
 We define the Lagrangian density $\underline{\mc{L}}$ such that $ L = \int' \dd^3 k \underline{\mc{L}}$, with the individual components\footnote{ The underline denotes that the quantity occupies the reciprocal space, defined as $
\mathbf{F}(\mathbf{r}, t)=\frac{1}{(2 \pi)^{3 / 2}} \int d^{3} k \underline{\mathbf{F}}(\mathbf{k}, t) e^{i \mathbf{k} \cdot \mathbf{r}}$.  While $\underline{\mathbf{F}}(\mathbf{k}, t)$ is complex, $\mathbf{F}(\mathbf{r}, t)$, being an observable, is constrained to be real such that $\underline{\mathbf{F}}(\mathbf{k}, t)=\underline{\mathbf{F}}^{*}(-\mathbf{k}, t)$. The integration is limited to half reciprocal space, as indicated by the prime over the integral sign, on account of the reality constraint.}:
\eqn{
&\underline{\mathcal{L}}_{\mathrm{em}}=\epsilon_{0}\left(|\dot{\underline{\mathbf{A}}}|^{2}-c^{2} k^{2}|\underline{\mathbf{A}}|^{2}\right) \label{emlagrangian} \\
&\underline{\mathcal{L}}_{\text {mat }} =\left[\rho\left|\dot{\underline{\mathbf{X}}}\right|^{2}-\rho \omega_{0}^{2}\left|\underline{\mathbf{X}}\right|^{2}\right]\label{mainpolarization}\\
&\underline{\mathcal{L}}_{\mathrm{res}}=\int_{0}^{\infty} d \omega\left(\rho\left|\underline{\dot{\mathbf{Y}}}_{\omega}\right|^{2}-\rho \omega^{2}\left|\underline{\mathbf{Y}}_{\omega}\right|^{2}\right) \label{reslang}\\
&\underline{\mathcal{L}}_{\mathrm{int}}=-\left[\alpha \underline{\mathbf{A}} \cdot \dot{\underline{\mathbf{X}}}^{*}+\int_{0}^{\infty}d\omega\,v(\omega, t) \underline{\mathbf{X}}^{ *} \cdot \underline{\dot{\mathbf{Y}}}_{\omega}+\text { c.c. }\right].\label{interactionlagrangian}
}
The EM Lagrangian density (Eq.\,\eqref{emlagrangian}) governs the free evolution of the EM field where $\underline{\mathbf{A}}$ denotes the magnetic vector potential. The Lagrangian density for the matter field (Eq.\,\eqref{mainpolarization}) depends on a harmonic dipole moment density field $\underline{\mathbf{X}}$ resonant at frequency $\omega_0$ with $\rho$ serving as the effective charge of the polarization. The reservoir Lagrangian density (Eq. \eqref{reslang}) can be considered as a continuum of harmonic polarization fields $\{Y_\omega\}_\omega$ coupled to the polarization field as depicted in Fig.\,\ref{Fig:sch}\,(a). The interaction Lagrangian (Eq.\,\eqref{interactionlagrangian}) accounts for the interaction between the magnetic vector potential $\underline{\mathbf{A}}$ and the polarization field via the coupling constant $\alpha$. It also includes the time-varying coupling between the polarization field and the reservoir via the coupling factor $v(\omega, t)$ that engenders the time-varying nature of the dielectric.

In order to make the model as general as possible we restrict ourselves to the minimal assumptions required to ensure causality and the existence of a finite frequency cutoff of the material response (see \ref{diagonalizehamiltonian}). These assumptions are that the analytical continuation of $|v(\omega, t)|^2$ to negative frequencies be an even function of frequency  and that $v(\omega, t)$ be square integrable over frequency at all times. Besides these two constraints, we make no assumptions about the timescales or the physical origin of the temporal dependence of the coupling factor, thereby allowing the model to be as general as possible.

\subsection{Hamiltonian and quantization}
\label{quantization}
In order to derive the Hamiltonian for the system, we start by defining the conjugate momenta corresponding to the EM and medium degrees of freedom. The EM and medium field vectors can be decomposed along any two orthogonal polarization directions denoted by $\lambda$, such that:
\begin{align}
-\epsilon_{0} \underline{E}^{\lambda} & \equiv \frac{\partial \underline{\mathcal{L}}}{\partial\underline{\dot{A}}^{\lambda *}}=\epsilon_{0} \underline{\dot{A}}^{\lambda}, \\
\underline{P}^{\lambda} & \equiv \frac{\partial \underline{\mathcal{L}}}{\partial \underline{\dot{X}}^{\lambda *}}=\rho \underline{\dot{X}}^{\lambda}-\alpha \underline{A}^{\lambda}, \\
\underline{Q}_{\omega}^{\lambda} & \equiv \frac{\partial \underline{\mathcal{L}}}{\partial \underline{\dot{Y}}_{\omega}^{\lambda *}}=\rho \underline{\dot{Y}}_{\omega}^{\lambda}-v(\omega, t) \underline{X}^{\lambda} .
\end{align}
Subjecting the Lagrangian in Eq.\,\eqref{lagrangian} to the Legendre transform with the aid of the conjugate variables, we arrive at the Hamiltonian.
\begin{align}
H=\int^{\prime} d^{3} k\left(\underline{\mathcal{H}}_{\mathrm{em}}+\underline{\mathcal{H}}_{\text {mat }}+\underline{\mathcal{H}}_{\mathrm{int}}\right)\label{hamiltonian}
\end{align}
where
\eqn{
&\underline{\mathcal{H}}_{\mathrm{em}}=\left[\epsilon_{0}|\underline{\mathbf{E}}|^{2}+\epsilon_{0} c^{2} \tilde{k}^{2}|\underline{\mathbf{A}}|^{2}\right]\\
&\underline{\mathcal{H}}_{\text {mat }}=\frac{|\underline{\mathbf{P}}|^{2}}{\rho}+\rho \tilde{\omega}_{0}^{2}(t)|\underline{\mathbf{X}}|^{2}+\int_{0}^{\infty} d \omega\non \\
&\sbkt{\frac{\left|\underline{\mathbf{Q}}_{\omega}\right|^{2}}{\rho}+\rho \omega^{2}\left|\underline{\mathbf{Y}}_{\omega}\right|^{2}+\frac{v(\omega, t)}{\rho}\bkt{\underline{\mathbf{X}}^{*} \cdot \underline{\mathbf{Q}}_{\omega}+\text { c.c. }}}\\
&\underline{\mathcal{H}}_{\text {int }}=\frac{\alpha}{\rho}\left(\underline{\mathbf{A}}^{*} \cdot \underline{\mathbf{P}}+\text{ c.c. }\right)
}
In the above set of equations, $\tilde{\omega}_0(t)$ is the time-dependent renormalized frequency of the harmonic polarization field, defined as
$\tilde{\omega}_{0}^{2}(t) \equiv \omega_{0}^{2}+\int_{0}^{\infty} d \omega \frac{|v(\omega,t)|^{2}}{\rho^{2}}$. The quantity, $\int_{0}^{\infty} d \omega \frac{|v(\omega,t)|^{2}}{\rho^{2}}$ is a dispersive shift arising from the coupling of the polarization field to the reservoir. Similarly, $\tilde{k}$ is the renormalized wave-vector of the EM field, defined as $\tilde{k}=\sqrt{k^{2}+\omega_c^{2}/c^2}$ where $\omega_{c}=\sqrt{\alpha^{2} / \rho \epsilon_{0}}$ is the dielectric induced dispersive shift.

We impose equal time commutation relations (ETCR) on the three pairs of conjugate variables, thereby elevating them to operators, with the result:
\begin{align}
   & \left[\underline{\hat{A}}^{\lambda}(\mathbf{k}, t), \underline{\hat{E}}^{\lambda^{\prime} *}\left(\mathbf{k}^{\prime}, t\right)\right]=-\frac{i \hbar}{\epsilon_{0}} \delta_{\lambda \lambda^{\prime}} \delta\left(\mathbf{k}-\mathbf{k}^{\prime}\right)\\
      & \left[\underline{\hat{X}}^{\lambda}(\mathbf{k}, t), \underline{\hat{P}}^{\lambda^{\prime} *}\left(\mathbf{k}^{\prime}, t\right)\right]=i \hbar \delta_{\lambda \lambda^{\prime}} \delta\left(\mathbf{k}-\mathbf{k}^{\prime}\right) \\
&\left[\underline{\hat{Y}}_{\omega}^{\lambda}(\mathbf{k}, t), \underline{\hat{Q}}_{\omega^{\prime}}^{\lambda^{\prime} *}\left(\mathbf{k}^{\prime}, t\right)\right]=i \hbar \delta_{\lambda \lambda^{\prime}} \delta\left(\mathbf{k}-\mathbf{k}^{\prime}\right) \delta\left(\omega-\omega^{\prime}\right)
\end{align}
 We introduce bosonic annihilation and creation operators for each pair of conjugate operators, as follows:
\begin{align}
   &\hat{a}(\lambda, \mathbf{k}, t)=\sqrt{\frac{\epsilon_{0}}{2 \hbar \tilde{k} c}}\left[\tilde{k} c \underline{\hat{A}}^{\lambda}(\mathbf{k}, t)-i \underline{\hat{E}}^{\lambda}(\mathbf{k}, t)\right] \\
&\hat{b}(\lambda, \mathbf{k}, t)=\sqrt{\frac{\rho}{2 \hbar \tilde{\omega}_{0}(t)}}\left[\tilde{\omega}_{0}(t) \underline{\hat{X}}^{\lambda}(\mathbf{k}, t)+\frac{i}{\rho} \underline{\hat{P}}^{\lambda}(\mathbf{k}, t)\right] \label{timedependent}\\
&\hat{b}_{\omega}(\lambda, \mathbf{k}, t)=\sqrt{\frac{\rho}{2 \hbar \omega}}\left[-i \omega \underline{\hat{Y}}_{\omega}^{\lambda}(\mathbf{k}, t)+\frac{1}{\rho} \underline{\hat{Q}}_{\omega}^{\lambda}(\mathbf{k}, t)\right]  
\end{align}
Note the explicit time dependence of the annihilation operator in Eq.\,\eqref{timedependent}, corresponding to the harmonic polarization field of the dielectric. Having imposed ETCR, the Hamiltonian in 
Eq.\,\eqref{hamiltonian} is now an operator and can be expressed in terms of the annihilation and creation operators. 
\begin{widetext}
\eqn{
   \hat{H}_{\mathrm{em}}=& \int d^{3} k \sum_{\lambda=1,2} \hbar \tilde{k} c \hat{a}^{\dagger}(\lambda, \mathbf{k}, t) \hat{a}(\lambda, \mathbf{k}, t)\\
   \hat{H}_{\text{mat}}=& \int d^{3} k \sum_{\lambda=1,2}\left[\hbar \tilde{\omega}_{0}(t) \hat{b}^{\dagger}(\lambda, \mathbf{k}, t) \hat{b}(\lambda, \mathbf{k}, t)+\int_{0}^{\infty} d \omega \hbar \omega \hat{b}_{\omega}^{\dagger}(\lambda, \mathbf{k}, t) \hat{b}_{\omega}(\lambda, \mathbf{k}, t)\right.\\
&\left.\quad+\frac{\hbar}{2} \int_{0}^{\infty} d \omega\, V(\omega,t)\left[\hat{b}^{\dagger}(\lambda,-\mathbf{k}, t)+\hat{b}(\lambda, \mathbf{k}, t)\right]\left[\hat{b}_{\omega}^{\dagger}(\lambda,-\mathbf{k}, t)+\hat{b}_{\omega}(\lambda, \mathbf{k}, t)\right]\right] \label{matterhamiltonian} \\
\hat{H}_{\text {int }}=&i \frac{\hbar}{2} \int d^{3} k \sum_{\lambda=1,2} \Lambda(k,t)\left[\hat{a}^{\dagger}(\lambda,-\mathbf{k}, t)+\hat{a}(\lambda, \mathbf{k}, t)\right]\left[\hat{b}^{\dagger}(\lambda,-\mathbf{k}, t)-\hat{b}(\lambda, \mathbf{k}, t)\right]
}
\end{widetext}
where $V(\omega,t) \equiv[v(\omega,t) / \rho] \sqrt{\omega / \tilde{\omega}_{0}(t)}$  and $\Lambda(k,t) \equiv \sqrt{\tilde{\omega}_{0}(t)  \omega_{c}^{2} / c\tilde{k}}$. In the following section, we drop the polarization index $\lambda$ and the summation over the index for convenience. It will be reintroduced in the final results of the section.
\subsection{Hamiltonian Diagonalization\label{diagonalizehamiltonian}}

\subsubsection{Diagonalizing the matter Hamiltonian}\label{sec:Diag_matterH}

The matter field of the dielectric medium was artificially separated into a primary polarization field and a reservoir in order to incorporate dispersion and time variance. However, we would like to express the dielectric in terms of a single set of normal modes, as depicted in Fig.\,\ref{Fig:sch}(b). This is achieved by Fano-diagonalizing  the matter Hamiltonian in Eq.\,\eqref{matterhamiltonian} \,\cite{PhysRev.103.1202}. We introduce the dressed operator $\hat{B}(\mathbf{k}, \omega,t)$:
\begin{align}
\hat{B}(\mathbf{k}, \omega,t)=& \alpha_{0}(\omega, t) \hat{b}(\mathbf{k},t)+\beta_{0}(\omega, t) \hat{b}^{\dagger}(-\mathbf{k},t)\nonumber \\
&+\int_{0}^{\infty} d \omega^{\prime}\left[\alpha_{1}\left(\omega, \omega^{\prime}, t\right) \hat{b}_{\omega^{\prime}}(\mathbf{k}, t)\right.\nonumber\\
&\left.+\beta_{1}\left(\omega, \omega^{\prime}, t\right) \hat{b}_{\omega^{\prime}}^{\dagger}(-\mathbf{k}, t)\right]\label{diagonalization}
\end{align}
requiring that it satisfies the following condition:
\eqn{
   \sbkt{\hat{B}(\mathbf{k}, \omega,t), \hat{H}_{\text {mat }}}=\hbar \omega \hat{B}(\mathbf{k}, \omega,t),\label{eigen}
}
thereby diagonalizing the matter Hamiltonian such that $\hat{H}_{\text {mat }}=\int d^{3} k \int_{0}^{\infty} d \omega\, \hbar \omega \hat{B}^{\dagger}(\mathbf{k}, \omega,t) \hat{B}(\mathbf{k}, \omega,t)$.

The dressed operators, being bosonic annihilation and creation operators, are also required to satisfy the equal time commutation relation (ETCR):
\begin{align}
\left[\hat{B}(\mathbf{k}, \omega,t), \hat{B}^{\dagger}\left(\mathbf{k}^{\prime}, \omega^{\prime},t\right)\right]=\delta\left(\mathbf{k}-\mathbf{k}^{\prime}\right) \delta\left(\omega-\omega^{\prime}\right)
\end{align}
The above ETCR coupled with Eq.\,\eqref{eigen} determines the time-varying coefficients  of expansion $\{\alpha_{0}\bkt{\omega, t},\beta_{0} \bkt{\omega, t}\}$ and $\{\alpha_1 (\omega, \omega', t), \beta_1 (\omega, \omega', t)\}$ as derived in Appendix\,\ref{App:Fano}. 
The total Hamiltonian thus bears the form:
\begin{widetext}
\begin{align}
    \hat{H}=\int d^{3} k\bigg\{\hbar \tilde{k} c \hat{a}^{\dagger}(\mathbf{k},t) \hat{a}(\mathbf{k},t)+\int_{0}^{\infty} d \omega &\hbar \omega \hat{B}^{\dagger}(\mathbf{k}, \omega,t) \hat{B}(\mathbf{k}, \omega,t)\nonumber\\&+\frac{\hbar k_c}{2} \sqrt{\frac{c}{\tilde{k}}} \int_{0}^{\infty} d \omega\left\{\zeta(\omega,t) \hat{B}^{\dagger}(\mathbf{k}, \omega,t)\left[\hat{a}(\mathbf{k},t)+\hat{a}^{\dagger}(-\mathbf{k},t)\right]+\text { H.c. }\right\}\bigg\}
\end{align}
\end{widetext}
where the time-varying coupling between the light and matter field is defined by the factor \eqn{\zeta(\omega,t)=i\sqrt{\tilde{\omega}_0(t)}\left(\alpha_0(\omega, t)+\beta_0(\omega, t)\right).\label{zetadef}}

\subsubsection{Diagonalizing the total Hamiltonian\label{diagonalizetotal}}
The total Hamiltonian is subjected to the same Fano-diagonalization procedure as $\hat{H}_{\text{mat}}$. The ensuing dressed operator $\hat{C}(\mathbf{k},\omega,t)$ is a hybrid between the matter field and the EM field operator, as illustrated in Fig.\,\ref{Fig:sch}(c). We therefore, identify $\hat{C}(\mathbf{k},\omega,t)$ as a polariton field:
\begin{align}
    \hat{C}(\mathbf{k}, \omega, t)=& \tilde{\alpha}_{0}(k, \omega, t) \hat{a}(\mathbf{k},t)+\tilde{\beta}_{0}(k, \omega, t) \hat{a}^{\dagger}(-\mathbf{k},t) \nonumber\\
+& \int_{0}^{\infty} d \omega^{\prime}\left[\tilde{\alpha}_{1}\left(k, \omega, \omega^{\prime},t\right) \hat{B}\left(\mathbf{k}, \omega^{\prime},t\right)\right. \nonumber\\
&\left.+\tilde{\beta}_{1}\left(k, \omega, \omega^{\prime},t\right) \hat{B}^{\dagger}\left(-\mathbf{k}, \omega^{\prime},t\right)\right].
\end{align}
The time-varying coefficients are determined by imposing the bosonic ETCR on $\hat{C}(\mathbf{k},\omega,t)$ and its Hermitian conjugate and also demanding that the polariton field satisfy an equation analogous to Eq.\,\eqref{eigen}, as shown in Appendix \ref{App:Polariton}. This establishes the completeness of the polaritonic operators.

The total Hamiltonian is thus fully diagonalized and assumes the form:
\begin{align}
    \hat{H}=\int d^{3} k \int_{0}^{\infty} d \omega \hbar \omega \hat{C}^{\dagger}(\mathbf{k}, \omega, t) \hat{C}(\mathbf{k}, \omega, t).
\end{align}

Due to the time-varying nature of the dielectric, the temporal evolution of the polariton field is not harmonic, as is evident from the Heisenberg equation:
\begin{align}
 \frac{d\,\hat{C}(\mathbf{k},\omega, t)}{dt}
 &=\frac{1}{i\hbar}[\hat{C}(\mathbf{k},\omega, t), \hat{H}]+ \frac{\partial\,\hat{C}(\mathbf{k},\omega, t)}{\partial t}\nonumber.\\
\end{align}
This anharmonic temporal evolution precludes an  explicit single-frequency representation for the electromagnetic field operators, as is typical for time-varying systems\,\cite{claasen1982stationary, kozek1991time}. In the absence of an explicit frequency domain representation, the electric field and displacement field are only implicitly related, as will be  evident below. 
The annihilation operator for the bare field, $\hat a(\mathbf{k},t)$, can be expressed purely in terms of the polaritonic operators $\hat{C}$ and $\hat{C}^\dagger$, 
\begin{align}\hat{a}(\mathbf{k}, t)=\int_{0}^{\infty} d \omega\bigg[\tilde{\alpha}_{0}^{*}&(k, \omega, t) \hat{C}( \mathbf{k}, \omega, t)\nonumber\\&-\tilde{\beta}_{0}(k, \omega,t) \hat{C}^{\dagger}(-\mathbf{k}, \omega,t)\bigg],\end{align}
we arrive at the electric and magnetic field operators:
\begin{widetext}
\begin{align}
  \hat{\mathbf{A}}(\mathbf{r}, t)&=\frac{1}{(2 \pi)^{3 / 2}} \int d^{3} k \sqrt{\frac{\hbar \omega_c^2}{2 \epsilon_{0}}} \sum_{\lambda=1,2} \mathbf{e}_{\lambda}(\mathbf{k}) \int_{0}^{\infty} d \omega\left[\frac{\zeta^{*}(\omega,t)}{\omega^{2} \epsilon(\omega,t)-k^{2} c^{2}} \hat{C}(\mathbf{k}, \omega,t) e^{i\mathbf{k} \cdot \mathbf{r}}+\text { H.c. }\right]\label{eq0}\\
 \hat{\mathbf{E}}(\mathbf{r}, t)&=\frac{i}{(2 \pi)^{3 / 2}} \int d^{3} k \sqrt{\frac{\hbar \omega_c^2}{2 \epsilon_{0}}} \sum_{\lambda=1,2} \mathbf{e}_{\lambda}(\mathbf{k}) \int_{0}^{\infty} d \omega\left[\frac{\omega \zeta^{*}(\omega, t)}{\omega^{2} \epsilon(\omega, t)-k^{2} c^{2}} \hat{C}(\mathbf{k}, \omega, t) e^{i\mathbf{k} \cdot \mathbf{r}}-\text { H.c. }\right]\label{eq1}\\
 \hat{\mathbf{B}}(\mathbf{r}, t)&=\frac{i}{(2 \pi)^{3 / 2}} \int d^{3} k \sqrt{\frac{\hbar \omega_c^2}{2 \epsilon_{0}}} \sum_{\lambda=1,2} \mathbf{k} \times \mathbf{e}_{\lambda}(\mathbf{k}) \int_{0}^{\infty} d \omega\left[\frac{\zeta^{*}(\omega,t)}{\omega^{2} \epsilon(\omega,t)-k^{2} c^{2}} \hat{C}(\mathbf{k}, \omega,t) e^{i\mathbf{k} \cdot \mathbf{r}}-\text { H.c. }\right]\label{eq2}\\
  \hat{\mathbf{D}}(\mathbf{r}, t)&=\frac{i}{(2 \pi)^{3 / 2}} \int d^{3} k \sqrt{\frac{\hbar \omega_c^2}{2 \epsilon_{0}}} \sum_{\lambda=1,2} \mathbf{e}_{\lambda}(\mathbf{k}) \int_{0}^{\infty} d \omega\left[\epsilon_{0} \epsilon(\omega, t) \frac{\omega \zeta^{*}(\omega,t)}{\omega^{2} \epsilon(\omega,t)-k^{2} c^{2}} \hat{C}(\mathbf{k}, \omega,t) e^{i\mathbf{k} \cdot \mathbf{r}}\right.\nonumber \\
 & \left.-\epsilon_{0} \frac{\zeta^{*}(\omega,t)}{\omega} \hat{C}(\mathbf{k}, \omega,t) e^{i\mathbf{k} \cdot \mathbf{r}}-\text { H.c. }\right]\label{eq3}
\end{align}
\end{widetext}
Here $\mathbf{e}_{\lambda}(\mathbf{k})$ is the polarization unit vector. We see a quantity emerge canonically from the microscopic model that can be identified as the time-varying relative permittivity:

\begin{align}
\epsilon(\omega,t)=1+\frac{\omega_c^2}{2 \omega}\left[\mathrm{P} \int_{-\infty}^{\infty} d \omega^{\prime} \frac{|\zeta\left(\omega^{\prime},t\right)|^2}{(\omega^{\prime}-\omega)\omega^\prime}+i \pi \frac{|\zeta(\omega,t)|^2}{\omega}\right]. \label{epsilon}
\end{align}
 In order for our interpretation of $\epsilon(\omega,t)$ as the time-varying relative permittivity to hold it must satisfy the causality condition for time-varying systems\,\cite{solis2021functional}. The Fourier transform of $\epsilon(\omega,t)$ with respect to $\omega$ yields the relative permittivity in the time-domain, $\epsilon(\tau,t)$, which can be considered as the response of the medium at time $t$ to the polariton field at time $t-\tau$. For causality to hold, $\tau$ should be positive and $\epsilon(\omega,t)$ should be real. In the frequency domain, this implies that analytic continuation of $\epsilon(\omega,t)$ in the complex plane should be analytic (i.e. no poles) in the upper half plane and that the real and imaginary parts of $\epsilon(\omega,t)$ should be even and odd functions of $\omega$ respectively. The first condition is automatically satisfied since the real and imaginary parts are related through the Hilbert transform in Eq.\,\eqref{epsilon}. The reality constraint is fulfilled by demanding that the analytic continuation of the reservoir-polarization coupling factor, $v(\omega, t)$ to negative frequencies be an even function of frequency, as was previously  stated. 

  The scheme detailed in this section formally demonstrates the canonical quantization procedure for linear time-varying dielectrics. However, it is unwieldy and unnecessary to suffer the details of the microscopic model in most cases. The evolution of the field operators depends solely on the time-varying relative permittivity and derived quantities.  Thus, the effect of the time-varying dielectric on the EM fields can be distilled into the time-varying relative permittivity, as we will consider below.

\subsection{Correspondence to Drude-Lorentz model}
\label{Sec:DL}
Our model is general enough to support a time-varying complex susceptibility description of any linear dielectric medium. As a demonstration of our formalism, we apply it to a minimal model of a dispersive and dissipative medium in the form of a Drude-Lorentz dielectric with time-varying damping factor $\gamma (t)$:
\eqn{
\epsilon(\omega,t)=1+\frac{\omega_{p}^{2}}{\omega_{T}^{2}-\omega^2-i\gamma(t) \omega},
}
where  $\omega_{T}$ and $\omega_p$ are the transverse resonant frequency and plasma frequency, respectively.

We consider the coupling factor $v(\omega, t)$ in our model to assume the form: 
\begin{align}
    |v(\omega, t)|^2=\frac{2\gamma(t)\rho^2}{\pi}\Theta(\omega-\Lambda)\Theta(\omega+\Lambda),
\end{align}
where $\Theta$ is the Heaviside function and the coupling factor has a high frequency cutoff, $\Lambda$. The cutoff frequency is essential to ensure that intermediate quantities in the microscopic model such as $\tilde{\omega}_0(t)$ do not diverge.

The above coupling yields the following effective permittivity, as illustrated in Appendix\,\ref{App:DL}:
\begin{align}
\label{eq:epsdl}
\epsilon(\omega,t)=1+\frac{\omega_{c}^{2}}{\omega_{0}^{2}-\omega^2-i\gamma(t) \omega}\Theta(\omega-\Lambda)\Theta(\omega+\Lambda).
\end{align}
Ignoring the frequency cutoff, and identifying $ \omega_c $ as the plasma frequency and $\omega_0$ as the transverse resonant frequency of the dielectric,  we see that with an appropriate choice of the coupling factor, $v(\omega,t)$, our model can emulate the case of a time-varying Drude-Lorentz dielectric. 

\section{Spontaneous emission and Level Shift in a Time-Varying Dielectric}

\begin{figure}[t]
    \includegraphics[width = 0.4\textwidth]{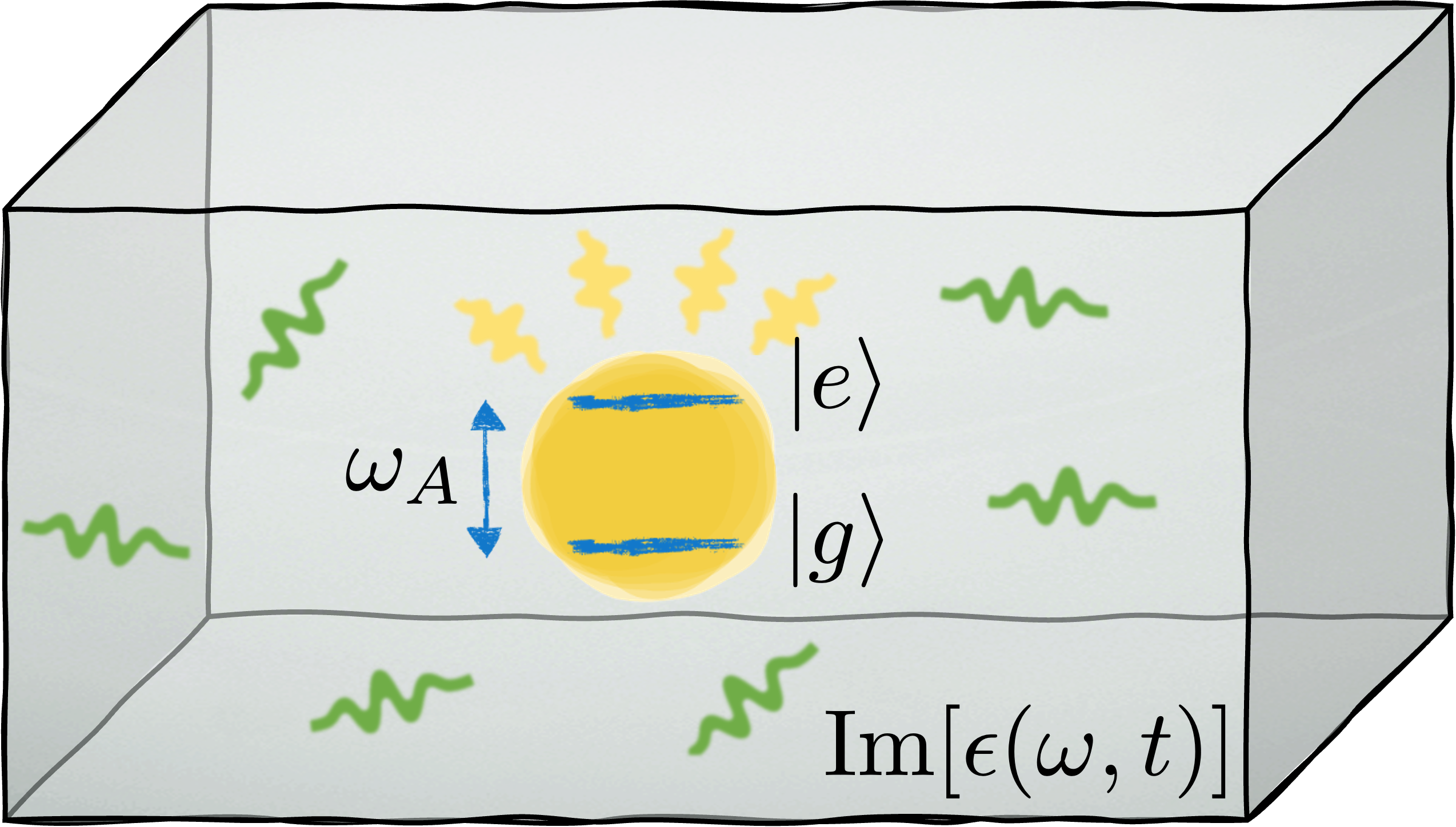}
    \caption{Schematic representation of an excited two-level atom spontaneously emitting in a time-varying dielectric medium.}
    \label{Fig:SpEm}
\end{figure}

 The spontaneous emission rate of an emitter is influenced by its electromagnetic environment\,\cite{purcell1995spontaneous}. The presence of a dielectric medium alters the electromagnetic environment of an emitter, thereby modifying its spontaneous emission rate.  It is  intuitive to anticipate a time-varying dielectric to modify the emission rates of any emitter in its vicinity in a non-trivial fashion depending on the nature of the temporal modulation of the dielectric.

 Since we are dealing with a time-varying system, we cannot rely on Fermi's golden rule to arrive at the spontaneous emission rate and instead, investigate the complete dynamics of the atom and field evolution\,\cite{dirac1927quantum}. The Hamiltonian for the atom-dielectric system is given by:

\begin{align}
\hat{H} =&\hbar\omega_A\hat{\sigma}^\dagger\hat{\sigma}\non\\
&+\sum_{\lambda=1,2}\int{d^3\mb{k}}\int_0^\infty d\omega\,\hbar\omega\,\hat{C}^\dagger_\lambda(\mathbf{k}, \omega, t)\hat{C}_\lambda(\mathbf{k}, \omega, t)\non\\
&-\frac{e}{m}\mathbf{\hat{p}}\cdot\mathbf{\hat{A}}(0,t)
\end{align}
We have dropped the $\mathbf{\hat{A}}^2$ term and adopted the minimal coupling form for the Hamiltonian \,\cite{PhysRevA.11.814}. The ground and excited states of the two level atom, denoted by $\ket{g}$ and $\ket{e}$ respectively are separated in energy by $\hbar\omega_A$. The atomic lowering  operator, $\hat{\sigma}$ is defined as $\hat{\sigma}\equiv\ket{g}\bra{e}$. The first term in the Hamiltonian governs the free evolution of the atom. The second term dictates the dynamics of the polariton field, i.e., the joint EM field and dielectric matter field system. The final term, which we will refer to as $\hat{H}_{AF}$ governs the interaction between the atom and the EM field under the dipole approximation, with $\hat{\mathbf{p}}$ denoting the linear momentum operator of the atom.

Expressing the magnetic vector potential operator in terms of the polaritonic operators as in Eq.\,\eqref{eq0}, we arrive at the following form for the atom-field coupling Hamiltonian.
\begin{align}
\hat{H}_{AF}=&\frac{i\omega_A}{(2\pi)^{3/2}}\sqrt{\frac{\hbar\omega_c^2}{2\epsilon_0}}\sum_{\lambda=1,2}\int d^3\mb{k}\,\mathbf{d}\cdot\mathbf{e}_{\mathbf{k}, \lambda}\int_0^\infty d\omega\, \non\\
&(\hat{\sigma}-\hat{\sigma}^\dagger)\left[\frac{\zeta^*(\omega,t)}{\omega^2\epsilon(\omega,t)-k^2c^2}\hat{C}_\lambda(\mathbf{k},\omega,t)+H.c.\right] 
\end{align}
Here, $\mb{d}$ is the transition dipole moment which has been assumed to be real and $\{\mathbf{e}_{\mathbf{k},\lambda}\}_\lambda$ are the unit polarization vectors in the transverse plane. We define the coupling strength between the atom and the polaritonic modes as:
\begin{align}
g_\lambda(\mathbf{k}, \omega, 
 t)&
 \equiv\frac{\omega_A}{\hbar(2\pi)^{3/2}}\sqrt{\frac{\hbar\omega_c^2}{2\epsilon_0}}\mb{d}\cdot\mathbf{e}_{\mathbf{k}, \lambda}\frac{\zeta(\omega,t)}{\omega^2\epsilon^*(\omega,t)-k^2c^2}\nonumber \\
 &=\frac{\omega_A}{2\sqrt{2\hbar\epsilon_0}\pi^2}\mb{d}\cdot\mathbf{e}_{\mathbf{k}, \lambda}\frac{\omega\sqrt{\text{Im}(\epsilon(\omega,t))}}{\omega^2\epsilon^*(\omega,t)-k^2c^2},\label{condensed} 
\end{align}

Applying the rotating wave approximation and using normal ordering of operators, the atom-field coupling Hamiltonian simplifies to:

\eqn{    H_{AF}=i\sum_{\lambda=1,2}\int d^{3}\mb{k}&\int_{0}^{\infty}d\omega\left[\hbar g_\lambda(\mb{k},\omega,t)\hat{C}_\lambda^\dagger(\mb{k},\omega,t)\hat{\sigma}\right.\non\\
    &\left.-\hbar g_\lambda^{*}(\mb{k},\omega,t)\hat{\sigma}^\dagger\hat{C}_\lambda(\mb{k},\omega,t)\right]
}

We examine the dynamics of the polaritonic operator:
\begin{align}
    \frac{d}{dt}\hat{C}_\lambda(\mathbf{k}, \omega, t)=&-i\omega\hat{C}_\lambda(\mathbf{k},\omega, t)+g_\lambda(\mathbf{k},\omega,t)\hat{\sigma}\non\\
    &+\frac{\partial}{\partial{t}}\hat{C}_\lambda(\mathbf{k},\omega,t)\label{eq60}
\end{align}
At this point, in order to account for physical realizability, we make certain assumptions regarding the relative rates of the different temporal processes involved. Assuming the two level atom to be resonant at an optical frequency, we are interested in the regime: $\omega_A \gg \dot{\epsilon}\gg \Gamma_A$, where $\Gamma_A$ is the spontaneous decay rate of the atom in vacuum. Dielectric modulation rate comparable or higher than optical frequencies cannot reasonably be achieved and for modulation rates slower than $\Gamma_A$, an adiabatic treatment is sufficient.  The temporal modulation of the dielectric of interest is, therefore, in the rapid adiabatic regime; rapid with respect to the free space spontaneous decay rate but adiabatic with respect to the resonant frequency of the atom. It can readily be shown that $\frac{\partial}{\partial{t}}\hat{C}_\lambda(\mathbf{k},\omega, t)\approx \frac{d\tilde{\omega}_0(t)/dt}{\tilde{\omega}_0(t)}\hat{C}_\lambda(\mathbf{k},\omega,t)$. In this rapid adiabatic regime, it follows that $\frac{\partial}{\partial{t}}\hat{C}_\lambda(\mathbf{k},\omega, t)\ll \omega \hat{C}_\lambda(\mathbf{k},\omega,t)$, allowing us to drop the partial derivative term in Eq.\,\eqref{eq60}, yielding:
\begin{align}
\hat{C}_\lambda(\mathbf{k},\omega,t)=&\hat{C}_\lambda(\mathbf{k},\omega,0)e^{-i\omega t}\nonumber\\
&+\hat{\sigma}(t)\int_0^t dt^\prime g_\lambda(\mathbf{k},\omega, t^\prime)e^{-i(\omega-\omega_A)(t-t^\prime)}\label{polaritonevolve}
\end{align}

The spontaneous emission rate can be determined by solving for the dynamics of the atomic operator $\sigma^z\equiv \sigma^\dagger\sigma-\sigma\sigma^\dagger$:

\begin{widetext}
\begin{align}
   &\frac{d}{dt}\hat{\sigma}^{z}(t)
   =-2\sum_{\lambda=1,2}\int {d^{3}\mb{k}}\int_{0}^{\infty}d\omega\left[ g_\lambda(\mb{k},\omega,t)\hat{C}_\lambda^\dagger(\mb{k},\omega,0)\hat{\sigma}\,e^{i\omega t}   +g_\lambda^{*}(\mb{k},\omega,t)\hat{\sigma}^\dagger\hat{C}_\lambda(\mb{k},\omega,0)\,e^{-i\omega t}\right]\nonumber\\
    &-(\hat{\mathbb{1}}+\hat{\sigma}^z)\sum_{\lambda=1,2}\int {d^3\mb{k}}\int_0^\infty\,d\omega\int_0^t dt^\prime\left[g_\lambda(\mathbf{k},\omega, t)g^*_\lambda(\mathbf{k},\omega, t^\prime)\,e^{i(\omega-\omega_A)(t-t^\prime)}+g^*_\lambda(\mathbf{k},\omega,t)g_\lambda(\mathbf{k},\omega,t^\prime)\,e^{-i(\omega-\omega_A)(t-t^\prime)}\right],
\end{align}
\end{widetext}
Assuming that at $ t = 0 $ the state of the total system is $ \ket{e}\otimes \ket{\cbkt{0}}$, where
$\ket{\cbkt{0}}$ denotes the vacuum state of the polaritonic operator. On taking the average over the initial state, the first two terms reduce to zero, yielding:
\begin{align}
\left\langle\frac{d}{dt}\hat{\sigma}^z(t)\right\rangle=-2\beta(t)(1+\left\langle\hat{\sigma}^z(t)\right\rangle)
\end{align} 
 Here $\beta(t)$ is the time-varying decay factor (see Appendix\,\ref{appendixD} for details):

\eqn{
\beta(t)
=\frac{\Gamma_A }{2\pi}\int_0^\infty d\omega \int_0^t dt^\prime\text{Re}\sbkt{i\bkt{\frac{\omega}{\omega_A}}e^{-i(\omega-\omega_A)(t-t^\prime)}\right. &\non\\
\left.\frac{\sqrt{\text{Im}(\epsilon(\omega,t))}\sqrt{\text{Im}(\epsilon(\omega,t^\prime))}}{\sqrt{\epsilon(\omega,t)}+\sqrt{\epsilon^*(\omega, t^\prime)}}} &\label{betat1},
}
where $ \Gamma_A \equiv \frac{\left|\mb{d}\cdot\mb{d}\right|^2\omega_A^3}{3\pi\hbar\epsilon_0c^3}$ is the free space  atomic spontaneous emission rate.

The radiative level shift for the excited state can be determined by investigating the dynamics of the atomic lowering operator $\hat{\sigma}$. 
\eqn{
\frac{d}{dt}\hat{\sigma}=&-i\omega_A\hat{\sigma}&\non\\
&+\sum_{\lambda=1,2}\int d^3\mb{k}\int_0^\infty d\omega\, g_\lambda^*(\mathbf{k},\omega, t)\hat{\sigma}_z\hat{C}_\lambda(\mathbf{k}, \omega,t)
}
Just as in the derivation for spontaneous emission rate, we substitute the result from Eq.\,\eqref{polaritonevolve} for the polariton field and take an expectation value assuming that the polariton field is in its ground or vacuum state at t=0 i.e. $\langle \hat{C}_\lambda(\mathbf{k},\omega,0)\rangle=0$

    \begin{align}
        \avg{\frac{d}{dt}\hat{\sigma}(t)}&=-i\left(\omega_A-\Delta(t)\right)\langle\hat{\sigma}(t)\rangle -\beta(t)\langle\hat{\sigma}(t)\rangle
    \end{align}

Where $\beta(t)$ is the time-varying decay rate derived in Eq.\,\eqref{betat1} and $\Delta(t)$ is the time-dependent level shift of the excited state.

\eqn{
\Delta(t)
=-\frac{\Gamma_A}{2\pi}\int_0^\infty d\omega \int_0^t dt^\prime\text{Im}\sbkt{i\bkt{\frac{\omega}{\omega_A}}e^{-i(\omega-\omega_A)(t-t^\prime)}\right. &\non\\
\left.\frac{\sqrt{\text{Im}(\epsilon(\omega,t))}\sqrt{\text{Im}(\epsilon(\omega,t^\prime))}}{\sqrt{\epsilon(\omega,t)}+\sqrt{\epsilon^*(\omega, t^\prime)}}}& \label{Lamb Shift}
}

The above expressions for spontaneous emission rate and level shift in Eq.\,\eqref{betat1} and Eq.\,\eqref{Lamb Shift} are applicable to any arbitrary two level atomic or atom-like emitter embedded in any time-varying dielectric with permittivity $\epsilon(\omega,t)$.  Due to the vastly disparate timescales involved and the resulting rapidly oscillating integral, numerically computing the decay rate and shift for an arbitrary dielectric is challenging, often prohibitively so. We instead try to gain some broad qualitative insights by analytically solving for the specific case of the time-varying Drude-Lorentz model discussed in Section\,\ref{Sec:DL}. 

 \begin{figure*}[t]
     \subfloat[]{\includegraphics[width=0.45\textwidth]{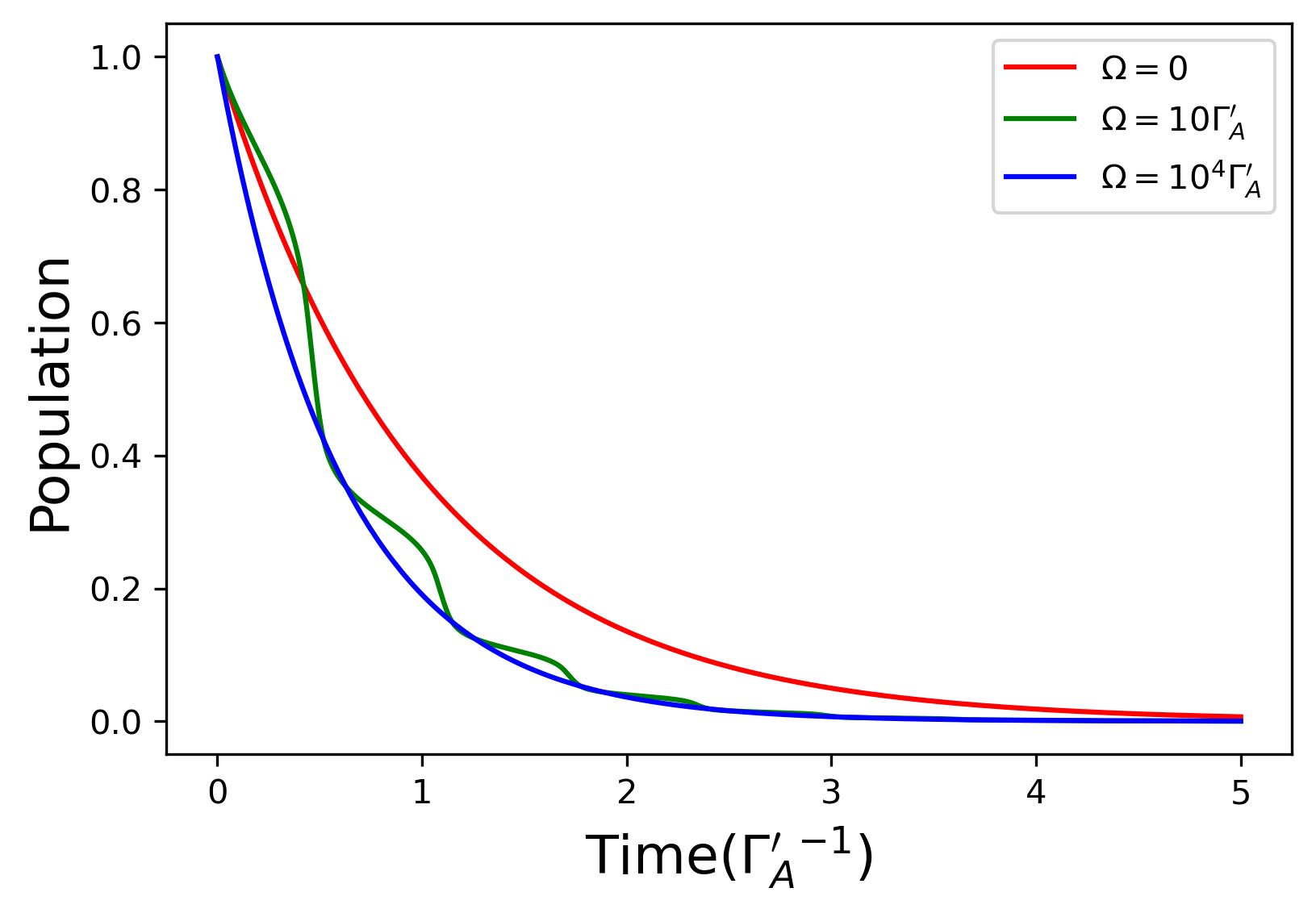}}         
    \subfloat[]
{\includegraphics[width=0.45\textwidth]{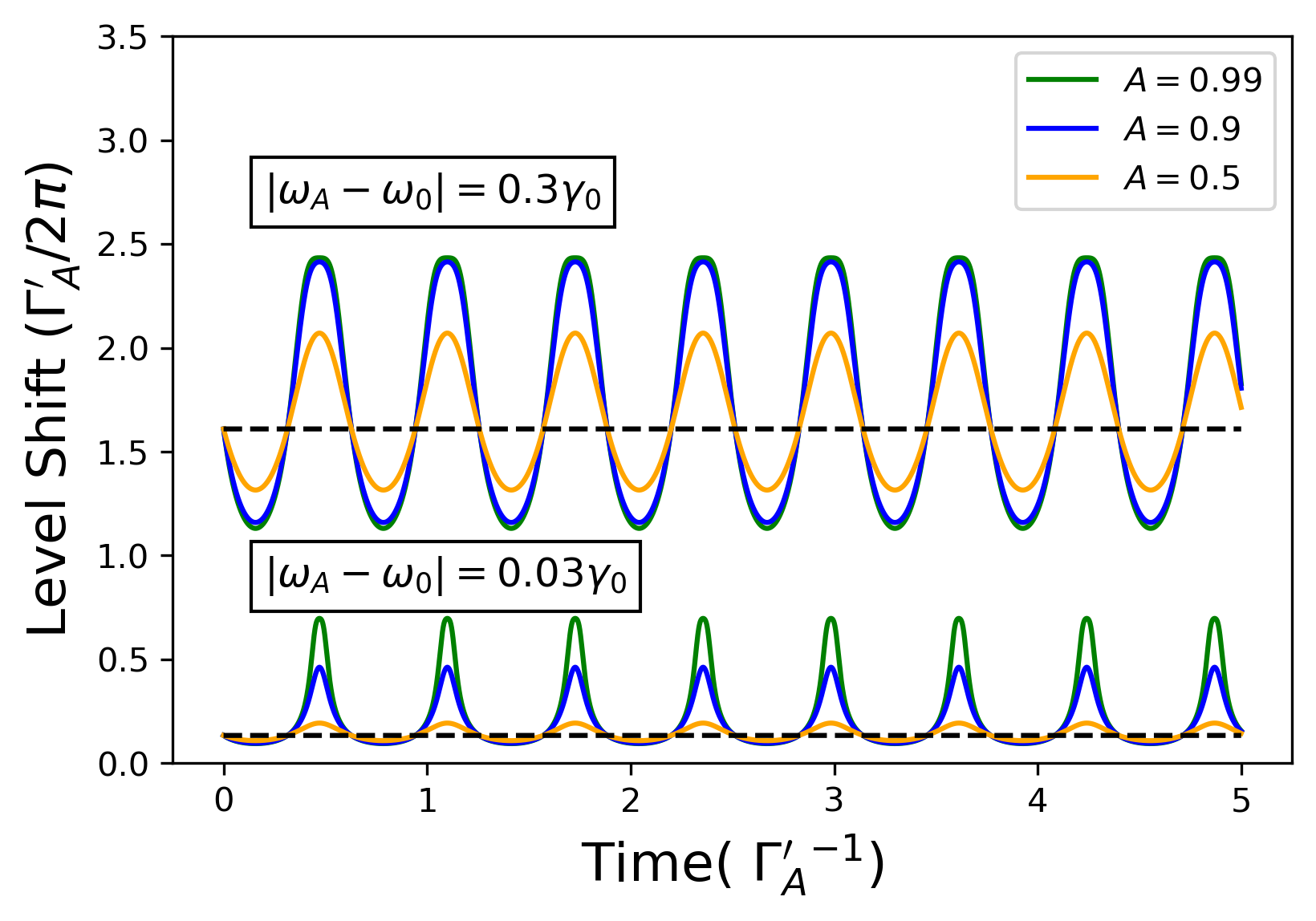}}\label{fig:lambshift}
         \caption{(a) Excited state population as a function of time for an initially excited atom deeply in the dissipative regime ($|\omega_A-\omega_0|=0.03\gamma_0$) in a Drude-Lorentz dielectric with sinusoidal modulation of $\gamma(t)=\gamma_0\left(1+A\sin\Omega t\right)$ with modulation amplitude $A=0.99$ and slow ($\Omega=10\Gamma^\prime_A$) and fast ($\Omega=10^4\Gamma^\prime_A$) modulation frequencies (note that $\Omega=0$ represents the unmodulated case). $\Gamma^\prime_A$ in (a) and (b) corresponds to the emission rate in the dielectric for each atom in the absence of modulation. (b) Time dependent component of excited state level shift for atoms both deeply (bottom curves) and weakly (top curves) in the dissipative regime in a Drude-Lorentz dielectric with sinusoidal modulation of $\gamma(t)=\gamma_0\left(1+A\sin\Omega t\right)$ where $\Omega=10\gamma_0$ for each case. Note that the offsets on the y-axis (dashed horizontal lines) are the shifts due to the dielectric in the absence of modulation.}
         \label{fig:poplamb}
 \end{figure*}
For a time-varying Drude-Lorentz dielectric resonant at $\omega_0$, with an unmodulated damping coefficient, $\gamma_0$, there are two frequency regimes of interest: the dispersive regime where $|\omega_A-\omega_0|\gg \gamma_0/2$ and the dissipative regime where $|\omega_A-\omega_0|\leq \gamma_0/2$. The dispersive regime is relevant for typical dielectric materials, which are generally not suitable for the type of temporal modulation we consider in this work. Furthermore, we will see below, that even if these materials could be modulated, it would have a minimal effect on the spontaneous emission rate or Lamb shift. The dissipative and near-dissipative regime is typical of ENZ and plasmonic meta-materials where proximity to the dielectric resonance enables large and dynamic material response\,\cite{yao2014plasmonic, boltasseva2011low} and atom arrays where interesting dynamics occurs near the atomic resonance. In each of the two regimes, the integral over frequency to evaluate $\beta(t)$ and $\Delta(t)$ can be broken into two parts, one being the integral over the frequency range spanning the absorption linewidth of the dielectric centered around $\omega_0$ and the other, the integral over the frequencies that are far removed from the dielectric resonance at $\omega_0$. We shall refer to the two frequency integral contributions as resonant and off-resonant contributions where the resonance here alludes to the dielectric resonant frequency, $\omega_0$ and not the atom transition frequency. For the frequency range responsible for the resonant contribution, the time-varying permittivity may be approximated as:
\begin{align}
\epsilon(\omega,t)&\approx 1+\frac{i\omega_c^2}{\omega_0\gamma(t)}.
\label{approxnearresonance}
\end{align}
On the other hand, for frequencies contributing to the off-resonant integral, the time-varying permittivity bears the following approximate form:
\begin{align}
\epsilon(\omega,t)&\approx 1 + \frac{\omega_c^2}{\omega_0^2-\omega^2}+i\omega_c^2\frac{\omega\gamma(t)}{(\omega_0^2-\omega^2)^2}.
\label{farresonanceapprox}
\end{align}
Aided by these approximate expressions, we evaluate $\beta(t)$ and $\Delta(t)$ by considering the resonant and off-resonant integrals separately in Appendix \ref{resonantappendix} and Appendix \ref{offresonantappendix}. The results, which are qualitatively distinct for the dissipative regime and the dispersive regime, have been summarized in the ensuing sections. 

\subsection{Dissipative regime}
The spontaneous emission rate bears the following form: 
\begin{align}
 \beta(t)=\frac{\Gamma_A}{2}\text{Re}\left(\sqrt{\epsilon(\omega_A,t)}\right)\equiv \beta_0(\omega_A, t)   
\end{align}
The recurring quantity, $\frac{\Gamma_A}{2}\frac{\omega}{\omega_A}\text{Re}\left(\sqrt{\epsilon(\omega,t)}\right)$ will henceforth be denoted by $\beta_0(\omega,t)$ and we can identify it as the vacuum spontaneous emission rate modified by the time-varying permittivity ${\epsilon(\omega_A,t)}$, which in this regime has a strong temporal dependence as evident from Eq.\,\eqref{approxnearresonance}. 

The level shift is as follows:
\begin{widetext}
\begin{align}
\Delta(t)=\mathrm{P}\int_{0}^\infty d\omega\frac{\beta_0(\omega, t)}{\pi(\omega-\omega_A)}+
\frac{\beta_0(\omega_A,t)}{2\pi}\mathrm{P}\int_{-\infty}^\infty d\omega^\prime\bigg[\frac{\text{sgn}(\omega^\prime)}{\omega_A-\omega_0+\gamma_0/2+\omega^\prime}
-\frac{\text{sgn}(\omega^\prime)}{\omega_A-\omega_0-\gamma_0/2+\omega^\prime}\bigg]
\end{align}
\end{widetext}
Here, $\mathrm{P}$ is the Cauchy principal value while sgn is the sign function. The first term stems from the off-resonant contribution to the frequency integral. As is evident from Eq. \eqref{farresonanceapprox}, the real part of the time-varying permittivity and consequently, the first term of the level shift, has a weak temporal dependence. The second term on the other hand arises from the resonant contribution and thus exhibits temporal variation. In the case of the dissipative regime, since $|\omega_A-\omega_0|\leq \gamma_0/2$, the second term is not negligible. In Fig.\,\ref{fig:lambshift}, we have plotted the time dependent component of the level shift, arising from the second term in the preceding equation, as a function of time for a specific time varying dielectric. For atoms resonant at optical frequencies, the time-invariant component is typically an order of magnitude larger.  
\subsection{Dispersive Regime}
The spontaneous emission rate and level shift for an atom in the dispersive regime are as follows:
\begin{align}
  \beta(t)&=\beta_0(\omega_A, t)   \\
  \Delta(t)&=\mathrm{P}\int_{0}^\infty d\omega\frac{\beta_0(\omega, t)}{\pi(\omega-\omega_A)}
\end{align}
Although they share the same expression (partially in the case of level shift) with the dissipative regime, the temporal behavior is qualitatively different. In the dispersive regime, the decay rate and the shift exhibit limited temporal variation.  In the dispersive as well as dissipative regime, $\beta(t)$ reduces to the expected expression for an atom in a dielectric in the absence of temporal modulation\,\cite{barnett1992spontaneous}. As a sanity check, we note that when we set the temporal modulation to zero, $\Delta(t)$ assumes the expected form for the level shift in a dispersive dielectric\,\cite{milonni1995field}. In our treatment, we have not accounted for local corrections to the electric field experienced by the atom. There exist several models for incorporating local field corrections and the choice of the model is informed by the details of the dielectric at hand and experimental results\,\cite{aubret2019understanding, glauber1991quantum, mie1976contributions, aspnes1982local, lorentz1916theory, bottcher1974theory, onsager1936electric,scheel1999spontaneous}.

We show the modification of the spontaneous emission rate and the level shift in the dissipative regime in Fig. \ref{fig:poplamb}. In all cases the plots are relative to the emitter properties in an identical unmodulated dielectric. We see that we can dynamically change the instantaneous spontaneous emission rate with slow modulation and we can speed up the emission (keeping the typical exponential decay) with fast modulation (Fig. \ref{fig:poplamb}a). The effect is larger for systems more deeply in the dissipative regime and for larger modulation amplitudes. We see the time-dependent component of the level shift directly follows modulation of the dielectric (Fig. \ref{fig:poplamb}b). This effect is largest in the near-dissipative regime where shifts of more than a full atomic linewidth can be achieved for large modulation depths.

\section{Discussion}

We have developed a self-consistent framework for quantization of the EM field in the presence of a time-varying dielectric. The dielectric is described analogously to the Huttner-Barnett model in terms of a macroscopic polarization field coupled to a reservoir, admitting both dispersive and dissipative behavior in the medium. We encode the time-variance of the dielectric into the coupling between the reservoir and the polarization field. We fully diagonalize the total matter + EM field system to arrive at an effective polaritonic description that can thus describe quantized EM fields in and near time-varying dielectrics. We observe the time dependence of the dielectric emerge encoded in a time-varying effective permittivity.  As an application of the framework, we calculate the radiative properties -- spontaneous emission and radiative level shift -- of a two-level atom interacting with the quantized field in a time-varying Drude-Lorentz dielectric. We observe substantial modification of these properties in the dissipative regime where the atomic frequency is near the dielectric resonance (Fig.\,\ref{fig:poplamb}). This can enable dynamic modification of quantum emitter properties via the dielectric without directly addressing the atom.

Quantum emitters in and near dielectric media are at the heart of many devices in quantum optics and other fields. In most cases the emitter properties are fixed by the environment, and a common task is to tune any photonic or other structures to atomic resonance. The modification of the QED radiative phenomena given here provide a knob to tune the emitter properties and must be accounted for in engineering time-varying photonic systems. The dissipative regime studied here is particularly applicable to state-of-the-art metamaterial platforms that support dynamic modulation of dielectric properties \cite{liu2011metamaterials, chen2016review, zheludev2012metamaterials} and atomic arrays, which have garnered significant recent attention \cite{masson2020atomic, srakaew2023subwavelength, nandi2021coherent, rui2020subradiant}. Using an effective dielectric description for the time-varying optical properties realizable by atomic arrays abrogates the need to follow the dynamics of the full many-body quantum system. 

More generally, our framework paves the way for studying quantum fluctuation phenomena such as Casimir forces in the presence of time-varying dielectric properties. Additionally, systems such as atomic arrays and metamaterials can allow one to change the boundary conditions on the EM field drastically, enabling the possibility of creating photons from the quantum vacuum in the presence of time-varying boundaries, akin to the dynamical Casimir effect\,\cite{Nation2012,Dodonov2010}. Such dynamical amplification of vacuum fluctuations bears correspondences to Hawking radiation and Unruh effect\,\cite{Nation2012}. Future work can include extending the current formalism to ultrafast materials where dielectric properties can be modulated on optically fast timescales \cite{Alqattan_ultrafast,Schultze_Controlling_dielectric,Schiffrin_optical_field_induced}.

\section*{Acknowledgments}
We are grateful to Peter W. Milonni for his thoughtful feedback on the manuscript. K.S. acknowledges support from the Quantum Collaborative, led by Arizona State University,  National Science Foundation under Grant No. PHY-2309341, and by the John Templeton Foundation under Award No. 62422. E.A.G acknowledges support from the National Science Foundation under Grant No. 2143172.
 
 \appendix
\begin{widetext}

\section{Fano-diagonalization of matter fields}
\label{App:Fano}

\subsection{Coefficients of dielectric matter field operator\label{appendixA1}}
The matter field operator, $\hat{B}(\mathbf{k},\omega,t)$ is required to obey the following equation:
\begin{align}
\left[\hat{B}(\mathbf{k}, \omega, t), \hat{H}_{\text{mat}}\right]=\hbar\omega\hat{B}(\mathbf{k},\omega,t)    
\end{align}
Substituting the expansion for the matter field operator in terms of the original operators as expressed in Eq.\,\eqref{diagonalization} gives us the following set of relations for the time dependent coefficients:
\begin{align}
& \alpha_{0}(\omega, t) \omega=\alpha_{0}(\omega,t) \tilde{\omega}_{0}(t)+\frac{1}{2} \int_{0}^{\infty} d \omega^{\prime}\left[\alpha_{1}\left(\omega, \omega^{\prime},t\right) V\left(\omega^{\prime},t\right)-\beta_{1}\left(\omega, \omega^{\prime},t\right) V^{*}\left(\omega^{\prime},t\right) \right]\nonumber\\
& \beta_{0}(\omega,t) \omega=-\beta_{0}(\omega,t) \tilde{\omega}_{0}(t)+\frac{1}{2} \int_{0}^{\infty} d \omega^{\prime}\left[\alpha_{1}\left(\omega, \omega^{\prime},t\right) V\left(\omega^{\prime},t\right)-\beta_{1}\left(\omega, \omega^{\prime},t\right) V^{*}\left(\omega^{\prime},t\right) \right]\nonumber\\
& \alpha_{1}\left(\omega, \omega^{\prime},t\right) \omega=\frac{1}{2}\left[\alpha_{0}(\omega,t)-\beta_{0}(\omega,t)\right] V^{*}\left(\omega^{\prime},t\right)+\alpha_{1}\left(\omega, \omega^{\prime},t\right) \omega^{\prime} \nonumber\\
& \beta_{1}\left(\omega, \omega^{\prime},t\right) \omega=\frac{1}{2}\left[\alpha_{0}(\omega,t)-\beta_{0}(\omega,t)\right] V\left(\omega^{\prime},t\right)-\beta_{1}\left(\omega, \omega^{\prime},t\right) \omega^{\prime}
\end{align}
Where $V(\omega,t)$ is as defined in \ref{quantization}. We may use the above set of equations to express all the coefficients in terms of one of the coefficients, say $\alpha_0(\omega,t)$. Then, $\alpha_0(\omega,t)$ itself may be determined using the ETCR that $\hat{B}(\mathbf{k},\omega, t)$ and its conjugate are required to satisfy. 
\begin{align}
    \left[\hat{B}(\mathbf{k},\omega,t), \hat{B}^\dagger(\mathbf{k^\prime}, \omega^\prime,t)\right]=\delta(\mathbf{k^\prime}-\mathbf{k})\delta(\omega-\omega^\prime)
\end{align}
After some algebraic manipulation, we have the following functional forms for the time dependent coefficients\,\cite{huttner1992quantization}:
\begin{align}
    \alpha_{0}(\omega,t)&=\left(\frac{\omega+\tilde{\omega}_{0}(t)}{2}\right) \frac{V(\omega,t)}{\omega^{2}-\tilde{\omega}_{0}^{2}(t) z(\omega,t)}\nonumber\\
    \beta_{0}(\omega,t)&=\left(\frac{\omega-\tilde{\omega}_{0}(t)}{2}\right) \frac{V(\omega,t)}{\omega^{2}-\tilde{\omega}_{0}^{2}(t) z(\omega,t)}\nonumber\\
    \alpha_{1}\left(\omega, \omega^{\prime},t\right)&=\delta\left(\omega-\omega^{\prime}\right)+\frac{\tilde{\omega}_{0}(t)}{2}\left(\frac{V^{*}\left(\omega^{\prime},t\right)}{\omega-\omega^{\prime}-i \varepsilon}\right)\frac{V(\omega,t)}{\omega^{2}-\tilde{\omega}_{0}^{2}(t) z(\omega,t)}\nonumber\\
\beta_{1}\left(\omega, \omega^{\prime},t\right)&=\frac{\tilde{\omega}_{0}(t)}{2}\left(\frac{V\left(\omega^{\prime},t\right)}{\omega+\omega^{\prime}}\right) \frac{V(\omega,t)}{\omega^{2}-\tilde{\omega}_{0}^{2}(t) z(\omega,t)}
\end{align}
Where the quantity, $z(\omega,t)$ is defined as follows:
\begin{align}
 z(\omega,t)=1-\frac{1}{2 \tilde{\omega}_{0}(t)}\left[\int_{-\infty}^{\infty} d \omega^{\prime} \frac{\left|V\left(\omega^{\prime},t\right)\right|^2}{\omega^{\prime}-\omega+i \varepsilon}\right]\label{zdefinition}   
\end{align}

\subsection{Preservation of Commutation Relations\label{appendixB}}

In order to verify that the dressed annihilation and creation operators form a complete set of basis operators, we invert Eq.\,\eqref{diagonalization} to obtain the following expression for the original bosonic operators:
\begin{align}
    \hat{b}(\mathbf{k},t)&=\int_{0}^{\infty} d \omega\left[\alpha_{0}^{*}(\omega,t) \hat{B}(\mathbf{k}, \omega,t)-\beta_{0}(\omega,t) \hat{B}^{\dagger}(-\mathbf{k}, \omega,t)\right]\\
    \hat{b}_{\omega}(\mathbf{k},t)&=\int_{0}^{\infty} d \omega^{\prime}\left[\alpha_{1}^{*}\left(\omega^{\prime}, \omega, t\right) \hat{B}\left(\mathbf{k}, \omega^{\prime},t\right)-\beta_{1}\left(\omega^{\prime}, \omega, t\right) \hat{B}^{\dagger}\left(-\mathbf{k}, \omega^{\prime},t\right)\right]
\end{align}
The ETCR for the original bosonic operators need to hold when they are expressed in terms of the dressed matter operators.

    \eqn{
    \sbkt{\hat{b}(\mathbf{k},t), \hat{b}^{\dagger}\left(\mathbf{k}^{\prime},t\right)}=&\delta\left(\mathbf{k}-\mathbf{k}^{\prime}\right)\non\int_{0}^{\infty} d \omega\left[\left|\alpha_{0}(\omega,t)\right|^{2}-\left|\beta_{0}(\omega,t)\right|^{2}\right]\label{inverse1}\\
    \sbkt{\hat{b}_{\omega}(\mathbf{k},t), \hat{b}_{\omega^{\prime},t}^{\dagger}\left(\mathbf{k}^{\prime}\right)}=&\delta\left(\mathbf{k}-\mathbf{k}^{\prime}\right) \int_{0}^{\infty} d \nu\left[\alpha_{1}^{*}(\nu, \omega,t) \alpha_{1}\left(\nu, \omega^{\prime},t\right) -\beta_{1}(\nu, \omega,t) \beta_{1}^{*}\left(\nu, \omega^{\prime},t\right)\right]
}
Thus, we require that 
\begin{align}
&\int_{0}^{\infty} d \omega\left[\left|\alpha_{0}(\omega,t)\right|^{2}-\left|\beta_{0}(\omega,t)\right|^{2}\right]=1\label{unity}\\
&\int_{0}^{\infty} d \nu\left[\alpha_{1}^{*}(\nu, \omega,t) \alpha_{1}\left(\nu, \omega^{\prime},t\right)-
\beta_{1}(\nu, \omega,t) \beta_{1}^{*}\left(\nu, \omega^{\prime},t\right)\right]=\delta(\omega-\omega^\prime)\label{unity2}.
\end{align}
The above equations can be shown to hold true for any time $t$.
\begin{align}
  \int_0^\infty d\omega\,|\alpha_0(\omega,t)|^2-|\beta_0(\omega,t)|^2&=\int_0^\infty d\omega\,\omega\tilde{\omega}_0(t)\frac{\left|V(\omega,t)\right|^2}{\left|\omega^2-\tilde{\omega}_0^2(t)z(\omega,t)\right|^2}\nonumber\\
  &=\frac{1}{i\pi}\int_{-\infty}^{\infty}d\omega\frac{\omega}{\omega^2-\tilde{\omega}_0^2(t)z(\omega,t)}\label{unityappendix}
\end{align}
In extending the integration to negative frequency, we have used the fact that $|\tilde V(\omega,t)|^2=\frac{1}{i\pi}\left[\left(\omega^2-\tilde{\omega}_0^2(t)z^*(\omega,t)\right)-\left(\omega^2-\tilde{\omega}_0^2(t)z(\omega,t)\right)\right]$ and that $z^*(-\omega,t)=z(\omega,t)$. Simplifying the integration further requires knowledge of the poles of the integrand in the complex plane or equivalently, the zeroes of $\omega^2-\tilde{\omega}_0^2(t)z(\omega,t)$. From the definition of $z(\omega,t)$ in Eq.\,\eqref{zdefinition}, it follows that the analytic continuation of $\omega^2-\tilde{\omega}_0^2(t)z(\omega,t)$ to complex frequency is analytic in lower half complex plane and that its zeroes lie solely on the imaginary axis. Equipped with this knowledge, we focus on the negative imaginary axis with the result that the complex frequency bears the form: $-i\upsilon$ where $\upsilon\geq 0$
\begin{align}
\omega^2-\tilde{\omega}_0^2(t)z(\omega,t)\vert_{\omega=-i\upsilon}&=-(\upsilon^2+\tilde{\omega}_0^2(t))+\omega_0(t)\int_0^\infty d\omega \frac{\omega|V(\omega,t)|^2}{\omega^2+(\upsilon+\epsilon)^2}
\end{align}
It is evident that the function is monotonically decreasing in $\upsilon$. Hence, if the function is less than zero at $\upsilon=0$, it will remain so for all values on the negative imaginary axis. This condition is satisfied, provided:

\begin{align}
    \int_0^\infty d\omega\frac{|V(\omega,t)|^2}{\omega}< \tilde{\omega}_0(t)
    \label{impinequality}
\end{align}

Expanding the quantities, $\left|V(\omega,t)\right|^2$ and $\tilde{\omega}_0(t)$ using their definitions from \ref{quantization}, the above inequality reduces to $\omega_0^2 > 0$, which is trivially satisfied. Thus, the integrand in Eq.\,\eqref{unityappendix} has no poles in the lower complex half plane. Using contour integration in the lower half plane, Eq.\,\eqref{unityappendix} can be shown to be equal to unity. The analyticity of $\omega^2-\tilde{\omega}_0^2(t)z(\omega,t)$ in the lower half plane coupled with the properties: $|\tilde V(\omega,t)|^2=\frac{1}{i\pi}\left[\left(\omega^2-\tilde{\omega}_0^2(t)z^*(\omega,t)\right)-\left(\omega^2-\tilde{\omega}_0^2(t)z(\omega,t)\right)\right]$ and $z^*(-\omega,t)=z(\omega,t)$, can be used to show that Eq.\,\eqref{unity2} holds.

\subsection{Dynamic Self Consistency}
The time-varying nature of the dielectric medium leads to an explicit temporal dependence for $\hat{b}(\mathbf{k},t)$ and $\hat{B}(\mathbf{k},\omega, t)$. In other words, $\frac{\partial \hat{B}}{\partial t}\neq 0$ and $\frac{\partial \hat{b}}{\partial t}\neq 0$. This explicit time dependence stems from the time dependent coupling factor, $v(\omega,t)$.

The evolution of the operators subject to $\hat{H}_{\text{mat}}$ is dictated by the following Heisenberg equations:
\begin{align}
    \frac{d\,\hat{b}(\mathbf{k},t)}{dt}&=\frac{1}{i\hbar}[\hat{b}(\mathbf{k},t), \hat{H}_{\text{mat}}]+ \frac{\partial\,\hat{b}(\mathbf{k},t)}{\partial t}\\
    \frac{d\,\hat{B}(\mathbf{k},\omega, t)}{dt}&=\frac{1}{i\hbar}[\hat{B}(\mathbf{k},\omega, t), \hat{H}_{\text{mat}}]+ \frac{\partial\,\hat{B}(\mathbf{k},\omega, t)}{\partial t}
\end{align}
The dynamical evolution owing to the first term in the Heisenberg equations is ensured to be self-consistent by the Fano-diagonalization procedure, particularly Eq.\,\eqref{eigen}. However, we must verify that the time evolution derived from the partial derivative term is self-consistent as well. We take the partial derivative of eqn. Eq.\,\eqref{inverse1} with respect to time with the result:

\begin{align}
\frac{\partial\,\hat{b}(\mathbf{k}, t)}{\partial t}&=\int_{0}^{\infty} d \omega\left[\frac{\partial\,\alpha_{0}^{*}(\omega,t)}{\partial t} \hat{B}(\mathbf{k}, \omega,t)-\frac{\partial\,\beta_{0}(\omega,t)}{\partial t} \hat{B}^{\dagger}(-\mathbf{k}, \omega,t)+ \alpha_{0}^{*}(\omega,t) \frac{\partial\,\hat{B}(\mathbf{k}, \omega,t)}{\partial t}-\beta_{0}(\omega,t)\frac{\partial\,\hat{B}^{\dagger}(-\mathbf{k}, \omega,t)}{\partial t}\right]\nonumber\\
&=\int_{0}^{\infty}d\omega\,\left\{\frac{\partial}{\partial t}\left[\left|\alpha_{0}(\omega,t)\right|^{2}-\left|\beta_{0}(\omega,t)\right|^{2}\right]\hat{b}(\mathbf{k},t)+\left[\left|\alpha_{0}(\omega,t)\right|^{2}-\left|\beta_{0}(\omega,t)\right|^{2}\right]\frac{\partial\,\hat{b}(\mathbf{k},t)}{\partial t}\right\}
\end{align}
The identity in Eq.\,\eqref{unity} holds at all times and thus, it follows that the dressed matter operators and the original bosonic operators are dynamically self-consistent. A similar check can be performed for $\frac{\partial \hat{b}_\omega(\mathbf{k},t) }{\partial t}$.
This should invoke little surprise since we had previously demonstrated that the dressed matter operators form a complete basis set {at all times}.

 \section{Coefficients of polaritonic operator\label{App:Polariton}}

The derivation for the time dependent coefficients of the polaritonic operator follows along similar lines as those for the diagonalized matter field. The coefficients have the following functional form:
\begin{align}
    \tilde{\alpha}_{0}(k, \omega,t)=&\sqrt{\frac{\omega_{c}^{2}}{\tilde{k} c}}\left(\frac{\omega+\tilde{k} c}{2}\right) \frac{\zeta(\omega,t)}{\epsilon^{*}(\omega,t) \omega^{2}-k^{2} c^{2}}\\
    \tilde{\beta}_{0}(k, \omega,t)=&\sqrt{\frac{\omega_{c}^{2}}{\tilde{k} c}}\left(\frac{\omega-\tilde{k} c}{2}\right) \frac{\zeta(\omega,t)}{\epsilon^{*}(\omega,t) \omega^{2}-k^{2} c^{2}}\\
\tilde{\alpha}_{1}\left(k, \omega, \omega^{\prime},t\right)=&\delta\left(\omega-\omega^{\prime}\right)+\frac{\omega_{c}^{2}}{2}\left(\frac{\zeta^{*}\left(\omega^{\prime},t\right)}{\omega-\omega^{\prime}-i \varepsilon}\right)\frac{\zeta(\omega,t)}{\epsilon^{*}(\omega,t) \omega^{2}-k^{2} c^{2}}\\
\tilde{\beta}_{1}\left(k, \omega, \omega^{\prime},t\right)=&\frac{\omega_{c}^{2}}{2}\left(\frac{\zeta^{*}\left(\omega^{\prime},t\right)}{\omega-\omega^{\prime}-i \varepsilon}\right) \frac{\zeta(\omega,t)}{\epsilon^{*}(\omega,t) \omega^{2}-k^{2} c^{2}} 
\end{align}
Where $\epsilon(\omega,t)$ has been identified as the time dependent dielectric constant.
\begin{align}
    \epsilon(\omega,t)=1+\frac{\omega_{c}^{2}}{2 \omega} \int_{-\infty}^{\infty} d \omega^{\prime} \frac{\left|\zeta\left(\omega^{\prime},t\right)\right|^2}{\omega^\prime\left(\omega^{\prime}-\omega-i \varepsilon\right)}
\end{align}

In the same vein of structural analogy,
the polaritonic annihilation and creation operators form a complete basis set provided that the coupling factor $\sqrt{\frac{k_c^2c}{\tilde{k}}}\zeta(\omega, t)$ fulfills an inequality analogous to Eq.\,\eqref{impinequality}:
\begin{align}
\frac{ck_c^2}{\tilde{k}} \int_{0}^{\infty} d \omega \frac{|\zeta(\omega, t)|^{2}}{\omega}<\tilde{k} c.
\end{align}

In light of Eq.\,\eqref{zetadef} and the definition of $\tilde{k}$ ($\tilde{k}=\sqrt{k^{2}+\omega_c^{2}/c^2}$) we see that the inequality is always fulfilled.

\section{Time-varying Drude-Lorentz model}
\label{App:DL}
From Eq.\,\eqref{epsilon}, we see that the imaginary part of the time varying permittivity can be expressed in terms of the coupling factor, $\zeta(\omega,t)$ which in turn can be expressed in terms of the polarization-reservoir coupling factor, $v(\omega,t)$ as obtained by expanding the coefficients in Eq.\,\eqref{zetadef}, detailed in Appendix\,\ref{appendixA1}:
\begin{align}
\label{eq:Imeps1}
    \text{Im}(\epsilon(\omega, t))=\frac{\pi\omega_c^2}{2\rho^2}\frac{\omega \left|v(\omega, t)\right|^2}{\left|\omega^2-{\tilde{\omega} _0(t)}^2z(\omega,t)\right|^2} 
\end{align}
We first compute the denominator separately by expanding the condensed quantities, $\tilde{\omega} _0(t)$ and $z(\omega,t)$. 
\begin{align}
   |\omega^2-{\tilde{\omega} _0(t)}^2z(\omega,t)|^2 =&\underset{\epsilon\rightarrow 0}{\text{lim}}\left|\omega^2-\omega_0^2+ 
   \frac{\omega-i\epsilon}{2\rho^2}\int_{-\infty}^{\infty}d\omega^\prime\frac{|v(\omega^\prime,t)|^2}{\omega^\prime-\omega+i\epsilon}\right|^2\\
   =&\underset{\epsilon\rightarrow 0}{\text{lim}}\left|\omega^2-\omega_0^2+\omega\gamma(t)\int_{ -\Lambda}^{\Lambda}d\omega^\prime\frac{1}{\omega^\prime-\omega+i\epsilon}\right|^2\\
   =&\left|\omega^2-\omega_0^2-i\omega\gamma(t)\right|^2
   \label{eq:den}
\end{align}
We have assumed that the high frequency cutoff is significantly larger than the frequency regime of interest and can be approximated to be infinite. In the third line, we have used contour integration in the upper half complex frequency plane to arrive at the final form for the denominator.

Thus substituting Eq.\,\eqref{eq:den} in Eq.\,\eqref{eq:Imeps1}, we arrive at:

\eqn{\im\sbkt{{\epsilon (\omega, t)}} = \omega_c^2\frac{\omega\gamma(t)\Theta(\omega-\Lambda)\Theta(\omega+\Lambda)}{(\omega^2-\omega_0^2)^2+\omega^2\gamma(t)^2}. 
}

The real part follows immediately from the imaginary part owing to the time-varying Kramers-Kronig relations \,\cite{solis2021functional}, leading to the full expression for the permittivity in Eq.\,\eqref{eq:epsdl}.

\section{Spontaneous Emission and Level Shift\label{appendixD}}

\subsection{$ \beta(t)$ simplification}

\begin{align}
&\beta(t)=\frac{1}{2}\sum_{\lambda=1,2}\int \mathbf{d^{3}k}\int_{0}^{\infty}d\omega\int_0^t dt^\prime\left[g(\mb{k},\omega,t)g^{*}(\mb{k},\omega,t^\prime)e^{i(\omega-\omega_{A})(t-t^\prime)}+g^{*}(\mb{k},\omega,t)g(\mb{k},\omega,t^\prime)e^{-i(\omega-\omega_{A})(t-t^\prime)}\right]\nonumber\\
&=\frac{\omega_A^2}{16\pi^4\hbar\epsilon_0}\sum_{\lambda=1,2}\int_0^\infty d\omega \int_0^t dt^\prime \int \mathbf{d^{3}k}\left|\mb{d}\cdot\mb{e}_{\lambda}(\mb{k})\right|^2\left [\frac{\omega^2\sqrt{\text{Im}(\epsilon(\omega,t))}\sqrt{\text{Im}(\epsilon(\omega,t^\prime))}e^{-i(\omega-\omega_A)(t-t^\prime)}}{\left(\omega^{2}\epsilon(\omega,t)-k^{2}c^{2}\right)\left(\omega^{2}\epsilon^*(\omega,t^\prime)-k^{2}c^{2}\right)}\right] + c.c\nonumber\\
&=\frac{\omega_A^2\left|\mb{d}\cdot\mb{d}\right|^2}{6\pi^3\hbar\epsilon_0}\int_0^\infty d\omega \int_0^t dt^\prime \int_0^\infty k^2 dk \left [\frac{\omega^2\sqrt{\text{Im}(\epsilon(\omega,t))}\sqrt{\text{Im}(\epsilon(\omega,t^\prime))}e^{-i(\omega-\omega_A)(t-t^\prime)}}{\left(\omega^{2}\epsilon(\omega,t)-k^{2}c^{2}\right)\left(\omega^{2}\epsilon^*(\omega,t^\prime)-k^{2}c^{2}\right)}\right] + c.c\nonumber\\
&=\frac{\omega_A^2\left|\mb{d}\cdot\mb{d}\right|^2}{12\pi^3\hbar\epsilon_0}\int_0^\infty d\omega \int_0^t dt^\prime \int_{-\infty}^\infty k^2 dk\left [\frac{\omega^2\sqrt{\text{Im}(\epsilon(\omega,t))}\sqrt{\text{Im}(\epsilon(\omega,t^\prime))}e^{-i(\omega-\omega_A)(t-t^\prime)}}{\left(\omega^{2}\epsilon(\omega,t)-k^{2}c^{2}\right)\left(\omega^{2}\epsilon^*(\omega,t^\prime)-k^{2}c^{2}\right)}\right] + c.c\nonumber\\
&=\frac{\omega_A^2\left|\mb{d}\cdot\mb{d}\right|^2}{12\pi^2\hbar\epsilon_0c^3}\int_0^\infty d\omega \int_0^t dt^\prime  \left [\frac{i\omega\sqrt{\text{Im}(\epsilon(\omega,t))}\sqrt{\text{Im}(\epsilon(\omega,t^\prime))}e^{-i(\omega-\omega_A)(t-t^\prime)}}{\sqrt{\epsilon(\omega,t)}+\sqrt{\epsilon^*(\omega, t^\prime)}}\right] + c.c \label{betat}
\end{align}
In the second line, we have expanded the quantity $g(\mathbf{k},\omega, t)$ and switched the order of integration. In the fourth line, since the integrand is an even function of $k$, we can extend the domain of integration to the negative real axis. We then use the residue theorem and contour integration in the upper half complex plane to arrive at the expression in the fifth line after some algebraic manipulation. The quantities $\sqrt{\epsilon(\omega,t)}$ and $\sqrt{\epsilon^*(\omega, t^\prime)}$ denote square roots with positive imaginary part. Note that $\sqrt{\text{Im}\left(\epsilon(\omega,t)\right)}$ is a complex quantity and has the same phase as $\zeta(\omega,t)$.

As a quick check, we demonstrate that the expression reduces to the expected form\,\cite{barnett1992spontaneous} when there is no temporal variation. In the absence of temporal modulation, we adopt the familiar Wigner-Weisskopf approximation to integrate over time and frequency, that yields the following result:
\begin{equation}
    \beta=\frac{\left|\mb{d}\cdot\mb{d}\right|^2\omega_A^3}{6\pi\hbar\epsilon_0c^3}\frac{i\text{Im}\left(\epsilon(\omega_A)\right)}{\sqrt{\epsilon(\omega_A)}+\sqrt{\epsilon^*(\omega_A)}} \label{eq:notimedepenence}
\end{equation}
We can rewrite the above expression in terms of the complex refractive index, $n(\omega)=\eta(\omega)+i\kappa(\omega)$ where $n^2(\omega)=\epsilon(\omega)$ such that $\text{Re}\left(\epsilon(\omega)\right)=\eta^2(\omega)-\kappa^2(\omega)$ and $\text{Im}\left(\epsilon(\omega)\right)=2\kappa(\omega)\eta(\omega)$. The phase of $\sqrt{\epsilon(\omega)}$ and $\sqrt{\epsilon^*(\omega)}$ are determined by the choice of our contour for $k$ integration. Since the contour lies in the upper half complex plane, $\sqrt{\epsilon(\omega)}=\eta(\omega)+i\kappa(\omega)$ and $\sqrt{\epsilon^*(\omega)}=-\eta(\omega)+i\kappa(\omega)$. Substituting these expressions in \eqref{eq:notimedepenence} yields $\beta=\Gamma_A\eta(\omega_A)/2$ which is in agreement with previously established results. 

\subsection{Resonant Contribution\label{resonantappendix}}
Consistent with our approximation in Eq.\eqref{approxnearresonance}, we adopt the following approximations for $\sqrt{\epsilon(\omega,t)}$ and $\sqrt{\epsilon^*(\omega,t^\prime)}$ in the frequency range: $|\omega-\omega_0|\leq \gamma_0/2$. 
\begin{align}
\sqrt{\epsilon(\omega,t)}&\approx \frac{\omega_c}{\sqrt{2\omega_0\gamma(t)}}(1+i)\nonumber\\
\sqrt{\epsilon^*(\omega,t^\prime)}&\approx \frac{\omega_c}{\sqrt{2\omega_0\gamma(t^\prime)}}(-1+i)
\end{align}

We arrive at the expression for the spontaneous emission rate given in the main text in the following manner:

\begin{align}
\beta(t)=&\frac{\left|\mb{d}\cdot\mb{d}\right|^2\omega_A^2}{6\sqrt{2}\pi^2\hbar\epsilon_0c^3}\int_0^t dt^\prime\int_{\omega_0-\gamma_0/2}^{\omega_0+\gamma_0/2}d\omega\left [\frac{\omega_c}{\sqrt{\omega_0}}\frac{i\omega e^{-i(\omega-\omega_A)(t-t^\prime)}}{\left(\sqrt{\gamma(t^\prime)}-\sqrt{\gamma(t)}\right)+i\left(\sqrt{\gamma(t)}+\sqrt{\gamma(t^\prime)}\right)}\right] +     c.c\nonumber\\
\approx&\frac{\left|\mb{d}\cdot\mb{d}\right|^2\omega_A^3}{6\sqrt{2}\pi^2\hbar\epsilon_0c^3}\int_0^t dt^\prime\int_{\omega_0-\gamma_0/2}^{\omega_0+\gamma_0/2}d\omega\left [\frac{\omega_c}{\sqrt{\omega_0}}\frac{ie^{-i(\omega-\omega_A)(t-t^\prime)}}{\left(\sqrt{\gamma(t^\prime)}-\sqrt{\gamma(t)}\right)+i\left(\sqrt{\gamma(t)}+\sqrt{\gamma(t^\prime)}\right)}\right] +     c.c\nonumber\\
=&\frac{\left|\mb{d}\cdot\mb{d}\right|^2\omega_A^3}{6\sqrt{2}\pi^2\hbar\epsilon_0c^3}\int_0^t d\tau\,\left [\frac{\omega_c}{\sqrt{\omega_0}}\frac{e^{i(\omega_A-\omega_0+\gamma_0/2)\tau}-e^{i(\omega_A-\omega_0-\gamma_0/2)\tau}}{\left[\left(\sqrt{\gamma(t-\tau)}-\sqrt{\gamma(t)}\right)+i\left(\sqrt{\gamma(t)}+\sqrt{\gamma(t-\tau)}\right)\right]\tau}\right] +     c.c\nonumber\\
\end{align}

For convenience, we consider the above integral in two parts, $\beta_1(t)$ and $\beta_2(t)$ comprising the terms $e^{i(\omega_A-\omega_0+\gamma_0/2)\tau}$ and $e^{i(\omega_A-\omega_0-\gamma_0/2)\tau}$ respectively such that $\beta(t)=\beta_1(t)-\beta_2(t)$. Since the timescale of interest, $t$ is much larger than $1/\omega_0$ or $1/\omega_A$, we can replace the limits of integration.

\begin{align}
 &\beta_1(t)\equiv\frac{\left|\mb{d}\cdot\mb{d}\right|^2}{6\sqrt{2}\pi^2\hbar\epsilon_0c^3}\int_0^\infty d\tau\,\omega_A^3\left [\frac{\omega_c}{\sqrt{\omega_0}}\frac{e^{i(\omega_A-\omega_0+\gamma_0/2)\tau}}{\left[\left(\sqrt{\gamma(t-\tau)}-\sqrt{\gamma(t)}\right)+i\left(\sqrt{\gamma(t)}+\sqrt{\gamma(t-\tau)}\right)\right]\tau}\right] +     c.c  \nonumber\\
 =&\frac{\left|\mb{d}\cdot\mb{d}\right|^2\omega_A^3}{6\sqrt{2}\pi^2\hbar\epsilon_0c^3}\frac{\omega_c}{\sqrt{\omega_0}}\int_{-\infty}^\infty d\tau\Theta(\tau)\left[\frac{e^{i(\omega_A-\omega_0+\gamma_0/2)\tau}}{\left[\left(\sqrt{\gamma(t-\tau)}-\sqrt{\gamma(t)}\right)+i\left(\sqrt{\gamma(t)}+\sqrt{\gamma(t-\tau)}\right)\right]\tau}\right] + c.c \nonumber\\
 =&\frac{\left|\mb{d}\cdot\mb{d}\right|^2\omega_A^3}{6\sqrt{2}\pi^2\hbar\epsilon_0c^3}\frac{\omega_c}{\sqrt{\omega_0}}\int_{-\infty}^\infty d\omega^\prime\mathcal{F}\left[\frac{1}{\left[\left(\sqrt{\gamma(t-\tau)}-\sqrt{\gamma(t)}\right)+i\left(\sqrt{\gamma(t)}+\sqrt{\gamma(t-\tau)}\right)\right]\tau}\right]\left(\omega^\prime\right)\frac{\delta\left(\omega_0-\omega_A-\gamma_0/2-\omega^\prime\right)}{2}\nonumber\\
 &-\frac{i\left|\mb{d}\cdot\mb{d}\right|^2\omega_A^3}{12\sqrt{2}\pi^3\hbar\epsilon_0c^3}\frac{\omega_c}{\sqrt{\omega_0}}\mathrm{P}\int_{-\infty}^\infty d\omega^\prime\frac{1}{\omega_0-\omega_A-\gamma_0/2-\omega^\prime}\mathcal{F}\left[\frac{1}{\left[\left(\sqrt{\gamma(t-\tau)}-\sqrt{\gamma(t)}\right)+i\left(\sqrt{\gamma(t)}+\sqrt{\gamma(t-\tau)}\right)\right]\tau}\right]\left(\omega^\prime\right)\non\\
 &+c.c. \label{beta1t}
 \end{align}
In the second line, we introduce the Heaviside step function, $\Theta(\tau)$. $\mathcal{F}\left[\frac{1}{\left[\left(\sqrt{\gamma(t-\tau)}-\sqrt{\gamma(t)}\right)+i\left(\sqrt{\gamma(t)}+\sqrt{\gamma(t-\tau)}\right)\right]\tau}\right]\left(\omega^\prime\right)$ in the third line denotes the Fourier transform of $\frac{1}{\left[\left(\sqrt{\gamma(t-\tau)}-\sqrt{\gamma(t)}\right)+i\left(\sqrt{\gamma(t)}+\sqrt{\gamma(t-\tau)}\right)\right]\tau}$ with respect to $\tau$ evaluated at $\omega^\prime$.

\begin{align}
\mathcal{F}\left[\frac{1}{\left[\left(\sqrt{\gamma(t-\tau)}-\sqrt{\gamma(t)}\right)+i\left(\sqrt{\gamma(t)}+\sqrt{\gamma(t-\tau)}\right)\right]\tau}\right]\left(\omega^\prime\right)=&\int_{-\infty}^{\infty}d\tau \frac{e^{-i\omega^\prime\tau}}{\left[\left(\sqrt{\gamma(t-\tau)}-\sqrt{\gamma(t)}\right)+i\left(\sqrt{\gamma(t)}+\sqrt{\gamma(t-\tau)}\right)\right]\tau}\nonumber\\
=&-\text{sgn}(\omega^\prime)\frac{\pi}{2\sqrt{\gamma(t)}}
\end{align}
Here $\text{sgn}(\cdot)$ denotes the signum function. Substituting the expression for the Fourier transform in Eq.\,\eqref{beta1t} yields the following:
\begin{align}
\beta_1(t)=&\frac{\left|\mb{d}\cdot\mb{d}\right|^2\omega_A^3}{24\pi\hbar\epsilon_0c^3\sqrt{2\gamma(t)}}\frac{\omega_c}{\sqrt{\omega_0}}\text{sgn}(\omega_A-\omega_0+\gamma_0/2)-\frac{i\left|\mb{d}\cdot\mb{d}\right|^2\omega_A^3}{24\pi^2\hbar\epsilon_0c^3\sqrt{2\gamma(t)}}\frac{\omega_c}{\sqrt{\omega_0}}\mathrm{P}\int_{-\infty}^\infty d\omega^\prime\frac{\text{sgn}(\omega^\prime)}{\omega_A-\omega_0+\gamma_0/2+\omega^\prime} + c.c.\\
=&\frac{\left|\mb{d}\cdot\mb{d}\right|^2\omega_A^3}{12\pi\hbar\epsilon_0c^3\sqrt{2\gamma(t)}}\frac{\omega_c}{\sqrt{\omega_0}}\text{sgn}(\omega_A-\omega_0+\gamma_0/2)
\end{align}
The quantity, $\beta_2(t)$ can be evaluated along similar lines with the result that $\beta_2(t)=\frac{\left|\mb{d}\cdot\mb{d}\right|^2\omega_A^3}{12\pi\hbar\epsilon_0c^3\sqrt{2\gamma(t)}}\frac{\omega_c}{\sqrt{\omega_0}}\text{sgn}(\omega_A-\omega_0-\gamma_0/2)$.
\begin{align}
\beta(t)=&\beta_1(t)-\beta_2(t)\nonumber\\
=&\frac{\left|\mb{d}\cdot\mb{d}\right|^2\omega_A^3}{12\pi\hbar\epsilon_0c^3\sqrt{2\gamma(t)}}\frac{\omega_c}{\sqrt{\omega_0}}[\text{sgn}(\omega_A-\omega_0+\gamma_0/2)-\text{sgn}(\omega_A-\omega_0-\gamma_0/2)]\nonumber\\
=&\frac{\left|\mb{d}\cdot\mb{d}\right|^2\omega_A^3}{6\pi\hbar\epsilon_0c^3\sqrt{2\gamma(t)}}\frac{\omega_c}{\sqrt{\omega_0}}\Theta(\gamma_0/2-|\omega_A-\omega_0|)
\end{align}
We have rewritten $\text{sgn}(\omega_A-\omega_0+\gamma_0/2)-\text{sgn}(\omega_A-\omega_0-\gamma_0/2)$ more succinctly as $2\Theta(\gamma_0/2-|\omega_A-\omega_0|)$. In the process of evaluating $\beta(t)$, we have also incidentally furnished the information necessary for computing the resonant contribution to the level shift since the level shift term is merely the imaginary counterpart of spontaneous decay rate.
\begin{align}
-i\Delta(t)=&\frac{\left|\mb{d}\cdot\mb{d}\right|^2\omega_A^2}{6\sqrt{2}\pi^2\hbar\epsilon_0c^3}\int_0^t dt^\prime\int_{\omega_0-\gamma_0/2}^{\omega_0+\gamma_0/2}d\omega\left [\frac{\omega_c}{\sqrt{\omega_0}}\frac{i\omega e^{-i(\omega-\omega_A)(t-t^\prime)}}{\left(\sqrt{\gamma(t^\prime)}-\sqrt{\gamma(t)}\right)+i\left(\sqrt{\gamma(t)}+\sqrt{\gamma(t^\prime)}\right)}\right]- c.c
\end{align}
Going through a similar series of steps as involved in the evaluation of $\beta(t)$, we arrive at the following expression for the frequency shift
\begin{align}
\Delta(t)=&\frac{\left|\mb{d}\cdot\mb{d}\right|^2\omega_A^3}{12\pi^2\hbar\epsilon_0c^3\sqrt{2\gamma(t)}}\frac{\omega_c}{\sqrt{\omega_0}}\mathrm{P}\int_{-\infty}^\infty d\omega^\prime\bigg[\frac{\text{sgn}(\omega^\prime)}{\omega_A-\omega_0+\gamma_0/2+\omega^\prime}-\frac{\text{sgn}(\omega^\prime)}{\omega_A-\omega_0-\gamma_0/2+\omega^\prime}\bigg]
\end{align}

\subsection{Off-Resonant Contribution\label{offresonantappendix}}
Along the lines of the approximation adopted in Eq. \eqref{farresonanceapprox}, we use the following approximate forms for  $\sqrt{\epsilon(\omega,t)}$ and $\sqrt{\epsilon^*(\omega,t^\prime)}$ to derive the off-resonant contribution.
\begin{align}
\sqrt{\epsilon(\omega,t)}&\approx \frac{|\sqrt{\omega_0^2-\omega^2+\omega_c^2}|}{|\sqrt{\omega_0^2-\omega^2}|}\left[1+\frac{i\omega_c^2\omega\gamma(t)}{2|(\omega_0^2-\omega^2)(\omega_0^2-\omega^2+\omega_c^2)|}\right]\nonumber\\
\sqrt{\epsilon^*(\omega,t^\prime)}&\approx \frac{-|\sqrt{\omega_0^2-\omega^2+\omega_c^2}|}{|\sqrt{\omega_0^2-\omega^2}|}\left[1-\frac{i\omega_c^2\omega\gamma(t^\prime)}{2|(\omega_0^2-\omega^2)(\omega_0^2-\omega^2+\omega_c^2)|}\right]    
\end{align}

The off-resonant contribution to the spontaneous decay rate is as follows:
\begin{align}
\beta(t)=&\frac{|\mathbf{d}\cdot\mathbf{d}|^2\omega_A^2}{6\pi^2\hbar\epsilon_0c^3}\int_0^\infty d\omega\frac{\left|\sqrt{\omega_0^2-\omega^2+\omega_c^2}\right|\,\omega}{\left|\sqrt{\omega_0^2-\omega^2}\right|} \int_0^t dt^\prime\left[\frac{\sqrt{\gamma(t)}\sqrt{\gamma(t^\prime)} e^{-i(\omega-\omega_A)(t-t^\prime)}}{\gamma(t)+\gamma(t^\prime)}\right] + c.c\nonumber\\
\approx&\frac{|\mathbf{d}\cdot\mathbf{d}|^2\omega_A^2}{6\pi^2\hbar\epsilon_0c^3}\int_0^\infty d\omega\frac{\left|\sqrt{\omega_0^2-\omega^2+\omega_c^2}\right|\,\omega}{\left|\sqrt{\omega_0^2-\omega^2}\right|} \int_0^\infty d\tau\left[\frac{\sqrt{\gamma(t)}\sqrt{\gamma(t-\tau)} e^{-i(\omega-\omega_A)\tau}}{\gamma(t)+\gamma(t-\tau)}\right] + c.c\nonumber\\
=&\frac{|\mathbf{d}\cdot\mathbf{d}|^2\omega_A^2}{6\pi^2\hbar\epsilon_0c^3}\int_0^\infty d\omega\frac{\left|\sqrt{\omega_0^2-\omega^2+\omega_c^2}\right|\,\omega}{\left|\sqrt{\omega_0^2-\omega^2}\right|} \underbrace{\int_{-\infty}^\infty d\tau\Theta(\tau)\left[\frac{\sqrt{\gamma(t)}\sqrt{\gamma(t-\tau)} e^{-i(\omega-\omega_A)\tau}}{\gamma(t)+\gamma(t-\tau)}\right]}_{I(\omega,t)} + c.c\nonumber\\
\end{align}

Just as in the preceding subsection, we replace the limits of integration and introduce the Heaviside step function. 

\begin{align}
I(\omega, t)\equiv &\int_{-\infty}^\infty d\tau\Theta(\tau)\left[\frac{\sqrt{\gamma(t)}\sqrt{\gamma(t-\tau)} e^{-i(\omega-\omega_A)\tau}}{\gamma(t)+\gamma(t-\tau)}\right]\nonumber\\
=&\int_{-\infty}^\infty d\omega^\prime \mathcal{F}\left[\frac{\sqrt{\gamma(t)}\sqrt{\gamma(t-\tau)}}{\gamma(t)+\gamma(t-\tau)}\right](\omega^\prime)\frac{\delta(\omega-\omega_A-\omega^\prime)}{2}- \frac{i}{2\pi}\mathrm{P}\int_{-\infty}^\infty d\omega^\prime \frac{\mathcal{F}\left[\frac{\sqrt{\gamma(t)}\sqrt{\gamma(t-\tau)}}{\gamma(t)+\gamma(t-\tau)}\right](\omega^\prime)}{\omega-\omega_A-\omega^\prime}\label{Iomegat}
\end{align}

Where $\mathcal{F}\left[\sqrt{\gamma(t)}\sqrt{\gamma(t-\tau)}/(\gamma(t)+\gamma(t-\tau))\right](\omega^\prime)$ denotes the Fourier transform of $\sqrt{\gamma(t)}\sqrt{\gamma(t-\tau)}/(\gamma(t)+\gamma(t-\tau))$ with respect to $\tau$ evaluated at $\omega^\prime$. In order to compute the Fourier transform, we consider the Taylor series expansion of $\sqrt{\gamma(t)}\sqrt{\gamma(t-\tau)}/(\gamma(t)+\gamma(t-\tau))$ about $\tau=0$

\begin{align}
\frac{\sqrt{\gamma(t)}\sqrt{\gamma(t-\tau)}}{\gamma(t)+\gamma(t-\tau)}&=\frac{1}{2}-\frac{1}{16{\gamma(t)}^2}\left[\frac{d\gamma(t-\tau)}{d\tau}\bigg|_{\tau=0}\right]^2\tau^2+O(\tau^3)\nonumber\\
&=\frac{1}{2}-\frac{1}{16{\gamma(t)}^2}\left[\frac{d\gamma(t)}{dt}\right]^2\tau^2+O(\tau^3)\nonumber
\end{align}

With the aid of the Taylor series expansion, we may now compute its Fourier transform.

\begin{align}
\mathcal{F}\left[\frac{\sqrt{\gamma(t)}\sqrt{\gamma(t-\tau)}}{\gamma(t)+\gamma(t-\tau)}\right](\omega^\prime)&=\frac{2\pi\delta(\omega^\prime)}{2}+\frac{2\pi}{16{\gamma(t)}^2}\left[\frac{d\gamma(t)}{dt}\right]^2\frac{d^2\delta(\omega^\prime)}{d{\omega^\prime}^2}+ O\left(\mathcal{F}[\tau^4](\omega^\prime)\right)\nonumber
\end{align}

Substituting the above expression in Eq.\,\eqref{Iomegat} yields the following:

    \begin{align}
      &I(\omega, t)=\frac{\pi\delta(\omega-\omega_A)}{2}+\frac{\pi}{16{\gamma(t)}^2}\left[\frac{d\gamma(t)}{dt}\right]^2\frac{d^2\delta(\omega-\omega_A)}{d{\omega}^2}+ O\left(\mathcal{F}[\tau^3](\omega-\omega_A)\right)\nonumber\\
      &-i\mathrm{P}\int_{-\infty}^{\infty}\frac{d\omega^\prime}{\omega^\prime}\bigg[\frac{\delta(\omega-\omega_A-\omega^\prime)}{2}+\frac{1}{16{\gamma(t)}^2}\left[\frac{d\gamma(t)}{dt}\right]^2\frac{d^2\delta(\omega-\omega_A-\omega^\prime)}{d{\omega}^2}+O\left(\mathcal{F}[\tau^3](\omega-\omega_A-\omega^\prime)\right)\bigg]
    \end{align}

We may now evaluate $\beta(t)$.

\begin{align}
\beta(t)=&\frac{|\mathbf{d}\cdot\mathbf{d}|^2}{6\pi\hbar\epsilon_0c^3}\left[\eta(\omega_A,t)\frac{\omega_A^3}{2}+\frac{\omega_A^2}{16\gamma(t)^2}\left[\frac{d\gamma(t)}{dt}\right]^2\frac{d\eta_g(\omega,t)}{d\omega}\bigg|_{\omega={\omega_A}} + O\left(\frac{d^2\eta_g(\omega,t)}{d\omega^2}\left[\frac{d\text{ln}(\gamma(t))}{dt}\right]^3\right)\right]\times\nonumber\\
&\int_0^\infty d\omega\,\delta(\omega-\omega_A) -\frac{i|\mathbf{d}\cdot\mathbf{d}|^2\omega_A^2}{6\pi^2\hbar\epsilon_0c^3}\mathrm{P}\int_{0}^\infty \frac{d\omega}{\omega-\omega_A}\Bigg[\eta(\omega,t)\frac{\omega}{2}+\frac{1}{16\gamma(t)^2}\left[\frac{d\gamma(t)}{dt}\right]^2\frac{d\eta_g(\omega,t)}{d\omega}+\nonumber\\
&+O\left(\frac{d^2\eta_g(\omega,t)}{d\omega^2}\left[\frac{d\text{ln}(\gamma(t))}{dt}\right]^3\right)\Bigg]+c.c. \label{betaanddelta}\\
\implies\beta(t)=&\frac{|\mathbf{d}\cdot\mathbf{d}|^2\omega_A^3\eta(\omega_A,t)}{6\pi\hbar\epsilon_0c^3}\int_0^\infty d\omega\,\delta(\omega-\omega_A)
\end{align}

In the preceding calculations, we recognize the quantity $\frac{\left|\sqrt{\omega_0^2-\omega_A^2+\omega_c^2}\right|}{\left|\sqrt{\omega_0^2-\omega_A^2}\right|}$ to be the real part of the complex refractive index i.e. $\eta(\omega,t)$ and $\eta_g(\omega,t)$ is the group index such that $\eta_g(\omega,t)=\eta(\omega,t)+\omega\frac{d\eta(\omega,t)}{d\omega}$. The higher order terms are proportional to $\frac{d^{n-1}\eta_g(\omega,t)}{\omega_A d\omega^{n-1}}\left[\frac{d\text{ln}(\gamma(t))}{dt}\right]^n$ with $n\geq 2$ and are thus, negligible compared to the first term. We drop the higher order terms. Although not explicitly stated, the frequency integral for the off-resonant contribution doesn't include the frequency range spanning the absorption profile of the dielectric. Therefore, depending on whether the atom lies in the dispersive or dissipative regime, the frequency integration domain may or may not include the peak of the delta function, $\delta(\omega-\omega_A)$. We now focus on the imaginary counterpart of $\beta(t)$ in Eq. \eqref{betaanddelta} to derive the level shift. 

\begin{align}
    \Delta(t)=\frac{|\mathbf{d}\cdot\mathbf{d}|^2\omega_A^2}{6\pi^2\hbar\epsilon_0c^3}\mathrm{P}\int_{0}^\infty d\omega\frac{\eta(\omega,t)\omega}{\omega-\omega_A}
\end{align}

Just as in the derivation for $\beta(t)$, we have dropped the higher order terms. 
\end{widetext}

\bibliography{refs}

\begin{thebibliography}{77}%
\makeatletter
\providecommand \@ifxundefined [1]{%
 \@ifx{#1\undefined}
}%
\providecommand \@ifnum [1]{%
 \ifnum #1\expandafter \@firstoftwo
 \else \expandafter \@secondoftwo
 \fi
}%
\providecommand \@ifx [1]{%
 \ifx #1\expandafter \@firstoftwo
 \else \expandafter \@secondoftwo
 \fi
}%
\providecommand \natexlab [1]{#1}%
\providecommand \enquote  [1]{``#1''}%
\providecommand \bibnamefont  [1]{#1}%
\providecommand \bibfnamefont [1]{#1}%
\providecommand \citenamefont [1]{#1}%
\providecommand \href@noop [0]{\@secondoftwo}%
\providecommand \href [0]{\begingroup \@sanitize@url \@href}%
\providecommand \@href[1]{\@@startlink{#1}\@@href}%
\providecommand \@@href[1]{\endgroup#1\@@endlink}%
\providecommand \@sanitize@url [0]{\catcode `\\12\catcode `\$12\catcode `\&12\catcode `\#12\catcode `\^12\catcode `\_12\catcode `\%12\relax}%
\providecommand \@@startlink[1]{}%
\providecommand \@@endlink[0]{}%
\providecommand \url  [0]{\begingroup\@sanitize@url \@url }%
\providecommand \@url [1]{\endgroup\@href {#1}{\urlprefix }}%
\providecommand \urlprefix  [0]{URL }%
\providecommand \Eprint [0]{\href }%
\providecommand \doibase [0]{https://doi.org/}%
\providecommand \selectlanguage [0]{\@gobble}%
\providecommand \bibinfo  [0]{\@secondoftwo}%
\providecommand \bibfield  [0]{\@secondoftwo}%
\providecommand \translation [1]{[#1]}%
\providecommand \BibitemOpen [0]{}%
\providecommand \bibitemStop [0]{}%
\providecommand \bibitemNoStop [0]{.\EOS\space}%
\providecommand \EOS [0]{\spacefactor3000\relax}%
\providecommand \BibitemShut  [1]{\csname bibitem#1\endcsname}%
\let\auto@bib@innerbib\@empty
\bibitem [{\citenamefont {Novotny}\ and\ \citenamefont {Hecht}(2012)}]{NovotnyBook}%
  \BibitemOpen
  \bibfield  {author} {\bibinfo {author} {\bibfnamefont {L.}~\bibnamefont {Novotny}}\ and\ \bibinfo {author} {\bibfnamefont {B.}~\bibnamefont {Hecht}},\ }\href {https://doi.org/10.1017/CBO9780511794193} {\emph {\bibinfo {title} {Principles of Nano-Optics}}},\ \bibinfo {edition} {2nd}\ ed.\ (\bibinfo  {publisher} {Cambridge University Press},\ \bibinfo {year} {2012})\BibitemShut {NoStop}%
\bibitem [{\citenamefont {D'Amico}\ \emph {et~al.}(2019)\citenamefont {D'Amico}, \citenamefont {Angelakis}, \citenamefont {Bussi{\`e}res}, \citenamefont {Caglayan}, \citenamefont {Couteau}, \citenamefont {Durt}, \citenamefont {Kolaric}, \citenamefont {Maletinsky}, \citenamefont {Pfeiffer}, \citenamefont {Rabl}, \citenamefont {Xuereb},\ and\ \citenamefont {Agio}}]{Damico2019}%
  \BibitemOpen
  \bibfield  {author} {\bibinfo {author} {\bibfnamefont {I.}~\bibnamefont {D'Amico}}, \bibinfo {author} {\bibfnamefont {D.~G.}\ \bibnamefont {Angelakis}}, \bibinfo {author} {\bibfnamefont {F.}~\bibnamefont {Bussi{\`e}res}}, \bibinfo {author} {\bibfnamefont {H.}~\bibnamefont {Caglayan}}, \bibinfo {author} {\bibfnamefont {C.}~\bibnamefont {Couteau}}, \bibinfo {author} {\bibfnamefont {T.}~\bibnamefont {Durt}}, \bibinfo {author} {\bibfnamefont {B.}~\bibnamefont {Kolaric}}, \bibinfo {author} {\bibfnamefont {P.}~\bibnamefont {Maletinsky}}, \bibinfo {author} {\bibfnamefont {W.}~\bibnamefont {Pfeiffer}}, \bibinfo {author} {\bibfnamefont {P.}~\bibnamefont {Rabl}}, \bibinfo {author} {\bibfnamefont {A.}~\bibnamefont {Xuereb}},\ and\ \bibinfo {author} {\bibfnamefont {M.}~\bibnamefont {Agio}},\ }\bibfield  {title} {\bibinfo {title} {Nanoscale quantum optics},\ }\href {https://doi.org/10.1393/ncr/i2019-10158-0} {\bibfield  {journal} {\bibinfo  {journal} {La Rivista del Nuovo Cimento}\ }\textbf {\bibinfo {volume} {42}},\
  \bibinfo {pages} {153} (\bibinfo {year} {2019})}\BibitemShut {NoStop}%
\bibitem [{\citenamefont {Tame}\ \emph {et~al.}(2013)\citenamefont {Tame}, \citenamefont {McEnery}, \citenamefont {{\"O}zdemir}, \citenamefont {Lee}, \citenamefont {Maier},\ and\ \citenamefont {Kim}}]{Tame2013}%
  \BibitemOpen
  \bibfield  {author} {\bibinfo {author} {\bibfnamefont {M.~S.}\ \bibnamefont {Tame}}, \bibinfo {author} {\bibfnamefont {K.~R.}\ \bibnamefont {McEnery}}, \bibinfo {author} {\bibfnamefont {{\c S}.~K.}\ \bibnamefont {{\"O}zdemir}}, \bibinfo {author} {\bibfnamefont {J.}~\bibnamefont {Lee}}, \bibinfo {author} {\bibfnamefont {S.~A.}\ \bibnamefont {Maier}},\ and\ \bibinfo {author} {\bibfnamefont {M.~S.}\ \bibnamefont {Kim}},\ }\bibfield  {title} {\bibinfo {title} {Quantum plasmonics},\ }\href {https://doi.org/10.1038/nphys2615} {\bibfield  {journal} {\bibinfo  {journal} {Nature Physics}\ }\textbf {\bibinfo {volume} {9}},\ \bibinfo {pages} {329} (\bibinfo {year} {2013})}\BibitemShut {NoStop}%
\bibitem [{\citenamefont {Bozhevolnyi}\ and\ \citenamefont {Khurgin}(2017)}]{Bozhevolnyi2017}%
  \BibitemOpen
  \bibfield  {author} {\bibinfo {author} {\bibfnamefont {S.~I.}\ \bibnamefont {Bozhevolnyi}}\ and\ \bibinfo {author} {\bibfnamefont {J.~B.}\ \bibnamefont {Khurgin}},\ }\bibfield  {title} {\bibinfo {title} {The case for quantum plasmonics},\ }\href {https://doi.org/10.1038/nphoton.2017.103} {\bibfield  {journal} {\bibinfo  {journal} {Nature Photonics}\ }\textbf {\bibinfo {volume} {11}},\ \bibinfo {pages} {398} (\bibinfo {year} {2017})}\BibitemShut {NoStop}%
\bibitem [{\citenamefont {Milonni}(1994)}]{Milonni}%
  \BibitemOpen
  \bibfield  {author} {\bibinfo {author} {\bibfnamefont {P.~W.}\ \bibnamefont {Milonni}},\ }\href {https://books.google.com/books?id=P83vAAAAMAAJ} {\emph {\bibinfo {title} {{The Quantum Vacuum: An Introduction to Quantum Electrodynamics}}}}\ (\bibinfo  {publisher} {Elsevier Science},\ \bibinfo {year} {1994})\BibitemShut {NoStop}%
\bibitem [{\citenamefont {Buhmann}(2012{\natexlab{a}})}]{Buhmann1}%
  \BibitemOpen
  \bibfield  {author} {\bibinfo {author} {\bibfnamefont {S.~Y.}\ \bibnamefont {Buhmann}},\ }\href@noop {} {\emph {\bibinfo {title} {{Dispersion Forces I}}}}\ (\bibinfo  {publisher} {Springer-Verlag, Berlin, Heidelberg},\ \bibinfo {year} {2012})\BibitemShut {NoStop}%
\bibitem [{\citenamefont {Buhmann}(2012{\natexlab{b}})}]{Buhmann2}%
  \BibitemOpen
  \bibfield  {author} {\bibinfo {author} {\bibfnamefont {S.~Y.}\ \bibnamefont {Buhmann}},\ }\href@noop {} {\emph {\bibinfo {title} {{Dispersion Forces II}}}}\ (\bibinfo  {publisher} {Springer-Verlag, Berlin, Heidelberg},\ \bibinfo {year} {2012})\BibitemShut {NoStop}%
\bibitem [{\citenamefont {Flick}\ \emph {et~al.}(2018)\citenamefont {Flick}, \citenamefont {Rivera},\ and\ \citenamefont {Narang}}]{Flick18}%
  \BibitemOpen
  \bibfield  {author} {\bibinfo {author} {\bibfnamefont {J.}~\bibnamefont {Flick}}, \bibinfo {author} {\bibfnamefont {N.}~\bibnamefont {Rivera}},\ and\ \bibinfo {author} {\bibfnamefont {P.}~\bibnamefont {Narang}},\ }\bibfield  {title} {\bibinfo {title} {Strong light-matter coupling in quantum chemistry and quantum photonics},\ }\href {https://doi.org/doi:10.1515/nanoph-2018-0067} {\bibfield  {journal} {\bibinfo  {journal} {Nanophotonics}\ }\textbf {\bibinfo {volume} {7}},\ \bibinfo {pages} {1479} (\bibinfo {year} {2018})}\BibitemShut {NoStop}%
\bibitem [{\citenamefont {Galego}\ \emph {et~al.}(2016)\citenamefont {Galego}, \citenamefont {Garcia-Vidal},\ and\ \citenamefont {Feist}}]{Galego16}%
  \BibitemOpen
  \bibfield  {author} {\bibinfo {author} {\bibfnamefont {J.}~\bibnamefont {Galego}}, \bibinfo {author} {\bibfnamefont {F.~J.}\ \bibnamefont {Garcia-Vidal}},\ and\ \bibinfo {author} {\bibfnamefont {J.}~\bibnamefont {Feist}},\ }\bibfield  {title} {\bibinfo {title} {Suppressing photochemical reactions with quantized light fields},\ }\href {https://doi.org/10.1038/ncomms13841} {\bibfield  {journal} {\bibinfo  {journal} {Nature Communications}\ }\textbf {\bibinfo {volume} {7}},\ \bibinfo {pages} {13841} (\bibinfo {year} {2016})}\BibitemShut {NoStop}%
\bibitem [{\citenamefont {Hertzog}\ \emph {et~al.}(2019)\citenamefont {Hertzog}, \citenamefont {Wang}, \citenamefont {Mony},\ and\ \citenamefont {B{\"o}rjesson}}]{Hertzog19}%
  \BibitemOpen
  \bibfield  {author} {\bibinfo {author} {\bibfnamefont {M.}~\bibnamefont {Hertzog}}, \bibinfo {author} {\bibfnamefont {M.}~\bibnamefont {Wang}}, \bibinfo {author} {\bibfnamefont {J.}~\bibnamefont {Mony}},\ and\ \bibinfo {author} {\bibfnamefont {K.}~\bibnamefont {B{\"o}rjesson}},\ }\bibfield  {title} {\bibinfo {title} {Strong light--matter interactions: a new direction within chemistry},\ }\href {https://doi.org/10.1039/C8CS00193F} {\bibfield  {journal} {\bibinfo  {journal} {Chem. Soc. Rev.}\ }\textbf {\bibinfo {volume} {48}},\ \bibinfo {pages} {937} (\bibinfo {year} {2019})}\BibitemShut {NoStop}%
\bibitem [{\citenamefont {Senellart}\ \emph {et~al.}(2017)\citenamefont {Senellart}, \citenamefont {Solomon},\ and\ \citenamefont {White}}]{Senellart2017}%
  \BibitemOpen
  \bibfield  {author} {\bibinfo {author} {\bibfnamefont {P.}~\bibnamefont {Senellart}}, \bibinfo {author} {\bibfnamefont {G.}~\bibnamefont {Solomon}},\ and\ \bibinfo {author} {\bibfnamefont {A.}~\bibnamefont {White}},\ }\bibfield  {title} {\bibinfo {title} {High-performance semiconductor quantum-dot single-photon sources},\ }\href {https://doi.org/10.1038/nnano.2017.218} {\bibfield  {journal} {\bibinfo  {journal} {Nature Nanotechnology}\ }\textbf {\bibinfo {volume} {12}},\ \bibinfo {pages} {1026} (\bibinfo {year} {2017})}\BibitemShut {NoStop}%
\bibitem [{\citenamefont {Lodahl}\ \emph {et~al.}(2015)\citenamefont {Lodahl}, \citenamefont {Mahmoodian},\ and\ \citenamefont {Stobbe}}]{LodahlRMP2015}%
  \BibitemOpen
  \bibfield  {author} {\bibinfo {author} {\bibfnamefont {P.}~\bibnamefont {Lodahl}}, \bibinfo {author} {\bibfnamefont {S.}~\bibnamefont {Mahmoodian}},\ and\ \bibinfo {author} {\bibfnamefont {S.}~\bibnamefont {Stobbe}},\ }\bibfield  {title} {\bibinfo {title} {Interfacing single photons and single quantum dots with photonic nanostructures},\ }\href {https://doi.org/10.1103/RevModPhys.87.347} {\bibfield  {journal} {\bibinfo  {journal} {Rev. Mod. Phys.}\ }\textbf {\bibinfo {volume} {87}},\ \bibinfo {pages} {347} (\bibinfo {year} {2015})}\BibitemShut {NoStop}%
\bibitem [{\citenamefont {O'Shea}\ \emph {et~al.}(2013)\citenamefont {O'Shea}, \citenamefont {Junge}, \citenamefont {Volz},\ and\ \citenamefont {Rauschenbeutel}}]{Oshea2013}%
  \BibitemOpen
  \bibfield  {author} {\bibinfo {author} {\bibfnamefont {D.}~\bibnamefont {O'Shea}}, \bibinfo {author} {\bibfnamefont {C.}~\bibnamefont {Junge}}, \bibinfo {author} {\bibfnamefont {J.}~\bibnamefont {Volz}},\ and\ \bibinfo {author} {\bibfnamefont {A.}~\bibnamefont {Rauschenbeutel}},\ }\bibfield  {title} {\bibinfo {title} {Fiber-optical switch controlled by a single atom},\ }\href {https://doi.org/10.1103/PhysRevLett.111.193601} {\bibfield  {journal} {\bibinfo  {journal} {Phys. Rev. Lett.}\ }\textbf {\bibinfo {volume} {111}},\ \bibinfo {pages} {193601} (\bibinfo {year} {2013})}\BibitemShut {NoStop}%
\bibitem [{\citenamefont {J{\"o}ns}\ \emph {et~al.}(2017)\citenamefont {J{\"o}ns}, \citenamefont {Schweickert}, \citenamefont {Versteegh}, \citenamefont {Dalacu}, \citenamefont {Poole}, \citenamefont {Gulinatti}, \citenamefont {Giudice}, \citenamefont {Zwiller},\ and\ \citenamefont {Reimer}}]{Johns2017}%
  \BibitemOpen
  \bibfield  {author} {\bibinfo {author} {\bibfnamefont {K.~D.}\ \bibnamefont {J{\"o}ns}}, \bibinfo {author} {\bibfnamefont {L.}~\bibnamefont {Schweickert}}, \bibinfo {author} {\bibfnamefont {M.~A.~M.}\ \bibnamefont {Versteegh}}, \bibinfo {author} {\bibfnamefont {D.}~\bibnamefont {Dalacu}}, \bibinfo {author} {\bibfnamefont {P.~J.}\ \bibnamefont {Poole}}, \bibinfo {author} {\bibfnamefont {A.}~\bibnamefont {Gulinatti}}, \bibinfo {author} {\bibfnamefont {A.}~\bibnamefont {Giudice}}, \bibinfo {author} {\bibfnamefont {V.}~\bibnamefont {Zwiller}},\ and\ \bibinfo {author} {\bibfnamefont {M.~E.}\ \bibnamefont {Reimer}},\ }\bibfield  {title} {\bibinfo {title} {Bright nanoscale source of deterministic entangled photon pairs violating bell's inequality},\ }\href {https://doi.org/10.1038/s41598-017-01509-6} {\bibfield  {journal} {\bibinfo  {journal} {Scientific Reports}\ }\textbf {\bibinfo {volume} {7}},\ \bibinfo {pages} {1700} (\bibinfo {year} {2017})}\BibitemShut {NoStop}%
\bibitem [{\citenamefont {Lodahl}\ \emph {et~al.}(2017)\citenamefont {Lodahl}, \citenamefont {Mahmoodian}, \citenamefont {Stobbe}, \citenamefont {Rauschenbeutel}, \citenamefont {Schneeweiss}, \citenamefont {Volz}, \citenamefont {Pichler},\ and\ \citenamefont {Zoller}}]{Lodahl2017}%
  \BibitemOpen
  \bibfield  {author} {\bibinfo {author} {\bibfnamefont {P.}~\bibnamefont {Lodahl}}, \bibinfo {author} {\bibfnamefont {S.}~\bibnamefont {Mahmoodian}}, \bibinfo {author} {\bibfnamefont {S.}~\bibnamefont {Stobbe}}, \bibinfo {author} {\bibfnamefont {A.}~\bibnamefont {Rauschenbeutel}}, \bibinfo {author} {\bibfnamefont {P.}~\bibnamefont {Schneeweiss}}, \bibinfo {author} {\bibfnamefont {J.}~\bibnamefont {Volz}}, \bibinfo {author} {\bibfnamefont {H.}~\bibnamefont {Pichler}},\ and\ \bibinfo {author} {\bibfnamefont {P.}~\bibnamefont {Zoller}},\ }\bibfield  {title} {\bibinfo {title} {Chiral quantum optics},\ }\href {https://doi.org/10.1038/nature21037} {\bibfield  {journal} {\bibinfo  {journal} {Nature}\ }\textbf {\bibinfo {volume} {541}},\ \bibinfo {pages} {473} (\bibinfo {year} {2017})}\BibitemShut {NoStop}%
\bibitem [{\citenamefont {Vahala}(2003)}]{Vahala2003}%
  \BibitemOpen
  \bibfield  {author} {\bibinfo {author} {\bibfnamefont {K.~J.}\ \bibnamefont {Vahala}},\ }\bibfield  {title} {\bibinfo {title} {Optical microcavities},\ }\href {https://doi.org/10.1038/nature01939} {\bibfield  {journal} {\bibinfo  {journal} {Nature}\ }\textbf {\bibinfo {volume} {424}},\ \bibinfo {pages} {839} (\bibinfo {year} {2003})}\BibitemShut {NoStop}%
\bibitem [{\citenamefont {Yu}\ \emph {et~al.}(2021)\citenamefont {Yu}, \citenamefont {von Kugelgen}, \citenamefont {Laorenza},\ and\ \citenamefont {Freedman}}]{Yu2021}%
  \BibitemOpen
  \bibfield  {author} {\bibinfo {author} {\bibfnamefont {C.-J.}\ \bibnamefont {Yu}}, \bibinfo {author} {\bibfnamefont {S.}~\bibnamefont {von Kugelgen}}, \bibinfo {author} {\bibfnamefont {D.~W.}\ \bibnamefont {Laorenza}},\ and\ \bibinfo {author} {\bibfnamefont {D.~E.}\ \bibnamefont {Freedman}},\ }\bibfield  {title} {\bibinfo {title} {A molecular approach to quantum sensing},\ }\href {https://doi.org/10.1021/acscentsci.0c00737} {\bibfield  {journal} {\bibinfo  {journal} {ACS Central Science}\ }\textbf {\bibinfo {volume} {7}},\ \bibinfo {pages} {712} (\bibinfo {year} {2021})}\BibitemShut {NoStop}%
\bibitem [{\citenamefont {Wang}\ \emph {et~al.}(2022)\citenamefont {Wang}, \citenamefont {Xia},\ and\ \citenamefont {Ho}}]{Wang2022}%
  \BibitemOpen
  \bibfield  {author} {\bibinfo {author} {\bibfnamefont {L.}~\bibnamefont {Wang}}, \bibinfo {author} {\bibfnamefont {Y.}~\bibnamefont {Xia}},\ and\ \bibinfo {author} {\bibfnamefont {W.}~\bibnamefont {Ho}},\ }\bibfield  {title} {\bibinfo {title} {Atomic-scale quantum sensing based on the ultrafast coherence of an $\mr{H}_2$ molecule in an $\mr{STM}$ cavity},\ }\href {https://doi.org/10.1126/science.abn9220} {\bibfield  {journal} {\bibinfo  {journal} {Science}\ }\textbf {\bibinfo {volume} {376}},\ \bibinfo {pages} {401} (\bibinfo {year} {2022})}\BibitemShut {NoStop}%
\bibitem [{\citenamefont {Szigeti}\ \emph {et~al.}(2020)\citenamefont {Szigeti}, \citenamefont {Nolan}, \citenamefont {Close},\ and\ \citenamefont {Haine}}]{Szigetti2020}%
  \BibitemOpen
  \bibfield  {author} {\bibinfo {author} {\bibfnamefont {S.~S.}\ \bibnamefont {Szigeti}}, \bibinfo {author} {\bibfnamefont {S.~P.}\ \bibnamefont {Nolan}}, \bibinfo {author} {\bibfnamefont {J.~D.}\ \bibnamefont {Close}},\ and\ \bibinfo {author} {\bibfnamefont {S.~A.}\ \bibnamefont {Haine}},\ }\bibfield  {title} {\bibinfo {title} {{High-Precision Quantum-Enhanced Gravimetry with a Bose-Einstein Condensate}},\ }\href {https://doi.org/10.1103/PhysRevLett.125.100402} {\bibfield  {journal} {\bibinfo  {journal} {Phys. Rev. Lett.}\ }\textbf {\bibinfo {volume} {125}},\ \bibinfo {pages} {100402} (\bibinfo {year} {2020})}\BibitemShut {NoStop}%
\bibitem [{\citenamefont {Blanco}\ \emph {et~al.}(2023)\citenamefont {Blanco}, \citenamefont {Essig}, \citenamefont {Fernandez-Serra}, \citenamefont {Ramani},\ and\ \citenamefont {Slone}}]{Blanco2023}%
  \BibitemOpen
  \bibfield  {author} {\bibinfo {author} {\bibfnamefont {C.}~\bibnamefont {Blanco}}, \bibinfo {author} {\bibfnamefont {R.}~\bibnamefont {Essig}}, \bibinfo {author} {\bibfnamefont {M.}~\bibnamefont {Fernandez-Serra}}, \bibinfo {author} {\bibfnamefont {H.}~\bibnamefont {Ramani}},\ and\ \bibinfo {author} {\bibfnamefont {O.}~\bibnamefont {Slone}},\ }\bibfield  {title} {\bibinfo {title} {Dark matter direct detection with quantum dots},\ }\href {https://doi.org/10.1103/PhysRevD.107.095035} {\bibfield  {journal} {\bibinfo  {journal} {Phys. Rev. D}\ }\textbf {\bibinfo {volume} {107}},\ \bibinfo {pages} {095035} (\bibinfo {year} {2023})}\BibitemShut {NoStop}%
\bibitem [{\citenamefont {Carney}\ \emph {et~al.}(2021)\citenamefont {Carney}, \citenamefont {Krnjaic}, \citenamefont {Moore}, \citenamefont {Regal}, \citenamefont {Afek}, \citenamefont {Bhave}, \citenamefont {Brubaker}, \citenamefont {Corbitt}, \citenamefont {Cripe}, \citenamefont {Crisosto}, \citenamefont {Geraci}, \citenamefont {Ghosh}, \citenamefont {Harris}, \citenamefont {Hook}, \citenamefont {Kolb}, \citenamefont {Kunjummen}, \citenamefont {Lang}, \citenamefont {Li}, \citenamefont {Lin}, \citenamefont {Liu}, \citenamefont {Lykken}, \citenamefont {Magrini}, \citenamefont {Manley}, \citenamefont {Matsumoto}, \citenamefont {Monte}, \citenamefont {Monteiro}, \citenamefont {Purdy}, \citenamefont {Riedel}, \citenamefont {Singh}, \citenamefont {Singh}, \citenamefont {Sinha}, \citenamefont {Taylor}, \citenamefont {Qin}, \citenamefont {Wilson},\ and\ \citenamefont {Zhao}}]{Carney2021}%
  \BibitemOpen
  \bibfield  {author} {\bibinfo {author} {\bibfnamefont {D.}~\bibnamefont {Carney}}, \bibinfo {author} {\bibfnamefont {G.}~\bibnamefont {Krnjaic}}, \bibinfo {author} {\bibfnamefont {D.~C.}\ \bibnamefont {Moore}}, \bibinfo {author} {\bibfnamefont {C.~A.}\ \bibnamefont {Regal}}, \bibinfo {author} {\bibfnamefont {G.}~\bibnamefont {Afek}}, \bibinfo {author} {\bibfnamefont {S.}~\bibnamefont {Bhave}}, \bibinfo {author} {\bibfnamefont {B.}~\bibnamefont {Brubaker}}, \bibinfo {author} {\bibfnamefont {T.}~\bibnamefont {Corbitt}}, \bibinfo {author} {\bibfnamefont {J.}~\bibnamefont {Cripe}}, \bibinfo {author} {\bibfnamefont {N.}~\bibnamefont {Crisosto}}, \bibinfo {author} {\bibfnamefont {A.}~\bibnamefont {Geraci}}, \bibinfo {author} {\bibfnamefont {S.}~\bibnamefont {Ghosh}}, \bibinfo {author} {\bibfnamefont {J.~G.~E.}\ \bibnamefont {Harris}}, \bibinfo {author} {\bibfnamefont {A.}~\bibnamefont {Hook}}, \bibinfo {author} {\bibfnamefont {E.~W.}\ \bibnamefont {Kolb}}, \bibinfo {author} {\bibfnamefont {J.}~\bibnamefont
  {Kunjummen}}, \bibinfo {author} {\bibfnamefont {R.~F.}\ \bibnamefont {Lang}}, \bibinfo {author} {\bibfnamefont {T.}~\bibnamefont {Li}}, \bibinfo {author} {\bibfnamefont {T.}~\bibnamefont {Lin}}, \bibinfo {author} {\bibfnamefont {Z.}~\bibnamefont {Liu}}, \bibinfo {author} {\bibfnamefont {J.}~\bibnamefont {Lykken}}, \bibinfo {author} {\bibfnamefont {L.}~\bibnamefont {Magrini}}, \bibinfo {author} {\bibfnamefont {J.}~\bibnamefont {Manley}}, \bibinfo {author} {\bibfnamefont {N.}~\bibnamefont {Matsumoto}}, \bibinfo {author} {\bibfnamefont {A.}~\bibnamefont {Monte}}, \bibinfo {author} {\bibfnamefont {F.}~\bibnamefont {Monteiro}}, \bibinfo {author} {\bibfnamefont {T.}~\bibnamefont {Purdy}}, \bibinfo {author} {\bibfnamefont {C.~J.}\ \bibnamefont {Riedel}}, \bibinfo {author} {\bibfnamefont {R.}~\bibnamefont {Singh}}, \bibinfo {author} {\bibfnamefont {S.}~\bibnamefont {Singh}}, \bibinfo {author} {\bibfnamefont {K.}~\bibnamefont {Sinha}}, \bibinfo {author} {\bibfnamefont {J.~M.}\ \bibnamefont {Taylor}}, \bibinfo
  {author} {\bibfnamefont {J.}~\bibnamefont {Qin}}, \bibinfo {author} {\bibfnamefont {D.~J.}\ \bibnamefont {Wilson}},\ and\ \bibinfo {author} {\bibfnamefont {Y.}~\bibnamefont {Zhao}},\ }\bibfield  {title} {\bibinfo {title} {Mechanical quantum sensing in the search for dark matter},\ }\href {https://doi.org/10.1088/2058-9565/abcfcd} {\bibfield  {journal} {\bibinfo  {journal} {Quantum Science and Technology}\ }\textbf {\bibinfo {volume} {6}},\ \bibinfo {pages} {024002} (\bibinfo {year} {2021})}\BibitemShut {NoStop}%
\bibitem [{\citenamefont {Casola}\ \emph {et~al.}(2018)\citenamefont {Casola}, \citenamefont {van~der Sar},\ and\ \citenamefont {Yacoby}}]{Casola2018}%
  \BibitemOpen
  \bibfield  {author} {\bibinfo {author} {\bibfnamefont {F.}~\bibnamefont {Casola}}, \bibinfo {author} {\bibfnamefont {T.}~\bibnamefont {van~der Sar}},\ and\ \bibinfo {author} {\bibfnamefont {A.}~\bibnamefont {Yacoby}},\ }\bibfield  {title} {\bibinfo {title} {Probing condensed matter physics with magnetometry based on nitrogen-vacancy centres in diamond},\ }\href {https://doi.org/10.1038/natrevmats.2017.88} {\bibfield  {journal} {\bibinfo  {journal} {Nature Reviews Materials}\ }\textbf {\bibinfo {volume} {3}},\ \bibinfo {pages} {17088} (\bibinfo {year} {2018})}\BibitemShut {NoStop}%
\bibitem [{\citenamefont {Liu}\ and\ \citenamefont {Zhang}(2011)}]{liu2011metamaterials}%
  \BibitemOpen
  \bibfield  {author} {\bibinfo {author} {\bibfnamefont {Y.}~\bibnamefont {Liu}}\ and\ \bibinfo {author} {\bibfnamefont {X.}~\bibnamefont {Zhang}},\ }\bibfield  {title} {\bibinfo {title} {Metamaterials: a new frontier of science and technology},\ }\href {https://doi.org/10.1039/C0CS00184H} {\bibfield  {journal} {\bibinfo  {journal} {Chem. Soc. Rev.}\ }\textbf {\bibinfo {volume} {40}},\ \bibinfo {pages} {2494} (\bibinfo {year} {2011})}\BibitemShut {NoStop}%
\bibitem [{\citenamefont {Chen}\ \emph {et~al.}(2016)\citenamefont {Chen}, \citenamefont {Taylor},\ and\ \citenamefont {Yu}}]{chen2016review}%
  \BibitemOpen
  \bibfield  {author} {\bibinfo {author} {\bibfnamefont {H.-T.}\ \bibnamefont {Chen}}, \bibinfo {author} {\bibfnamefont {A.~J.}\ \bibnamefont {Taylor}},\ and\ \bibinfo {author} {\bibfnamefont {N.}~\bibnamefont {Yu}},\ }\bibfield  {title} {\bibinfo {title} {A review of metasurfaces: physics and applications},\ }\href {https://doi.org/10.1088/0034-4885/79/7/076401} {\bibfield  {journal} {\bibinfo  {journal} {Reports on Progress in Physics}\ }\textbf {\bibinfo {volume} {79}},\ \bibinfo {pages} {076401} (\bibinfo {year} {2016})}\BibitemShut {NoStop}%
\bibitem [{\citenamefont {Zheludev}\ and\ \citenamefont {Kivshar}(2012)}]{zheludev2012metamaterials}%
  \BibitemOpen
  \bibfield  {author} {\bibinfo {author} {\bibfnamefont {N.~I.}\ \bibnamefont {Zheludev}}\ and\ \bibinfo {author} {\bibfnamefont {Y.~S.}\ \bibnamefont {Kivshar}},\ }\bibfield  {title} {\bibinfo {title} {From metamaterials to metadevices},\ }\href {https://doi.org/10.1038/nmat3431} {\bibfield  {journal} {\bibinfo  {journal} {Nature Materials}\ }\textbf {\bibinfo {volume} {11}},\ \bibinfo {pages} {917} (\bibinfo {year} {2012})}\BibitemShut {NoStop}%
\bibitem [{\citenamefont {Xiao}\ \emph {et~al.}(2020)\citenamefont {Xiao}, \citenamefont {Wang}, \citenamefont {Liu}, \citenamefont {Zhou}, \citenamefont {Jiang},\ and\ \citenamefont {Zhang}}]{xiao2020active}%
  \BibitemOpen
  \bibfield  {author} {\bibinfo {author} {\bibfnamefont {S.}~\bibnamefont {Xiao}}, \bibinfo {author} {\bibfnamefont {T.}~\bibnamefont {Wang}}, \bibinfo {author} {\bibfnamefont {T.}~\bibnamefont {Liu}}, \bibinfo {author} {\bibfnamefont {C.}~\bibnamefont {Zhou}}, \bibinfo {author} {\bibfnamefont {X.}~\bibnamefont {Jiang}},\ and\ \bibinfo {author} {\bibfnamefont {J.}~\bibnamefont {Zhang}},\ }\bibfield  {title} {\bibinfo {title} {Active metamaterials and metadevices: a review},\ }\href {https://doi.org/10.1088/1361-6463/abaced} {\bibfield  {journal} {\bibinfo  {journal} {Journal of Physics D: Applied Physics}\ }\textbf {\bibinfo {volume} {53}},\ \bibinfo {pages} {503002} (\bibinfo {year} {2020})}\BibitemShut {NoStop}%
\bibitem [{\citenamefont {Engheta}(2023)}]{engheta2023four}%
  \BibitemOpen
  \bibfield  {author} {\bibinfo {author} {\bibfnamefont {N.}~\bibnamefont {Engheta}},\ }\bibfield  {title} {\bibinfo {title} {Four-dimensional optics using time-varying metamaterials},\ }\href {https://doi.org/10.1126/science.adf1094} {\bibfield  {journal} {\bibinfo  {journal} {Science}\ }\textbf {\bibinfo {volume} {379}},\ \bibinfo {pages} {1190} (\bibinfo {year} {2023})}\BibitemShut {NoStop}%
\bibitem [{\citenamefont {Galiffi}\ \emph {et~al.}(2022)\citenamefont {Galiffi}, \citenamefont {Tirole}, \citenamefont {Yin}, \citenamefont {Li}, \citenamefont {Vezzoli}, \citenamefont {Huidobro}, \citenamefont {Silveirinha}, \citenamefont {Sapienza}, \citenamefont {Al{\`u}},\ and\ \citenamefont {Pendry}}]{galiffi2022photonics}%
  \BibitemOpen
  \bibfield  {author} {\bibinfo {author} {\bibfnamefont {E.}~\bibnamefont {Galiffi}}, \bibinfo {author} {\bibfnamefont {R.}~\bibnamefont {Tirole}}, \bibinfo {author} {\bibfnamefont {S.}~\bibnamefont {Yin}}, \bibinfo {author} {\bibfnamefont {H.}~\bibnamefont {Li}}, \bibinfo {author} {\bibfnamefont {S.}~\bibnamefont {Vezzoli}}, \bibinfo {author} {\bibfnamefont {P.~A.}\ \bibnamefont {Huidobro}}, \bibinfo {author} {\bibfnamefont {M.~G.}\ \bibnamefont {Silveirinha}}, \bibinfo {author} {\bibfnamefont {R.}~\bibnamefont {Sapienza}}, \bibinfo {author} {\bibfnamefont {A.}~\bibnamefont {Al{\`u}}},\ and\ \bibinfo {author} {\bibfnamefont {J.~B.}\ \bibnamefont {Pendry}},\ }\bibfield  {title} {\bibinfo {title} {{Photonics of time-varying media}},\ }\href {https://doi.org/10.1117/1.AP.4.1.014002} {\bibfield  {journal} {\bibinfo  {journal} {Advanced Photonics}\ }\textbf {\bibinfo {volume} {4}},\ \bibinfo {pages} {014002} (\bibinfo {year} {2022})}\BibitemShut {NoStop}%
\bibitem [{\citenamefont {Maas}\ \emph {et~al.}(2013)\citenamefont {Maas}, \citenamefont {Parsons}, \citenamefont {Engheta},\ and\ \citenamefont {Polman}}]{maas2013experimental}%
  \BibitemOpen
  \bibfield  {author} {\bibinfo {author} {\bibfnamefont {R.}~\bibnamefont {Maas}}, \bibinfo {author} {\bibfnamefont {J.}~\bibnamefont {Parsons}}, \bibinfo {author} {\bibfnamefont {N.}~\bibnamefont {Engheta}},\ and\ \bibinfo {author} {\bibfnamefont {A.}~\bibnamefont {Polman}},\ }\bibfield  {title} {\bibinfo {title} {Experimental realization of an epsilon-near-zero metamaterial at visible wavelengths},\ }\href {https://doi.org/10.1038/nphoton.2013.256} {\bibfield  {journal} {\bibinfo  {journal} {Nature Photonics}\ }\textbf {\bibinfo {volume} {7}},\ \bibinfo {pages} {907} (\bibinfo {year} {2013})}\BibitemShut {NoStop}%
\bibitem [{\citenamefont {Neira}\ \emph {et~al.}(2018)\citenamefont {Neira}, \citenamefont {Wurtz},\ and\ \citenamefont {Zayats}}]{neira2018all}%
  \BibitemOpen
  \bibfield  {author} {\bibinfo {author} {\bibfnamefont {A.~D.}\ \bibnamefont {Neira}}, \bibinfo {author} {\bibfnamefont {G.~A.}\ \bibnamefont {Wurtz}},\ and\ \bibinfo {author} {\bibfnamefont {A.~V.}\ \bibnamefont {Zayats}},\ }\bibfield  {title} {\bibinfo {title} {All-optical switching in silicon photonic waveguides with an epsilon-near-zero resonant cavity},\ }\href {https://doi.org/10.1364/PRJ.6.0000B1} {\bibfield  {journal} {\bibinfo  {journal} {Photon. Res.}\ }\textbf {\bibinfo {volume} {6}},\ \bibinfo {pages} {B1} (\bibinfo {year} {2018})}\BibitemShut {NoStop}%
\bibitem [{\citenamefont {Diroll}\ \emph {et~al.}(2016)\citenamefont {Diroll}, \citenamefont {Guo}, \citenamefont {Chang},\ and\ \citenamefont {Schaller}}]{diroll2016large}%
  \BibitemOpen
  \bibfield  {author} {\bibinfo {author} {\bibfnamefont {B.~T.}\ \bibnamefont {Diroll}}, \bibinfo {author} {\bibfnamefont {P.}~\bibnamefont {Guo}}, \bibinfo {author} {\bibfnamefont {R.~P.~H.}\ \bibnamefont {Chang}},\ and\ \bibinfo {author} {\bibfnamefont {R.~D.}\ \bibnamefont {Schaller}},\ }\bibfield  {title} {\bibinfo {title} {Large transient optical modulation of epsilon-near-zero colloidal nanocrystals},\ }\href {https://doi.org/10.1021/acsnano.6b05116} {\bibfield  {journal} {\bibinfo  {journal} {ACS Nano}\ }\textbf {\bibinfo {volume} {10}},\ \bibinfo {pages} {10099} (\bibinfo {year} {2016})}\BibitemShut {NoStop}%
\bibitem [{\citenamefont {Xie}\ \emph {et~al.}(2020)\citenamefont {Xie}, \citenamefont {Wu}, \citenamefont {Fu},\ and\ \citenamefont {Li}}]{xie2020tunable}%
  \BibitemOpen
  \bibfield  {author} {\bibinfo {author} {\bibfnamefont {Z.~T.}\ \bibnamefont {Xie}}, \bibinfo {author} {\bibfnamefont {J.}~\bibnamefont {Wu}}, \bibinfo {author} {\bibfnamefont {H.~Y.}\ \bibnamefont {Fu}},\ and\ \bibinfo {author} {\bibfnamefont {Q.}~\bibnamefont {Li}},\ }\bibfield  {title} {\bibinfo {title} {Tunable electro- and all-optical switch based on epsilon-near-zero metasurface},\ }\href {https://doi.org/10.1109/JPHOT.2020.3010284} {\bibfield  {journal} {\bibinfo  {journal} {IEEE Photonics Journal}\ }\textbf {\bibinfo {volume} {12}},\ \bibinfo {pages} {1} (\bibinfo {year} {2020})}\BibitemShut {NoStop}%
\bibitem [{\citenamefont {Bohn}\ \emph {et~al.}(2021)\citenamefont {Bohn}, \citenamefont {Luk}, \citenamefont {Tollerton}, \citenamefont {Hutchings}, \citenamefont {Brener}, \citenamefont {Horsley}, \citenamefont {Barnes},\ and\ \citenamefont {Hendry}}]{bohn2021all}%
  \BibitemOpen
  \bibfield  {author} {\bibinfo {author} {\bibfnamefont {J.}~\bibnamefont {Bohn}}, \bibinfo {author} {\bibfnamefont {T.~S.}\ \bibnamefont {Luk}}, \bibinfo {author} {\bibfnamefont {C.}~\bibnamefont {Tollerton}}, \bibinfo {author} {\bibfnamefont {S.~W.}\ \bibnamefont {Hutchings}}, \bibinfo {author} {\bibfnamefont {I.}~\bibnamefont {Brener}}, \bibinfo {author} {\bibfnamefont {S.}~\bibnamefont {Horsley}}, \bibinfo {author} {\bibfnamefont {W.~L.}\ \bibnamefont {Barnes}},\ and\ \bibinfo {author} {\bibfnamefont {E.}~\bibnamefont {Hendry}},\ }\bibfield  {title} {\bibinfo {title} {All-optical switching of an epsilon-near-zero plasmon resonance in indium tin oxide},\ }\href {https://doi.org/10.1038/s41467-021-21332-y} {\bibfield  {journal} {\bibinfo  {journal} {Nature Communications}\ }\textbf {\bibinfo {volume} {12}},\ \bibinfo {pages} {1017} (\bibinfo {year} {2021})}\BibitemShut {NoStop}%
\bibitem [{\citenamefont {Sounas}\ and\ \citenamefont {Al{\`u}}(2017)}]{sounas2017non}%
  \BibitemOpen
  \bibfield  {author} {\bibinfo {author} {\bibfnamefont {D.~L.}\ \bibnamefont {Sounas}}\ and\ \bibinfo {author} {\bibfnamefont {A.}~\bibnamefont {Al{\`u}}},\ }\bibfield  {title} {\bibinfo {title} {Non-reciprocal photonics based on time modulation},\ }\href {https://doi.org/10.1038/s41566-017-0051-x} {\bibfield  {journal} {\bibinfo  {journal} {Nature Photonics}\ }\textbf {\bibinfo {volume} {11}},\ \bibinfo {pages} {774} (\bibinfo {year} {2017})}\BibitemShut {NoStop}%
\bibitem [{\citenamefont {Zhou}\ \emph {et~al.}(2020)\citenamefont {Zhou}, \citenamefont {Alam}, \citenamefont {Karimi}, \citenamefont {Upham}, \citenamefont {Reshef}, \citenamefont {Liu}, \citenamefont {Willner},\ and\ \citenamefont {Boyd}}]{Zhou20}%
  \BibitemOpen
  \bibfield  {author} {\bibinfo {author} {\bibfnamefont {Y.}~\bibnamefont {Zhou}}, \bibinfo {author} {\bibfnamefont {M.~Z.}\ \bibnamefont {Alam}}, \bibinfo {author} {\bibfnamefont {M.}~\bibnamefont {Karimi}}, \bibinfo {author} {\bibfnamefont {J.}~\bibnamefont {Upham}}, \bibinfo {author} {\bibfnamefont {O.}~\bibnamefont {Reshef}}, \bibinfo {author} {\bibfnamefont {C.}~\bibnamefont {Liu}}, \bibinfo {author} {\bibfnamefont {A.~E.}\ \bibnamefont {Willner}},\ and\ \bibinfo {author} {\bibfnamefont {R.~W.}\ \bibnamefont {Boyd}},\ }\bibfield  {title} {\bibinfo {title} {Broadband frequency translation through time refraction in an epsilon-near-zero material},\ }\href {https://doi.org/10.1038/s41467-020-15682-2} {\bibfield  {journal} {\bibinfo  {journal} {Nature Communications}\ }\textbf {\bibinfo {volume} {11}},\ \bibinfo {pages} {2180} (\bibinfo {year} {2020})}\BibitemShut {NoStop}%
\bibitem [{\citenamefont {Masson}\ and\ \citenamefont {Asenjo-Garcia}(2020)}]{masson2020atomic}%
  \BibitemOpen
  \bibfield  {author} {\bibinfo {author} {\bibfnamefont {S.~J.}\ \bibnamefont {Masson}}\ and\ \bibinfo {author} {\bibfnamefont {A.}~\bibnamefont {Asenjo-Garcia}},\ }\bibfield  {title} {\bibinfo {title} {Atomic-waveguide quantum electrodynamics},\ }\href {https://doi.org/10.1103/PhysRevResearch.2.043213} {\bibfield  {journal} {\bibinfo  {journal} {Phys. Rev. Res.}\ }\textbf {\bibinfo {volume} {2}},\ \bibinfo {pages} {043213} (\bibinfo {year} {2020})}\BibitemShut {NoStop}%
\bibitem [{\citenamefont {Srakaew}\ \emph {et~al.}(2023)\citenamefont {Srakaew}, \citenamefont {Weckesser}, \citenamefont {Hollerith}, \citenamefont {Wei}, \citenamefont {Adler}, \citenamefont {Bloch},\ and\ \citenamefont {Zeiher}}]{srakaew2023subwavelength}%
  \BibitemOpen
  \bibfield  {author} {\bibinfo {author} {\bibfnamefont {K.}~\bibnamefont {Srakaew}}, \bibinfo {author} {\bibfnamefont {P.}~\bibnamefont {Weckesser}}, \bibinfo {author} {\bibfnamefont {S.}~\bibnamefont {Hollerith}}, \bibinfo {author} {\bibfnamefont {D.}~\bibnamefont {Wei}}, \bibinfo {author} {\bibfnamefont {D.}~\bibnamefont {Adler}}, \bibinfo {author} {\bibfnamefont {I.}~\bibnamefont {Bloch}},\ and\ \bibinfo {author} {\bibfnamefont {J.}~\bibnamefont {Zeiher}},\ }\bibfield  {title} {\bibinfo {title} {{A subwavelength atomic array switched by a single Rydberg atom}},\ }\href {https://doi.org/10.1038/s41567-023-01959-y} {\bibfield  {journal} {\bibinfo  {journal} {Nature Physics}\ }\textbf {\bibinfo {volume} {19}},\ \bibinfo {pages} {714} (\bibinfo {year} {2023})}\BibitemShut {NoStop}%
\bibitem [{\citenamefont {Nandi}\ \emph {et~al.}(2021)\citenamefont {Nandi}, \citenamefont {An},\ and\ \citenamefont {Hosseini}}]{nandi2021coherent}%
  \BibitemOpen
  \bibfield  {author} {\bibinfo {author} {\bibfnamefont {A.}~\bibnamefont {Nandi}}, \bibinfo {author} {\bibfnamefont {H.}~\bibnamefont {An}},\ and\ \bibinfo {author} {\bibfnamefont {M.}~\bibnamefont {Hosseini}},\ }\bibfield  {title} {\bibinfo {title} {Coherent atomic mirror formed by randomly distributed ions inside a crystal},\ }\href {https://doi.org/10.1364/OL.423092} {\bibfield  {journal} {\bibinfo  {journal} {Opt. Lett.}\ }\textbf {\bibinfo {volume} {46}},\ \bibinfo {pages} {1880} (\bibinfo {year} {2021})}\BibitemShut {NoStop}%
\bibitem [{\citenamefont {Rui}\ \emph {et~al.}(2020)\citenamefont {Rui}, \citenamefont {Wei}, \citenamefont {Rubio-Abadal}, \citenamefont {Hollerith}, \citenamefont {Zeiher}, \citenamefont {Stamper-Kurn}, \citenamefont {Gross},\ and\ \citenamefont {Bloch}}]{rui2020subradiant}%
  \BibitemOpen
  \bibfield  {author} {\bibinfo {author} {\bibfnamefont {J.}~\bibnamefont {Rui}}, \bibinfo {author} {\bibfnamefont {D.}~\bibnamefont {Wei}}, \bibinfo {author} {\bibfnamefont {A.}~\bibnamefont {Rubio-Abadal}}, \bibinfo {author} {\bibfnamefont {S.}~\bibnamefont {Hollerith}}, \bibinfo {author} {\bibfnamefont {J.}~\bibnamefont {Zeiher}}, \bibinfo {author} {\bibfnamefont {D.~M.}\ \bibnamefont {Stamper-Kurn}}, \bibinfo {author} {\bibfnamefont {C.}~\bibnamefont {Gross}},\ and\ \bibinfo {author} {\bibfnamefont {I.}~\bibnamefont {Bloch}},\ }\bibfield  {title} {\bibinfo {title} {A subradiant optical mirror formed by a single structured atomic layer},\ }\href {https://doi.org/10.1038/s41586-020-2463-x} {\bibfield  {journal} {\bibinfo  {journal} {Nature}\ }\textbf {\bibinfo {volume} {583}},\ \bibinfo {pages} {369} (\bibinfo {year} {2020})}\BibitemShut {NoStop}%
\bibitem [{\citenamefont {Corzo}\ \emph {et~al.}(2016)\citenamefont {Corzo}, \citenamefont {Gouraud}, \citenamefont {Chandra}, \citenamefont {Goban}, \citenamefont {Sheremet}, \citenamefont {Kupriyanov},\ and\ \citenamefont {Laurat}}]{corzo2016large}%
  \BibitemOpen
  \bibfield  {author} {\bibinfo {author} {\bibfnamefont {N.~V.}\ \bibnamefont {Corzo}}, \bibinfo {author} {\bibfnamefont {B.}~\bibnamefont {Gouraud}}, \bibinfo {author} {\bibfnamefont {A.}~\bibnamefont {Chandra}}, \bibinfo {author} {\bibfnamefont {A.}~\bibnamefont {Goban}}, \bibinfo {author} {\bibfnamefont {A.~S.}\ \bibnamefont {Sheremet}}, \bibinfo {author} {\bibfnamefont {D.~V.}\ \bibnamefont {Kupriyanov}},\ and\ \bibinfo {author} {\bibfnamefont {J.}~\bibnamefont {Laurat}},\ }\bibfield  {title} {\bibinfo {title} {{Large Bragg Reflection from One-Dimensional Chains of Trapped Atoms Near a Nanoscale Waveguide}},\ }\href {https://doi.org/10.1103/PhysRevLett.117.133603} {\bibfield  {journal} {\bibinfo  {journal} {Phys. Rev. Lett.}\ }\textbf {\bibinfo {volume} {117}},\ \bibinfo {pages} {133603} (\bibinfo {year} {2016})}\BibitemShut {NoStop}%
\bibitem [{\citenamefont {Su}\ \emph {et~al.}(2019)\citenamefont {Su}, \citenamefont {Liu}, \citenamefont {Ji}, \citenamefont {Qi}, \citenamefont {Song}, \citenamefont {Zhao}, \citenamefont {Xiao},\ and\ \citenamefont {Jia}}]{su2019observation}%
  \BibitemOpen
  \bibfield  {author} {\bibinfo {author} {\bibfnamefont {D.}~\bibnamefont {Su}}, \bibinfo {author} {\bibfnamefont {R.}~\bibnamefont {Liu}}, \bibinfo {author} {\bibfnamefont {Z.}~\bibnamefont {Ji}}, \bibinfo {author} {\bibfnamefont {X.}~\bibnamefont {Qi}}, \bibinfo {author} {\bibfnamefont {Z.}~\bibnamefont {Song}}, \bibinfo {author} {\bibfnamefont {Y.}~\bibnamefont {Zhao}}, \bibinfo {author} {\bibfnamefont {L.}~\bibnamefont {Xiao}},\ and\ \bibinfo {author} {\bibfnamefont {S.}~\bibnamefont {Jia}},\ }\bibfield  {title} {\bibinfo {title} {Observation of ladder-type electromagnetically induced transparency with atomic optical lattices near a nanofiber},\ }\href {https://doi.org/10.1088/1367-2630/ab172e} {\bibfield  {journal} {\bibinfo  {journal} {New Journal of Physics}\ }\textbf {\bibinfo {volume} {21}},\ \bibinfo {pages} {043053} (\bibinfo {year} {2019})}\BibitemShut {NoStop}%
\bibitem [{\citenamefont {Glauber}\ and\ \citenamefont {Lewenstein}(1991)}]{glauber1991quantum}%
  \BibitemOpen
  \bibfield  {author} {\bibinfo {author} {\bibfnamefont {R.~J.}\ \bibnamefont {Glauber}}\ and\ \bibinfo {author} {\bibfnamefont {M.}~\bibnamefont {Lewenstein}},\ }\bibfield  {title} {\bibinfo {title} {Quantum optics of dielectric media},\ }\href {https://doi.org/10.1103/PhysRevA.43.467} {\bibfield  {journal} {\bibinfo  {journal} {Phys. Rev. A}\ }\textbf {\bibinfo {volume} {43}},\ \bibinfo {pages} {467} (\bibinfo {year} {1991})}\BibitemShut {NoStop}%
\bibitem [{\citenamefont {Watson}\ and\ \citenamefont {Jauch}(1949)}]{watson1949phenomenological}%
  \BibitemOpen
  \bibfield  {author} {\bibinfo {author} {\bibfnamefont {K.~M.}\ \bibnamefont {Watson}}\ and\ \bibinfo {author} {\bibfnamefont {J.~M.}\ \bibnamefont {Jauch}},\ }\bibfield  {title} {\bibinfo {title} {{Phenomenological Quantum Electrodynamics. Part III. Dispersion}},\ }\href {https://doi.org/10.1103/PhysRev.75.1249} {\bibfield  {journal} {\bibinfo  {journal} {Phys. Rev.}\ }\textbf {\bibinfo {volume} {75}},\ \bibinfo {pages} {1249} (\bibinfo {year} {1949})}\BibitemShut {NoStop}%
\bibitem [{\citenamefont {Drummond}(1990)}]{drummond1990electromagnetic}%
  \BibitemOpen
  \bibfield  {author} {\bibinfo {author} {\bibfnamefont {P.~D.}\ \bibnamefont {Drummond}},\ }\bibfield  {title} {\bibinfo {title} {Electromagnetic quantization in dispersive inhomogeneous nonlinear dielectrics},\ }\href {https://doi.org/10.1103/PhysRevA.42.6845} {\bibfield  {journal} {\bibinfo  {journal} {Phys. Rev. A}\ }\textbf {\bibinfo {volume} {42}},\ \bibinfo {pages} {6845} (\bibinfo {year} {1990})}\BibitemShut {NoStop}%
\bibitem [{\citenamefont {Deutsch}\ and\ \citenamefont {Garrison}(1991)}]{deutsch1991paraxial}%
  \BibitemOpen
  \bibfield  {author} {\bibinfo {author} {\bibfnamefont {I.~H.}\ \bibnamefont {Deutsch}}\ and\ \bibinfo {author} {\bibfnamefont {J.~C.}\ \bibnamefont {Garrison}},\ }\bibfield  {title} {\bibinfo {title} {Paraxial quantum propagation},\ }\href {https://doi.org/10.1103/PhysRevA.43.2498} {\bibfield  {journal} {\bibinfo  {journal} {Phys. Rev. A}\ }\textbf {\bibinfo {volume} {43}},\ \bibinfo {pages} {2498} (\bibinfo {year} {1991})}\BibitemShut {NoStop}%
\bibitem [{\citenamefont {Raymer}(2020)}]{Raymer2020}%
  \BibitemOpen
  \bibfield  {author} {\bibinfo {author} {\bibfnamefont {M.~G.}\ \bibnamefont {Raymer}},\ }\bibfield  {title} {\bibinfo {title} {Quantum theory of light in a dispersive structured linear dielectric: a macroscopic hamiltonian tutorial treatment},\ }\href {https://doi.org/10.1080/09500340.2019.1706773} {\bibfield  {journal} {\bibinfo  {journal} {Journal of Modern Optics}\ }\textbf {\bibinfo {volume} {67}},\ \bibinfo {pages} {196} (\bibinfo {year} {2020})}\BibitemShut {NoStop}%
\bibitem [{\citenamefont {Fano}(1956)}]{PhysRev.103.1202}%
  \BibitemOpen
  \bibfield  {author} {\bibinfo {author} {\bibfnamefont {U.}~\bibnamefont {Fano}},\ }\bibfield  {title} {\bibinfo {title} {Atomic theory of electromagnetic interactions in dense materials},\ }\href {https://doi.org/10.1103/PhysRev.103.1202} {\bibfield  {journal} {\bibinfo  {journal} {Phys. Rev.}\ }\textbf {\bibinfo {volume} {103}},\ \bibinfo {pages} {1202} (\bibinfo {year} {1956})}\BibitemShut {NoStop}%
\bibitem [{\citenamefont {Hopfield}(1958)}]{PhysRev.112.1555}%
  \BibitemOpen
  \bibfield  {author} {\bibinfo {author} {\bibfnamefont {J.~J.}\ \bibnamefont {Hopfield}},\ }\bibfield  {title} {\bibinfo {title} {Theory of the contribution of excitons to the complex dielectric constant of crystals},\ }\href {https://doi.org/10.1103/PhysRev.112.1555} {\bibfield  {journal} {\bibinfo  {journal} {Phys. Rev.}\ }\textbf {\bibinfo {volume} {112}},\ \bibinfo {pages} {1555} (\bibinfo {year} {1958})}\BibitemShut {NoStop}%
\bibitem [{\citenamefont {Huttner}\ and\ \citenamefont {Barnett}(1992)}]{huttner1992quantization}%
  \BibitemOpen
  \bibfield  {author} {\bibinfo {author} {\bibfnamefont {B.}~\bibnamefont {Huttner}}\ and\ \bibinfo {author} {\bibfnamefont {S.~M.}\ \bibnamefont {Barnett}},\ }\bibfield  {title} {\bibinfo {title} {Quantization of the electromagnetic field in dielectrics},\ }\href {https://doi.org/10.1103/PhysRevA.46.4306} {\bibfield  {journal} {\bibinfo  {journal} {Phys. Rev. A}\ }\textbf {\bibinfo {volume} {46}},\ \bibinfo {pages} {4306} (\bibinfo {year} {1992})}\BibitemShut {NoStop}%
\bibitem [{\citenamefont {Suttorp}\ and\ \citenamefont {van Wonderen}(2004)}]{suttorp2004fano}%
  \BibitemOpen
  \bibfield  {author} {\bibinfo {author} {\bibfnamefont {L.~G.}\ \bibnamefont {Suttorp}}\ and\ \bibinfo {author} {\bibfnamefont {A.~J.}\ \bibnamefont {van Wonderen}},\ }\bibfield  {title} {\bibinfo {title} {Fano diagonalization of a polariton model for an inhomogeneous absorptive dielectric},\ }\href {https://doi.org/10.1209/epl/i2004-10131-8} {\bibfield  {journal} {\bibinfo  {journal} {Europhysics Letters}\ }\textbf {\bibinfo {volume} {67}},\ \bibinfo {pages} {766} (\bibinfo {year} {2004})}\BibitemShut {NoStop}%
\bibitem [{\citenamefont {Kheirandish}\ and\ \citenamefont {Soltani}(2008)}]{kheirandish2008extension}%
  \BibitemOpen
  \bibfield  {author} {\bibinfo {author} {\bibfnamefont {F.}~\bibnamefont {Kheirandish}}\ and\ \bibinfo {author} {\bibfnamefont {M.}~\bibnamefont {Soltani}},\ }\bibfield  {title} {\bibinfo {title} {{Extension of the Huttner-Barnett model to a magnetodielectric medium}},\ }\href {https://doi.org/10.1103/PhysRevA.78.012102} {\bibfield  {journal} {\bibinfo  {journal} {Phys. Rev. A}\ }\textbf {\bibinfo {volume} {78}},\ \bibinfo {pages} {012102} (\bibinfo {year} {2008})}\BibitemShut {NoStop}%
\bibitem [{\citenamefont {Bhat}\ and\ \citenamefont {Sipe}(2006)}]{bhat2006hamiltonian}%
  \BibitemOpen
  \bibfield  {author} {\bibinfo {author} {\bibfnamefont {N.~A.}\ \bibnamefont {Bhat}}\ and\ \bibinfo {author} {\bibfnamefont {J.}~\bibnamefont {Sipe}},\ }\bibfield  {title} {\bibinfo {title} {Hamiltonian treatment of the electromagnetic field in dispersive and absorptive structured media},\ }\href {https://doi.org/10.1103/PhysRevA.73.063808} {\bibfield  {journal} {\bibinfo  {journal} {Physical Review A}\ }\textbf {\bibinfo {volume} {73}},\ \bibinfo {pages} {063808} (\bibinfo {year} {2006})}\BibitemShut {NoStop}%
\bibitem [{\citenamefont {Sipe}(2009)}]{sipe2009photons}%
  \BibitemOpen
  \bibfield  {author} {\bibinfo {author} {\bibfnamefont {J.}~\bibnamefont {Sipe}},\ }\bibfield  {title} {\bibinfo {title} {Photons in dispersive dielectrics},\ }\href {https://doi.org/10.1088/1464-4258/11/11/114006} {\bibfield  {journal} {\bibinfo  {journal} {Journal of Optics A: Pure and Applied Optics}\ }\textbf {\bibinfo {volume} {11}},\ \bibinfo {pages} {114006} (\bibinfo {year} {2009})}\BibitemShut {NoStop}%
\bibitem [{\citenamefont {Gruner}\ and\ \citenamefont {Welsch}(1996)}]{Gruner96}%
  \BibitemOpen
  \bibfield  {author} {\bibinfo {author} {\bibfnamefont {T.}~\bibnamefont {Gruner}}\ and\ \bibinfo {author} {\bibfnamefont {D.-G.}\ \bibnamefont {Welsch}},\ }\bibfield  {title} {\bibinfo {title} {{Green-function approach to the radiation-field quantization for homogeneous and inhomogeneous Kramers-Kronig dielectrics}},\ }\href {https://doi.org/10.1103/PhysRevA.53.1818} {\bibfield  {journal} {\bibinfo  {journal} {Phys. Rev. A}\ }\textbf {\bibinfo {volume} {53}},\ \bibinfo {pages} {1818} (\bibinfo {year} {1996})}\BibitemShut {NoStop}%
\bibitem [{\citenamefont {Kn{\"o}ll}\ \emph {et~al.}(2001)\citenamefont {Kn{\"o}ll}, \citenamefont {Scheel},\ and\ \citenamefont {Welsch}}]{Knoll}%
  \BibitemOpen
  \bibfield  {author} {\bibinfo {author} {\bibfnamefont {L.}~\bibnamefont {Kn{\"o}ll}}, \bibinfo {author} {\bibfnamefont {S.}~\bibnamefont {Scheel}},\ and\ \bibinfo {author} {\bibfnamefont {D.-G.}\ \bibnamefont {Welsch}},\ }\href {https://books.google.com/books?id=nxMgtCNz_eUC} {\emph {\bibinfo {title} {{QED in dispersing and absorbing media}}}},\ Coherence and Statistics of Photons and Atoms,Perina, J.,Lasers and Applications Series\ (\bibinfo  {publisher} {Wiley},\ \bibinfo {year} {2001})\BibitemShut {NoStop}%
\bibitem [{\citenamefont {Claasen}\ and\ \citenamefont {Mecklenbrauker}(1982)}]{claasen1982stationary}%
  \BibitemOpen
  \bibfield  {author} {\bibinfo {author} {\bibfnamefont {T.}~\bibnamefont {Claasen}}\ and\ \bibinfo {author} {\bibfnamefont {W.}~\bibnamefont {Mecklenbrauker}},\ }\bibfield  {title} {\bibinfo {title} {On stationary linear time-varying systems},\ }\href {https://doi.org/10.1109/TCS.1982.1085130} {\bibfield  {journal} {\bibinfo  {journal} {IEEE Transactions on Circuits and Systems}\ }\textbf {\bibinfo {volume} {29}},\ \bibinfo {pages} {169} (\bibinfo {year} {1982})}\BibitemShut {NoStop}%
\bibitem [{\citenamefont {Kozek}\ and\ \citenamefont {Hlawatsch}(1991)}]{kozek1991time}%
  \BibitemOpen
  \bibfield  {author} {\bibinfo {author} {\bibfnamefont {W.}~\bibnamefont {Kozek}}\ and\ \bibinfo {author} {\bibfnamefont {F.}~\bibnamefont {Hlawatsch}},\ }\bibfield  {title} {\bibinfo {title} {{Time-frequency representation of linear time-varying systems using the Weyl symbol}},\ }in\ \href {https://ieeexplore.ieee.org/document/151896} {\emph {\bibinfo {booktitle} {1991 Sixth International Conference on Digital Processing of Signals in Communications}}}\ (\bibinfo {year} {1991})\ pp.\ \bibinfo {pages} {25--30}\BibitemShut {NoStop}%
\bibitem [{\citenamefont {Sol\'{\i}s}\ and\ \citenamefont {Engheta}(2021)}]{solis2021functional}%
  \BibitemOpen
  \bibfield  {author} {\bibinfo {author} {\bibfnamefont {D.~M.}\ \bibnamefont {Sol\'{\i}s}}\ and\ \bibinfo {author} {\bibfnamefont {N.}~\bibnamefont {Engheta}},\ }\bibfield  {title} {\bibinfo {title} {{Functional analysis of the polarization response in linear time-varying media: A generalization of the Kramers-Kronig relations}},\ }\href {https://doi.org/10.1103/PhysRevB.103.144303} {\bibfield  {journal} {\bibinfo  {journal} {Phys. Rev. B}\ }\textbf {\bibinfo {volume} {103}},\ \bibinfo {pages} {144303} (\bibinfo {year} {2021})}\BibitemShut {NoStop}%
\bibitem [{\citenamefont {Purcell}(1995)}]{purcell1995spontaneous}%
  \BibitemOpen
  \bibfield  {author} {\bibinfo {author} {\bibfnamefont {E.~M.}\ \bibnamefont {Purcell}},\ }\bibinfo {title} {Spontaneous emission probabilities at radio frequencies},\ in\ \href {https://doi.org/10.1007/978-1-4615-1963-8_40} {\emph {\bibinfo {booktitle} {{Confined Electrons and Photons: New Physics and Applications}}}},\ \bibinfo {editor} {edited by\ \bibinfo {editor} {\bibfnamefont {E.}~\bibnamefont {Burstein}}\ and\ \bibinfo {editor} {\bibfnamefont {C.}~\bibnamefont {Weisbuch}}}\ (\bibinfo  {publisher} {Springer US},\ \bibinfo {address} {Boston, MA},\ \bibinfo {year} {1995})\ pp.\ \bibinfo {pages} {839--839}\BibitemShut {NoStop}%
\bibitem [{\citenamefont {Dirac}\ and\ \citenamefont {Bohr}(1927)}]{dirac1927quantum}%
  \BibitemOpen
  \bibfield  {author} {\bibinfo {author} {\bibfnamefont {P.~A.~M.}\ \bibnamefont {Dirac}}\ and\ \bibinfo {author} {\bibfnamefont {N.~H.~D.}\ \bibnamefont {Bohr}},\ }\bibfield  {title} {\bibinfo {title} {The quantum theory of the emission and absorption of radiation},\ }\href {https://doi.org/10.1098/rspa.1927.0039} {\bibfield  {journal} {\bibinfo  {journal} {Proceedings of the Royal Society of London. Series A, Containing Papers of a Mathematical and Physical Character}\ }\textbf {\bibinfo {volume} {114}},\ \bibinfo {pages} {243} (\bibinfo {year} {1927})}\BibitemShut {NoStop}%
\bibitem [{\citenamefont {Milonni}\ and\ \citenamefont {Smith}(1975)}]{PhysRevA.11.814}%
  \BibitemOpen
  \bibfield  {author} {\bibinfo {author} {\bibfnamefont {P.~W.}\ \bibnamefont {Milonni}}\ and\ \bibinfo {author} {\bibfnamefont {W.~A.}\ \bibnamefont {Smith}},\ }\bibfield  {title} {\bibinfo {title} {Radiation reaction and vacuum fluctuations in spontaneous emission},\ }\href {https://doi.org/10.1103/PhysRevA.11.814} {\bibfield  {journal} {\bibinfo  {journal} {Phys. Rev. A}\ }\textbf {\bibinfo {volume} {11}},\ \bibinfo {pages} {814} (\bibinfo {year} {1975})}\BibitemShut {NoStop}%
\bibitem [{\citenamefont {Yao}\ and\ \citenamefont {Liu}(2014)}]{yao2014plasmonic}%
  \BibitemOpen
  \bibfield  {author} {\bibinfo {author} {\bibfnamefont {K.}~\bibnamefont {Yao}}\ and\ \bibinfo {author} {\bibfnamefont {Y.}~\bibnamefont {Liu}},\ }\href {https://doi.org/doi:10.1515/ntrev-2012-0071} {\bibfield  {journal} {\bibinfo  {journal} {Nanotechnology Reviews}\ }\textbf {\bibinfo {volume} {3}},\ \bibinfo {pages} {177} (\bibinfo {year} {2014})}\BibitemShut {NoStop}%
\bibitem [{\citenamefont {Boltasseva}\ and\ \citenamefont {Atwater}(2011)}]{boltasseva2011low}%
  \BibitemOpen
  \bibfield  {author} {\bibinfo {author} {\bibfnamefont {A.}~\bibnamefont {Boltasseva}}\ and\ \bibinfo {author} {\bibfnamefont {H.~A.}\ \bibnamefont {Atwater}},\ }\bibfield  {title} {\bibinfo {title} {Low-loss plasmonic metamaterials},\ }\href {https://doi.org/10.1126/science.1198258} {\bibfield  {journal} {\bibinfo  {journal} {Science}\ }\textbf {\bibinfo {volume} {331}},\ \bibinfo {pages} {290} (\bibinfo {year} {2011})}\BibitemShut {NoStop}%
\bibitem [{\citenamefont {Barnett}\ \emph {et~al.}(1992)\citenamefont {Barnett}, \citenamefont {Huttner},\ and\ \citenamefont {Loudon}}]{barnett1992spontaneous}%
  \BibitemOpen
  \bibfield  {author} {\bibinfo {author} {\bibfnamefont {S.~M.}\ \bibnamefont {Barnett}}, \bibinfo {author} {\bibfnamefont {B.}~\bibnamefont {Huttner}},\ and\ \bibinfo {author} {\bibfnamefont {R.}~\bibnamefont {Loudon}},\ }\bibfield  {title} {\bibinfo {title} {Spontaneous emission in absorbing dielectric media},\ }\href {https://doi.org/10.1103/PhysRevLett.68.3698} {\bibfield  {journal} {\bibinfo  {journal} {Phys. Rev. Lett.}\ }\textbf {\bibinfo {volume} {68}},\ \bibinfo {pages} {3698} (\bibinfo {year} {1992})}\BibitemShut {NoStop}%
\bibitem [{\citenamefont {Milonni}(1995)}]{milonni1995field}%
  \BibitemOpen
  \bibfield  {author} {\bibinfo {author} {\bibfnamefont {P.}~\bibnamefont {Milonni}},\ }\bibfield  {title} {\bibinfo {title} {Field quantization and radiative processes in dispersive dielectric media},\ }\href {https://doi.org/10.1080/09500349514551741} {\bibfield  {journal} {\bibinfo  {journal} {Journal of Modern Optics}\ }\textbf {\bibinfo {volume} {42}},\ \bibinfo {pages} {1991} (\bibinfo {year} {1995})}\BibitemShut {NoStop}%
\bibitem [{\citenamefont {Aubret}\ \emph {et~al.}(2019)\citenamefont {Aubret}, \citenamefont {Orrit},\ and\ \citenamefont {Kulzer}}]{aubret2019understanding}%
  \BibitemOpen
  \bibfield  {author} {\bibinfo {author} {\bibfnamefont {A.}~\bibnamefont {Aubret}}, \bibinfo {author} {\bibfnamefont {M.}~\bibnamefont {Orrit}},\ and\ \bibinfo {author} {\bibfnamefont {F.}~\bibnamefont {Kulzer}},\ }\bibfield  {title} {\bibinfo {title} {{Understanding Local-Field Correction Factors in the Framework of the Onsager-B{\"o}ttcher Model}},\ }\href {https://doi.org/https://doi.org/10.1002/cphc.201800923} {\bibfield  {journal} {\bibinfo  {journal} {ChemPhysChem}\ }\textbf {\bibinfo {volume} {20}},\ \bibinfo {pages} {345} (\bibinfo {year} {2019})}\BibitemShut {NoStop}%
\bibitem [{\citenamefont {Mie}(1908)}]{mie1976contributions}%
  \BibitemOpen
  \bibfield  {author} {\bibinfo {author} {\bibfnamefont {G.}~\bibnamefont {Mie}},\ }\bibfield  {title} {\bibinfo {title} {Beitr\"{a}ge zur optik tr\"{u}ber medien, speziell kolloidaler metall\"{o}sungen},\ }\href {https://doi.org/https://doi.org/10.1002/andp.19083300302} {\bibfield  {journal} {\bibinfo  {journal} {Annalen der Physik}\ }\textbf {\bibinfo {volume} {330}},\ \bibinfo {pages} {377} (\bibinfo {year} {1908})}\BibitemShut {NoStop}%
\bibitem [{\citenamefont {Aspnes}(1982)}]{aspnes1982local}%
  \BibitemOpen
  \bibfield  {author} {\bibinfo {author} {\bibfnamefont {D.~E.}\ \bibnamefont {Aspnes}},\ }\bibfield  {title} {\bibinfo {title} {{Local-field effects and effective-medium theory: A microscopic perspective}},\ }\href {https://doi.org/10.1119/1.12734} {\bibfield  {journal} {\bibinfo  {journal} {American Journal of Physics}\ }\textbf {\bibinfo {volume} {50}},\ \bibinfo {pages} {704} (\bibinfo {year} {1982})}\BibitemShut {NoStop}%
\bibitem [{\citenamefont {Lorentz}(1916)}]{lorentz1916theory}%
  \BibitemOpen
  \bibfield  {author} {\bibinfo {author} {\bibfnamefont {H.}~\bibnamefont {Lorentz}},\ }\href {https://books.google.com/books?id=dPRQAQAAMAAJ} {\emph {\bibinfo {title} {The Theory of Electrons and Its Applications to the Phenomena of Light and Radiant Heat}}}\ (\bibinfo  {publisher} {B.G. Teubner},\ \bibinfo {year} {1916})\BibitemShut {NoStop}%
\bibitem [{\citenamefont {B\"ottcher}\ \emph {et~al.}(1974)\citenamefont {B\"ottcher}, \citenamefont {van Belle}, \citenamefont {Bordewijk}, \citenamefont {Rip},\ and\ \citenamefont {Yue}}]{bottcher1974theory}%
  \BibitemOpen
  \bibfield  {author} {\bibinfo {author} {\bibfnamefont {C.~J.~F.}\ \bibnamefont {B\"ottcher}}, \bibinfo {author} {\bibfnamefont {O.~C.}\ \bibnamefont {van Belle}}, \bibinfo {author} {\bibfnamefont {P.}~\bibnamefont {Bordewijk}}, \bibinfo {author} {\bibfnamefont {A.}~\bibnamefont {Rip}},\ and\ \bibinfo {author} {\bibfnamefont {D.~D.}\ \bibnamefont {Yue}},\ }\bibfield  {title} {\bibinfo {title} {Theory of electric polarization},\ }\href {https://doi.org/10.1149/1.2402382} {\bibfield  {journal} {\bibinfo  {journal} {Journal of The Electrochemical Society}\ }\textbf {\bibinfo {volume} {121}},\ \bibinfo {pages} {211Ca} (\bibinfo {year} {1974})}\BibitemShut {NoStop}%
\bibitem [{\citenamefont {Onsager}(1936)}]{onsager1936electric}%
  \BibitemOpen
  \bibfield  {author} {\bibinfo {author} {\bibfnamefont {L.}~\bibnamefont {Onsager}},\ }\bibfield  {title} {\bibinfo {title} {Electric moments of molecules in liquids},\ }\href {https://doi.org/10.1021/ja01299a050} {\bibfield  {journal} {\bibinfo  {journal} {Journal of the American Chemical Society}\ }\textbf {\bibinfo {volume} {58}},\ \bibinfo {pages} {1486} (\bibinfo {year} {1936})}\BibitemShut {NoStop}%
\bibitem [{\citenamefont {Scheel}\ \emph {et~al.}(1999)\citenamefont {Scheel}, \citenamefont {Kn{\"o}ll},\ and\ \citenamefont {Welsch}}]{scheel1999spontaneous}%
  \BibitemOpen
  \bibfield  {author} {\bibinfo {author} {\bibfnamefont {S.}~\bibnamefont {Scheel}}, \bibinfo {author} {\bibfnamefont {L.}~\bibnamefont {Kn{\"o}ll}},\ and\ \bibinfo {author} {\bibfnamefont {D.-G.}\ \bibnamefont {Welsch}},\ }\bibfield  {title} {\bibinfo {title} {Spontaneous decay of an excited atom in an absorbing dielectric},\ }\href {https://doi.org/10.1103/PhysRevA.60.4094} {\bibfield  {journal} {\bibinfo  {journal} {Physical Review A}\ }\textbf {\bibinfo {volume} {60}},\ \bibinfo {pages} {4094} (\bibinfo {year} {1999})}\BibitemShut {NoStop}%
\bibitem [{\citenamefont {Nation}\ \emph {et~al.}(2012)\citenamefont {Nation}, \citenamefont {Johansson}, \citenamefont {Blencowe},\ and\ \citenamefont {Nori}}]{Nation2012}%
  \BibitemOpen
  \bibfield  {author} {\bibinfo {author} {\bibfnamefont {P.~D.}\ \bibnamefont {Nation}}, \bibinfo {author} {\bibfnamefont {J.~R.}\ \bibnamefont {Johansson}}, \bibinfo {author} {\bibfnamefont {M.~P.}\ \bibnamefont {Blencowe}},\ and\ \bibinfo {author} {\bibfnamefont {F.}~\bibnamefont {Nori}},\ }\bibfield  {title} {\bibinfo {title} {Colloquium: Stimulating uncertainty: Amplifying the quantum vacuum with superconducting circuits},\ }\href {https://doi.org/10.1103/RevModPhys.84.1} {\bibfield  {journal} {\bibinfo  {journal} {Rev. Mod. Phys.}\ }\textbf {\bibinfo {volume} {84}},\ \bibinfo {pages} {1} (\bibinfo {year} {2012})}\BibitemShut {NoStop}%
\bibitem [{\citenamefont {Dodonov}(2010)}]{Dodonov2010}%
  \BibitemOpen
  \bibfield  {author} {\bibinfo {author} {\bibfnamefont {V.~V.}\ \bibnamefont {Dodonov}},\ }\bibfield  {title} {\bibinfo {title} {Current status of the dynamical casimir effect},\ }\href {https://doi.org/10.1088/0031-8949/82/03/038105} {\bibfield  {journal} {\bibinfo  {journal} {Physica Scripta}\ }\textbf {\bibinfo {volume} {82}},\ \bibinfo {pages} {038105} (\bibinfo {year} {2010})}\BibitemShut {NoStop}%
\bibitem [{\citenamefont {Hui}\ \emph {et~al.}(2023)\citenamefont {Hui}, \citenamefont {Alqattan}, \citenamefont {Zhang}, \citenamefont {Pervak}, \citenamefont {Chowdhury},\ and\ \citenamefont {Hassan}}]{Alqattan_ultrafast}%
  \BibitemOpen
  \bibfield  {author} {\bibinfo {author} {\bibfnamefont {D.}~\bibnamefont {Hui}}, \bibinfo {author} {\bibfnamefont {H.}~\bibnamefont {Alqattan}}, \bibinfo {author} {\bibfnamefont {S.}~\bibnamefont {Zhang}}, \bibinfo {author} {\bibfnamefont {V.}~\bibnamefont {Pervak}}, \bibinfo {author} {\bibfnamefont {E.}~\bibnamefont {Chowdhury}},\ and\ \bibinfo {author} {\bibfnamefont {M.~T.}\ \bibnamefont {Hassan}},\ }\bibfield  {title} {\bibinfo {title} {Ultrafast optical switching and data encoding on synthesized light fields},\ }\href {https://doi.org/10.1126/sciadv.adf1015} {\bibfield  {journal} {\bibinfo  {journal} {Science Advances}\ }\textbf {\bibinfo {volume} {9}},\ \bibinfo {pages} {eadf1015} (\bibinfo {year} {2023})}\BibitemShut {NoStop}%
\bibitem [{\citenamefont {Schultze}\ \emph {et~al.}(2012)\citenamefont {Schultze}, \citenamefont {Bothschafter}, \citenamefont {Sommer}, \citenamefont {Holzner}, \citenamefont {Schweinberger}, \citenamefont {Fiess}, \citenamefont {Hofstetter}, \citenamefont {Kienberger}, \citenamefont {Apalkov}, \citenamefont {Yakovlev}, \citenamefont {Stockman},\ and\ \citenamefont {Krausz}}]{Schultze_Controlling_dielectric}%
  \BibitemOpen
  \bibfield  {author} {\bibinfo {author} {\bibfnamefont {M.}~\bibnamefont {Schultze}}, \bibinfo {author} {\bibfnamefont {E.~M.}\ \bibnamefont {Bothschafter}}, \bibinfo {author} {\bibfnamefont {A.}~\bibnamefont {Sommer}}, \bibinfo {author} {\bibfnamefont {S.}~\bibnamefont {Holzner}}, \bibinfo {author} {\bibfnamefont {W.}~\bibnamefont {Schweinberger}}, \bibinfo {author} {\bibfnamefont {M.}~\bibnamefont {Fiess}}, \bibinfo {author} {\bibfnamefont {M.}~\bibnamefont {Hofstetter}}, \bibinfo {author} {\bibfnamefont {R.}~\bibnamefont {Kienberger}}, \bibinfo {author} {\bibfnamefont {V.}~\bibnamefont {Apalkov}}, \bibinfo {author} {\bibfnamefont {V.~S.}\ \bibnamefont {Yakovlev}}, \bibinfo {author} {\bibfnamefont {M.~I.}\ \bibnamefont {Stockman}},\ and\ \bibinfo {author} {\bibfnamefont {F.}~\bibnamefont {Krausz}},\ }\bibfield  {title} {\bibinfo {title} {Controlling dielectrics with the electric field of light},\ }\href {https://doi.org/10.1038/nature11720} {\bibfield  {journal} {\bibinfo  {journal} {Nature}\ }\textbf
  {\bibinfo {volume} {493}},\ \bibinfo {pages} {75} (\bibinfo {year} {2012})}\BibitemShut {NoStop}%
\bibitem [{\citenamefont {Schiffrin}\ \emph {et~al.}(2012)\citenamefont {Schiffrin}, \citenamefont {Paasch-Colberg}, \citenamefont {Karpowicz}, \citenamefont {Apalkov}, \citenamefont {Gerster}, \citenamefont {M\"{u}hlbrandt}, \citenamefont {Korbman}, \citenamefont {Reichert}, \citenamefont {Schultze}, \citenamefont {Holzner}, \citenamefont {Barth}, \citenamefont {Kienberger}, \citenamefont {Ernstorfer}, \citenamefont {Yakovlev}, \citenamefont {Stockman},\ and\ \citenamefont {Krausz}}]{Schiffrin_optical_field_induced}%
  \BibitemOpen
  \bibfield  {author} {\bibinfo {author} {\bibfnamefont {A.}~\bibnamefont {Schiffrin}}, \bibinfo {author} {\bibfnamefont {T.}~\bibnamefont {Paasch-Colberg}}, \bibinfo {author} {\bibfnamefont {N.}~\bibnamefont {Karpowicz}}, \bibinfo {author} {\bibfnamefont {V.}~\bibnamefont {Apalkov}}, \bibinfo {author} {\bibfnamefont {D.}~\bibnamefont {Gerster}}, \bibinfo {author} {\bibfnamefont {S.}~\bibnamefont {M\"{u}hlbrandt}}, \bibinfo {author} {\bibfnamefont {M.}~\bibnamefont {Korbman}}, \bibinfo {author} {\bibfnamefont {J.}~\bibnamefont {Reichert}}, \bibinfo {author} {\bibfnamefont {M.}~\bibnamefont {Schultze}}, \bibinfo {author} {\bibfnamefont {S.}~\bibnamefont {Holzner}}, \bibinfo {author} {\bibfnamefont {J.~V.}\ \bibnamefont {Barth}}, \bibinfo {author} {\bibfnamefont {R.}~\bibnamefont {Kienberger}}, \bibinfo {author} {\bibfnamefont {R.}~\bibnamefont {Ernstorfer}}, \bibinfo {author} {\bibfnamefont {V.~S.}\ \bibnamefont {Yakovlev}}, \bibinfo {author} {\bibfnamefont {M.~I.}\ \bibnamefont {Stockman}},\ and\ \bibinfo
  {author} {\bibfnamefont {F.}~\bibnamefont {Krausz}},\ }\bibfield  {title} {\bibinfo {title} {Optical-field-induced current in dielectrics},\ }\href {https://doi.org/10.1038/nature11567} {\bibfield  {journal} {\bibinfo  {journal} {Nature}\ }\textbf {\bibinfo {volume} {493}},\ \bibinfo {pages} {70} (\bibinfo {year} {2012})}\BibitemShut {NoStop}%
\end{thebibliography}%

\end{document}